\documentclass[preprint,12pt]{elsarticle}

\usepackage{amssymb}
\usepackage{amsmath}
\usepackage{physics}
\usepackage{listings}
\newcommand{\kket}[1]{|#1\rangle\!\rangle}
\usepackage{bbold}
\usepackage{booktabs}
\usepackage{array}
\usepackage{comment}
\usepackage{underscore}     
\usepackage{bm}

\usepackage{xspace}
\newcommand{\SIsecxxz}{S1\xspace}                
\newcommand{\SIsectwolevel}{S2\xspace}            
\newcommand{\SIsectwolevelmodel}{S2.1\xspace}     
\newcommand{\SIsectwolevelpaths}{S2.2\xspace}     
\newcommand{\SIsectwolevelreadout}{S2.3\xspace}   
\newcommand{\SIsecderivation}{S3\xspace}          
\newcommand{\SIsecscaling}{S5\xspace}             
\newcommand{\SIsecbackends}{S6\xspace}            
\newcommand{\SIsecgrid}{S7\xspace}                
\newcommand{\SIsecresolved}{S8\xspace}            
\newcommand{\SItabscenarios}{S1\xspace}           
\newcommand{\SItabresolved}{S4\xspace}            

\journal{Computer Physics Communications}

\begin{document}

\begin{frontmatter}



\title{QuDPy-FDGF: A Python-Based Tool for Computing Ultrafast Nonlinear Optical Responses Using Frequency Domain Green's Functions}


\author[UdeM,IC]{Mathieu~Desmarais} 

\author[UdeM,IC]{Sim\'on~Paiva-Ortega}

\author[UdeM]{F\'elix Brunelle}

\author[IC,UH]{Eric~R.~Bittner}

\author[UdeM,IC]{Hao~Li}

\author[UdeM,IC]{Carlos~Silva-Acu\~{n}a}

\affiliation[UdeM]{organization={D\'epartement de physique, Universit\'e de Montr\'eal},Department and Organization
            addressline={1375 Avenue Th\'er\`ese-Lavoie-Roux}, 
            city={Montr\'eal},
            postcode={H2V~0B3}, 
            state={Qu\'ebec},
            country={Canada}}
\affiliation[IC]{organization={Institut Courtois, Universit\'e de Montr\'eal},Department and Organization
            addressline={1375 Avenue Th\'er\`ese-Lavoie-Roux}, 
            city={Montr\'eal},
            postcode={H2V~0B3}, 
            state={Qu\'ebec},
            country={Canada}}
\affiliation[UH]{organization={Department of Physics, University of Houston},
            addressline={3507 Cullen Boulevard}, 
            city={Houston},
            postcode={77204-5005}, 
            state={Texas},
            country={United~States}}

\begin{abstract}

We present \textbf{QuDPy-FDGF}, an open-source Python framework that evaluates Liouville-space pathways as hybrid chains of time propagators and frequency-domain Green's functions. Frequency-resolved intervals are computed directly through shifted linear systems, without time propagation followed by a Fourier transform, while other intervals may remain in the time domain. A model-agnostic block interface separates the physical model from pathway construction, numerical propagation, and detection. The released implementation provides an explicit dense Liouville-space backend for reference calculations and a sector-resolved sparse backend using matrix-free generator actions, Krylov time propagation, and generalized minimal residual (GMRES) resolvent solves. Its observable layer reuses propagated pathway responses to evaluate operator expectation values, action-detected readout interactions, instantaneous jump rates, and integrated mean jump counts.
The framework is assessed on an analytically solvable open two-level system and on two examples of increasing complexity: the complete third-order rephasing response, with its eight sequences, of a thermally initialized anisotropic Heisenberg (XXZ) spin chain with unequal excitation sectors, and two coupled dissipative anharmonic modes whose spectra depend on the detected observable, with the fifth-order double-quantum response of their closed limit. The calculations reproduce the analytical two-level response, check the matrix-free sparse evaluation against the dense assembly of the generator, and recover the harmonic-limit cancellation of the fifth-order signal when the required three-excitation manifold is retained. QuDPy-FDGF therefore provides a unified framework for pathway- and observable-resolved nonlinear spectroscopy in finite or truncated Hilbert spaces with unitary or Markovian Gorini--Kossakowski--Sudarshan--Lindblad (GKSL) dynamics.
\end{abstract}

\begin{highlights}
\item Nonlinear spectra from shifted resolvent solves: no time grid, no FFT
\item Model-agnostic block interface with dense and matrix-free sector backends
\item One propagation gives polarization, action-detected and jump-count signals
\item Exact integrated fluorescence, checked on analytic and harmonic limits
\item Open-source Python code, tested up to fifth-order double-quantum spectra
\end{highlights}

\begin{keyword}
Nonlinear spectroscopy \sep Liouville pathways \sep Green's functions \sep Open quantum systems \sep Lindblad equation \sep Action detection \sep Python



\end{keyword}

\end{frontmatter}

\section*{PROGRAM SUMMARY}
\begin{small}
\noindent
{\em Program title:} QuDPy-FDGF\\
{\em CPC Library link to program files:} (to be added by Technical Editor)\\
{\em Developer's repository link:} \texttt{https://github.com/SilvaScience/QuDPy-FDGF}\\
{\em Licensing provisions:} GNU General Public License 3 (GPL)\\
{\em Programming language:} Python (3.13 or later); requires NumPy and SciPy, and Matplotlib for the plotting layer; QuTiP and UFSS are optional and used by the examples\\
{\em Supplementary material:} Supplementary PDF (S1: XXZ spin chain and its bimagnon, S2: analysis of the open two-level model of the quickstart, S3: element-by-element derivation of its response, S4: the six scenarios of the Perspective, S5: cost versus Hilbert-space dimension, S6: agreement of the dense and sparse backends, S7: evaluation of a grid, S8: resolved maps at equal resolution); API reference manual (PDF); example, analysis and benchmark notebooks and test suite in the repository\\[0.5ex]
{\em Nature of problem:}\\
Multidimensional nonlinear spectroscopies, such as two-dimensional electronic, vibrational and terahertz spectroscopy and their action-detected variants, measure signals that are sums of Liouville pathways: ordered chains of light--matter interactions separated by the free evolution of a system coupled to its environment. Computing them from a microscopic model requires an open-system generator (unitary or Markovian, of Gorini--Kossakowski--Sudarshan--Lindblad (GKSL) form), a phase-matched set of pathways, initial states that may be thermal and spread over several excitation sectors, and a detection observable that may be a polarization, a population read out by an additional interaction, or a count of emitted photons. Time-domain codes obtain frequency axes by sampling a time trace and applying a Fourier transform, which ties the spectral resolution to the propagation window and makes narrow lines and high orders costly. A tool is needed that evaluates arbitrary pathways and observables for generic models directly in the frequency domain.\\[0.5ex]
{\em Solution method:}\\
QuDPy-FDGF represents a signal as an ordered chain of interaction superoperators, time propagators $\exp(\mathcal{L}t)$, frequency-domain resolvents $[(\eta-i\omega)\mathbb{1}-\mathcal{L}]^{-1}$ and a detection functional. For each interval the user chooses time or frequency evaluation. Frequency intervals are solved as shifted linear systems, so no time trace is Fourier transformed and the spectral grid is independent of any propagation window. The physical model is supplied through a block contract (sectors, Hamiltonian, transition and collapse operators, initial state). A dense Liouville backend serves as reference; a sector-resolved sparse backend applies the generator block by block, with Krylov propagation in time and generalized minimal residual (GMRES) iterations in frequency. Propagated pathway states are reused to evaluate polarization, action-detected populations, instantaneous jump rates and integrated jump counts, the latter exactly through a Heisenberg-picture detector. Pathways are written as lists of interaction labels, generated with UFSS, or enumerated from the phase-matching signature. A grid is evaluated by cutting each chain between two scanned intervals and contracting the two parts, which takes about $N_1+N_2$ instead of $2N_1N_2$ shifted solves for two axes.\\[0.5ex]
{\em Additional comments including restrictions and unusual features:}\\
Restrictions: the Hilbert space must be finite or truncated; the Hamiltonian and the generator are time independent and Markovian of GKSL form (no hierarchical equations of motion, no Redfield generators that are not of GKSL form); interactions are impulsive and treated in the rotating-wave approximation; the reference state must be stationary, which the code checks (warning), and the stationary state of the generator can be computed and imposed when it is unique; the sparse backend still stores density vectors of size $D^2$, where $D$ is the Hilbert-space dimension; integrated fluorescence needs a unique stationary state on an infinite window; a calculation contains only the pathways supplied (written by hand, generated with UFSS, or enumerated from a signature); a model that conserves no excitation number is a single sector, limited to about a dozen spins on a laptop. The number of GMRES iterations grows as the broadening $\eta$ decreases when no physical dissipation bounds it; when the generator is diagonal in the eigenbasis the shifted systems are solved by division, otherwise a diagonal preconditioner reduces the number of iterations. Unusual features: $\eta$ is both a numerical parameter and a uniform Lorentzian broadening; the readout interaction of action detection belongs to the observable, not to the pathway. The README gives the input and output of a complete test run.\\[0.5ex]
{\em References:} \cite{shah2023qudpy,rose2021ufss}
\end{small}

\section{Introduction}

Multidimensional coherent spectroscopy spreads a nonlinear optical response over several frequency axes, so that overlapping resonances, couplings between transitions, and energy-transfer pathways that are hidden in linear spectra become visible \cite{mukamel1995principles,abramavicius2009excitonreview,oliver2018multidimensional}. The method now extends from electronic and vibrational transitions to terahertz magnetic resonance \cite{lu2017thz2d}, and it is increasingly combined with indirect readouts, such as fluorescence detection, which change which pathways contribute \cite{schroeter2018fluorescence,kuhn2020interpreting}. Double-quantum coherence spectra, already available at third order, correlate two-exciton states with their one-exciton constituents and measure exciton--exciton interactions directly \cite{abramavicius2008dqc,kim2009dqc}; fifth-order responses go one step further and reach three-exciton states \cite{roslyak2013multiexcitons}. Assigning such spectra requires calculations from a microscopic model, because the number of overlapping contributions grows quickly with the order and with the number of states.

A calculation of this kind has four parts that must stay consistent. The dynamics of the system coupled to its environment, taken here as Markovian of Gorini--Kossakowski--Sudarshan--Lindblad (GKSL) form \cite{gorini1976gksl,lindblad1976gksl,manzano2020lindblad}, damp coherences and transfer populations. The interactions with the fields act in a fixed order on the ket or the bra side of the density operator, and a Liouville pathway is defined by this order. The phase-matched component of interest (rephasing, non-rephasing, double-quantum, \ldots) selects the pathways that contribute. Finally, a detection observable turns the final state into a number. Thermally populated initial states, Hilbert spaces organized in excitation sectors of different sizes, and high orders multiply both the bookkeeping and the cost.

In most codes the frequency axes are obtained by propagating the response on a grid of delays and Fourier transforming it. The spectral resolution is then tied to the length and the step of the time window, which have to be chosen for each model and whose cost grows as the lines narrow. A causal Green's function of the Liouvillian, its resolvent, gives the same quantity at any requested frequency by solving one shifted linear system. No truncation or leakage from a finite window enters the result at a requested frequency: it is the response of the retained model, with a broadening $\eta$ that regularizes the solve, up to the tolerance of that solve (the assumptions of the model are listed in Sec.~\ref{sec:scope}). The frequencies need not form a uniform grid, and the cost depends on the number of frequencies requested and on the conditioning of the system, not on the length of a time window; delays that are not transformed, such as population waiting times, remain in the time domain. A resolved map of narrow lines still has more points, and the cost of the evaluation is discussed in Sec.~\ref{sec:performance}.

\begin{sloppypar}
Several packages address parts of this problem. QuDPy, from the same group, computes Liouville-space pathways of open systems by time propagation and transforms the sampled signal \cite{shah2023qudpy,shah2022qudpysoftware}; the present work follows its pathway conventions. UFSS is complementary: it generates and selects double-sided Feynman diagrams automatically, up to high order and including pulse overlap, and its UF$^2$ method evaluates the signals of closed and Markovian open systems (Lindblad, Redfield) with finite pulses from a diagonalization of the Liouvillian propagator \cite{rose2019uf2,rose2021uf2open,rose2021ufss,rose2025ufsssoftware}; its diagrams can be translated into the pathways used here. NISE integrates the Schr\"odinger equation along trajectories of fluctuating Hamiltonians to simulate vibrational and electronic spectra \cite{jansen2021computational}; SPECTRON and its quantum-chemistry interface compute nonlinear signals of molecules and aggregates from sum-over-states and nonlinear-exciton formulations \cite{segatta2021ispectron}, and direct nonlinear exciton propagation has been applied, for instance, to the OH stretching band of liquid water \cite{falvo2009water}; and QuTiP provides master-equation solvers for open systems \cite{lambert2026qutip5}. Sum-over-states expressions are frequency-domain formulas that rely on a diagonalization: of the system Hamiltonian, with phenomenological dephasing rates, in the usual formulation, or of the Liouvillian itself, as in UF$^2$; the cost of the diagonalization then limits the Liouville-space dimension (Sec.~\ref{sec:perf-resolved}). QuDPy and UFSS already describe population-based detection by an additional interaction, for fluorescence or photocurrent readout \cite{shah2023qudpy,rose2021ufss}. 
\end{sloppypar}

What is not available in a single tool is the combination of the following. First, the pathway set, the open-system generator and the detected observable are specified independently, so that one propagation serves a coherent polarization, a projected population read out by an additional interaction, and an integrated photon count. Second, the model is organized in excitation sectors of unequal sizes connected by rectangular transition blocks, and the reference state may populate several of them, so that the pathways that only a thermal state activates are propagated like the others. Third, the frequency-resolved intervals are evaluated at the requested frequencies by shifted linear solves with a generator of GKSL form, so that no propagation window has to be chosen and the Liouvillian does not have to be diagonalized; the dense backend serves small models and the matrix-free sparse backend is checked against it. Fourth, the pathway set can be enumerated from the phase-matching signature, which keeps the sequences that relaxation or a thermal state make non-zero. We do not claim that the frequency domain is faster than the time domain in general. This matters for a recurrent problem in the interpretation of multidimensional spectra: the same dynamics read by different detectors gives different apparent lines \cite{paivaortega2026detection}, and assigning multi-exciton features requires pathways of third and fifth order from a microscopic model.

QuDPy-FDGF is an open-source, model-agnostic Python framework to compute spectral responses built for this combination. It represents a signal as an ordered chain of light--matter interactions, time propagators, frequency-domain resolvents, and a separately specified observable, and its two interchangeable backends are a dense Liouville backend and a matrix-free sector backend that uses Krylov actions in time and generalized minimal residual (GMRES) \cite{saad1986gmres} shifted solves in frequency. Pathway sets may be written by hand, generated with UFSS, or enumerated from the phase-matching signature. Recently developed PULSUS \cite{PULSUS2026} is a frequency domain Green's function method focusing to incorporate the counter rotating wave terms and finite Gaussian pulses into spectral response, rather than facilitating a broad range of spectral simulation.

We test the code on an open two-level system with a closed-form response and on two examples. Example~1 is a thermally initialized XXZ spin chain with unequal excitation sectors, whose complete third-order rephasing response has eight sequences, five of which only the thermal state activates. Example~2 is a model of two coupled dissipative modes taken from a published study of detection-dependent spectra \cite{paivaortega2026detection}; its Liouville diagrams are compared one by one with the data deposited with that study and with the released QuDPy, both from the same group, and its closed limit gives a fifth-order double-quantum response. The cost of the evaluation is compared with the released QuDPy in Sec.~\ref{sec:performance}, together with the limits of this comparison.

The paper is organized as follows. Section~\ref{sec:theory} summarizes the formalism, Sec.~\ref{sec:architecture} describes the software architecture and Sec.~\ref{sec:usage} walks through a complete calculation. Section~\ref{sec:examples} presents two examples, Sec.~\ref{sec:performance} the validation and the cost of the evaluation, and Secs.~\ref{sec:discussion} and~\ref{sec:conclusions} the limits and the conclusions. The appendices detail the sector-resolved realization and the analytical benchmark; the supplementary material and a reference manual document the examples, the cost measurements and the programming interface.

\section{Theoretical background and description of the algorithm}
\label{sec:theory}

Throughout this section we set $\hbar=1$, so energies, angular frequencies, and dissipative rates are expressed in the same units. Factors of $\hbar$ can be restored by replacing an energy argument by $\hbar\omega$ and retaining the corresponding factors in the light--matter interaction.

\subsection{Semiclassical light--matter Hamiltonian}

The material system is treated quantum mechanically and the applied field classically. In the electric-dipole approximation, the Hamiltonian is
\begin{equation}
    \hat{H}(t)=\hat{H}_0+\hat{V}(t),
    \qquad
    \hat{V}(t)=-\hat{\boldsymbol{\mu}}\cdot\boldsymbol{E}(\boldsymbol{r},t),
    \label{eq:hamiltonian-semiclassical}
\end{equation}
where $\hat{H}_0$ generates the field-free dynamics and $\hat{\boldsymbol{\mu}}=\sum_a q_a\boldsymbol{r}_a$ is the material electric-dipole operator. Equation~\eqref{eq:hamiltonian-semiclassical} is the leading term of the multipole expansion and is accurate when the characteristic system size $a$ is small compared with the optical wavelength,
\begin{equation}
    ka=\frac{2\pi a}{\lambda}\ll 1.
    \label{eq:dipole_app}
\end{equation}
The field can then be evaluated at a common position within the material subsystem. Terms beyond the dipole approximation, namely the electric-quadrupole and magnetic-dipole corrections to the electric coupling and the spatial variation of the field, become relevant for spatially extended systems, dipole-forbidden transitions, or chirality-sensitive measurements, and are outside the present implementation \cite{cohentannoudji1998atomphoton,abramavicius2009excitonreview}. The pathway algebra below uses only a transition operator coupled linearly to a classical field. A magnetic-dipole coupling $-\hat{\mathbf{m}}\cdot\boldsymbol{B}(t)$ is therefore handled in the same way, with the magnetization $\hat{\mathbf{m}}$ in place of $\hat{\boldsymbol{\mu}}$, as in Example~1.

\subsection{Open-system dynamics: from Liouville--von Neumann to GKSL}
\label{Opensystemdynamics}

An isolated density operator evolves according to the Liouville--von Neumann equation,
\begin{equation}
    \frac{d\hat{\rho}}{dt}=-i[\hat{H},\hat{\rho}].
    \label{eq:LiouvilleVonNeumann}
\end{equation}
Relaxation and dephasing are included through a Markovian open-system generator. To motivate its form, consider an infinitesimal completely positive, trace-preserving map written in a Kraus representation \cite{watrous2018theory},
\begin{equation}
    \hat{\rho}(t+dt)=\sum_m \hat{M}_m\hat{\rho}(t)\hat{M}_m^\dagger,
    \qquad
    \sum_m \hat{M}_m^\dagger \hat{M}_m=\mathbb{1}.
    \label{eq:Kraus_representation}
\end{equation}
Separating a no-jump branch from elementary jump events of probability $O(dt)$ gives \cite{dalibard1992wavefunction,dum1992montecarlo,carmichael1993open}
\begin{equation}
    \hat{M}_0=\mathbb{1}-i\hat{H}dt-\frac{1}{2}\sum_c \gamma_c \hat{L}_c^\dagger \hat{L}_c dt,
    \qquad
    \hat{M}_c=\sqrt{\gamma_c\,dt}\,\hat{L}_c.
\end{equation}
These operators satisfy the trace-preservation condition through first order in $dt$. Substitution into Eq.~\eqref{eq:Kraus_representation} gives
\begin{equation}
\begin{split}
    \hat{\rho}(t+dt)-\hat{\rho}(t)
    ={}&-i[\hat{H},\hat{\rho}]dt
    \\
    &+\sum_c\gamma_c
    \left(
        \hat{L}_c\hat{\rho} \hat{L}_c^\dagger
        -\frac{1}{2}\{\hat{L}_c^\dagger \hat{L}_c,\hat{\rho}\}
    \right)dt
    +O(dt^2).
\end{split}
\end{equation}
Dividing by $dt$ and taking the continuous-time limit yield the GKSL equation,
\begin{equation}
    \frac{d\hat{\rho}}{dt}
    =\mathcal{L}[\hat{\rho}]
    =-i[\hat{H},\hat{\rho}]
    +\sum_c\gamma_c\,\mathcal{D}_c[\hat{\rho}],
    \qquad
    \mathcal{D}_c[\hat{\rho}]
    =\hat{L}_c\hat{\rho} \hat{L}_c^\dagger
    -\frac{1}{2}\{\hat{L}_c^\dagger \hat{L}_c,\hat{\rho}\}.
    \label{eq:gksl-generator}
\end{equation}
Here $\hat{L}_c$ is the jump operator of channel $c$, $\mathcal{D}_c$ its dissipator, and $\gamma_c\geq0$ its rate; the rate is kept outside both the operator and the dissipator, as in the implementation. Because $\gamma_c$ multiplies $\mathcal{D}_c$ unchanged, its relation to a measured decay time depends on the normalization of $\hat{L}_c$: a diagonal channel $\hat{L}_c=\sum_x l_x\ket{x}\bra{x}$ damps the coherence $\ket{x}\bra{y}$ at the rate $\gamma_c(l_x-l_y)^2/2$. A pure-dephasing rate $\gamma_\phi=1/T_2^*$ of an optical coherence is therefore obtained with $\gamma_c=\gamma_\phi/2$ for $\hat{L}_c=\hat{\sigma}_z$, but with $\gamma_c=2\gamma_\phi$ for the projector $\hat{L}_c=\ket{e}\bra{e}$. The method \texttt{decay\_rates()} of the solver reports the population and coherence decay rates implied by the declared channels for a loaded model (Sec.~\ref{sec:usage}). Equation~\eqref{eq:gksl-generator} defines the field-free generator $\mathcal{L}$ used throughout the manuscript and by both numerical backends. The GKSL form guarantees completely positive, trace-preserving Markovian dynamics, but it does not by itself guarantee relaxation toward the Gibbs state of $\hat{H}$. A thermal interpretation requires collapse operators and rates that satisfy the appropriate balance conditions or follow from a compatible microscopic model.

This restriction is deliberate. More microscopic weak-coupling descriptions can retain additional bath information, but their approximations do not automatically preserve complete positivity in every parameter regime. QuDPy-FDGF therefore accepts a GKSL generator as the open-system contract and leaves the physical construction of its channels to the model.

\subsection{Vectorization and the Liouville-space generator}
\label{vectorization}

For numerical propagation, the density operator is vectorized in a fixed orthonormal basis $\{\ket{i}\}_{i=1}^{d}$ using column stacking,
\begin{equation}
    \hat{\rho}=\sum_{ij}\rho_{ij}\ket{i}\bra{j}
    \longrightarrow
    \kket{\hat{\rho}}
    =\operatorname{vec}(\hat{\rho})
    =\sum_{ij}\rho_{ij}\ket{j}\otimes\ket{i}.
\end{equation}
With this convention \cite{horn1991topics},
\begin{equation}
\begin{aligned}
    \operatorname{vec}(\hat{A}\hat{\rho} \hat{B})
    &=(\hat{B}^{T}\otimes \hat{A})\operatorname{vec}(\hat{\rho}),
    \\
    \hat{A}\hat{\rho}
    &\longrightarrow(\mathbb{1}\otimes \hat{A})\kket{\hat{\rho}},
    &
    \hat{\rho} \hat{A}
    &\longrightarrow(\hat{A}^{T}\otimes\mathbb{1})\kket{\hat{\rho}}.
\end{aligned}
    \label{eq:right-left-multiplication}
\end{equation}
The commutator superoperator $\hat{H}^{\times}=[\hat{H},\,\cdot\,]$ is represented by
\begin{equation}
    \hat{H}^{\times}
    =\mathbb{1}\otimes \hat{H}-\hat{H}^T\otimes\mathbb{1},
    \qquad
    \hat{H}^{\times}\kket{\hat{\rho}}
    =\kket{[\hat{H},\hat{\rho}]},
    \label{eq:Liouvillian}
\end{equation}
and each dissipation superoperator by
\begin{equation}
    \mathcal{D}_c
    =\hat{L}_c^*\otimes \hat{L}_c
    -\frac{1}{2}\mathbb{1}\otimes \hat{L}_c^\dagger \hat{L}_c
    -\frac{1}{2}(\hat{L}_c^\dagger \hat{L}_c)^T\otimes\mathbb{1}.
    \label{eq:vec_dissipation}
\end{equation}
The matrix representation of the GKSL generator is therefore
\begin{equation}
    \frac{d}{dt}\kket{\hat{\rho}}
    =\mathcal{L}\kket{\hat{\rho}},
    \qquad
    \mathcal{L}=-i\hat{H}^{\times}+\sum_c\gamma_c\mathcal{D}_c.
    \label{eq:vectorized-gksl}
\end{equation}
A superoperator and its matrix in this column-stacked basis are denoted by the same symbol: $\mathcal{L}[\hat{\rho}]$ acts on the density operator and $\mathcal{L}\kket{\hat{\rho}}$ on its vectorized form. The sparse (matrix-free) backend applies $\mathcal{L}$ without assembling this matrix, which is all that the iterative solver require; the dense backend constructs it explicitly.

$\hat\rho\mapsto -i[\hat H,\hat\rho]+\sum_c\gamma_c\mathcal{D}_c[\hat\rho]$
For a Hilbert space of dimension $d$, the vectorized state has dimension $d^2$ and an explicit dense generator contains $d^4$ entries. This scaling motivates the sector-resolved matrix-free realization described in \ref{app:sector-fdgf-realization}; it does not change the generator or the pathway algebra.

The reference state is assumed to be stationary before the pulse sequence,
\begin{equation}
    \mathcal{L}[\hat{\rho}_{\mathrm{eq}}]=0,
    \qquad\text{equivalently}\qquad
    \mathcal{L}\kket{\hat{\rho}_{\mathrm{eq}}}=0.
    \label{eq:stationary-state}
\end{equation}
This condition is a physical property of the supplied model and is not imposed by the pathway construction. In particular, when $\hat{\rho}_{\mathrm{eq}}$ is chosen as a Gibbs state, the dissipative channels must be compatible with that state. The solver evaluates the residual $\lVert\mathcal{L}\kket{\hat{\rho}_{\mathrm{eq}}}\rVert/\lVert\kket{\hat{\rho}_{\mathrm{eq}}}\rVert$ once the model is loaded. It has the dimension of a rate and is compared with an absolute tolerance expressed in the units of the Hamiltonian (default $10^{-8}$); a warning is emitted when it is exceeded, so that the tolerance must be set consistently with the unit chosen for the model. It can also compute the stationary state of $\mathcal{L}$, the trace-one solution of $(\mathcal{L}+\kket{\hat{r}}\langle\!\langle\mathbb{1}|)\kket{\hat{\rho}}=\kket{\hat{r}}$ with $\mathrm{Tr}\,\hat{r}=1$, which is unique when the kernel of $\mathcal{L}$ is one-dimensional (the dense backend tests this), and use it as the reference state.

\subsection{Two equivalent representations: propagator and resolvent}
\label{twoequivalent}

The same time-independent generator supplies two complementary free-evolution operators. A time-domain propagator reads
\begin{equation}
    \mathcal{U}(t)=\exp(\mathcal{L}t),
    \qquad t\geq0,
\end{equation}
whereas in the frequency domain it is characterized by the causal Fourier--Laplace resolvent
\begin{align}
    \mathcal{R}_\eta(\omega)
    &=
    \int_0^\infty dt\,
    \exp[(i\omega-\eta)t]\mathcal{U}(t)
    \nonumber\\
    &=
    \left[(\eta-i\omega)\mathbb{1}-\mathcal{L}\right]^{-1},
    \qquad \eta>0.
    \label{eq:resolvent}
\end{align}
Thus, applying the resolvent to a state $b$ is equivalent to solving
\begin{equation}
    \left[(\eta-i\omega)\mathbb{1}-\mathcal{L}\right]x=b.
\end{equation}
This is the convention implemented by QuDPy-FDGF. It carries no additional frequency-domain prefactor. If a conventional retarded Green's function is defined instead as
\begin{equation}
    \mathcal{G}^{R}(\omega)
    =-i\int_0^\infty dt\,
    \exp[(i\omega-\eta)t]\mathcal{U}(t),
\end{equation}
then $\mathcal{G}^{R}(\omega)=-i\mathcal{R}_\eta(\omega)$. Keeping this relation explicit prevents factors of $i$ from being counted both in the Green's function and in the perturbative pathway coefficient.

For a diagonalizable generator, let
\begin{equation}
    \mathcal{L}\kket{r_\nu}
    =\lambda_\nu\kket{r_\nu},
    \qquad
    \langle\!\langle l_\nu|\mathcal{L}
    =\lambda_\nu\langle\!\langle l_\nu|,
    \qquad
    \langle\!\langle l_\nu|r_{\nu'}\rangle\!\rangle=\delta_{\nu\nu'}.
\end{equation}
Writing the signed modal frequency and decay rate as
$\lambda_\nu=-\kappa_\nu-i\Omega_\nu$, with $\kappa_\nu\geq0$, gives
\begin{equation}
    \mathcal{R}_\eta(\omega)
    =
    \sum_\nu
    \frac{
        \kket{r_\nu}
        \langle\!\langle l_\nu|
    }{
        \eta+\kappa_\nu-i(\omega-\Omega_\nu)
    }.
    \label{eq:resolvent-spectral}
\end{equation}
The pole location is set by $\Omega_\nu$, its half-width by $\eta+\kappa_\nu$, and its residue by the overlap of the preparation and detection operators with the left and right modes. The parameter $\eta$ is first a causal convergence and regularization parameter; it represents additional homogeneous broadening only when that interpretation is deliberately assigned. Microscopic relaxation and dephasing already enter through $\kappa_\nu$. For a non-diagonalizable generator, Eq.~\eqref{eq:resolvent} remains the defining expression, while a Jordan decomposition produces higher-order pole terms rather than the simple sum in Eq.~\eqref{eq:resolvent-spectral} \cite{kato1995perturbation}. Such a generator occurs at the exceptional points of Liouvillians, where eigenvalues and eigenmodes merge and the generator has fewer independant eigenmodes than its dimension \cite{minganti2019exceptional}.

The interval-by-interval choice between $\mathcal{U}(t)$ and $\mathcal{R}_\eta(\omega)$ is the central computational strategy of the solver. Frequency-resolved intervals are evaluated directly at the requested frequencies instead of being reconstructed from a sampled time trace, while waiting or other explicitly time-resolved intervals remain in the time domain.

The two representations therefore answer different numerical versions of the same physical question. Time propagation is natural when a delay is scanned directly or when transient dynamics are sought. The resolvent is natural when only selected frequencies or a frequency grid are required: spectral resolution is then controlled by the requested frequency points and $\eta$, rather than by a finite propagation window and its discrete Fourier transform.

\subsection{Perturbative expansion and Liouville-space pathways}
\label{sec:perturbative-expansion}

For a field projected onto a fixed polarization, $\hat{V}(t)=-\hat{\mu} E(t)$, Eq.~\eqref{eq:gksl-generator} becomes
\begin{equation}
    \frac{d}{dt}\kket{\hat{\rho}(t)}
    =
    \mathcal{L}\kket{\hat{\rho}(t)}
    +iE(t)\hat{\mu}^\times\kket{\hat{\rho}(t)},
    \qquad
    \hat{\mu}^\times\hat{\rho}\equiv[\hat{\mu},\hat{\rho}].
    \label{eq:motion-equation}
\end{equation}
For a stationary state in the remote past, the integral equation is
\begin{equation}
    \kket{\hat{\rho}(t)}
    =
    \kket{\hat{\rho}_{\mathrm{eq}}}
    +i\int_{-\infty}^{t}d\tau\,
    E(\tau)\,
    \mathcal{U}(t-\tau)\hat{\mu}^\times
    \kket{\hat{\rho}(\tau)}.
    \label{eq:motion-integral}
\end{equation}
Iteration and collection of terms containing exactly $n$ field interactions give
\begin{equation}
\begin{split}
    \kket{\hat{\rho}^{(n)}(t)}
    ={}&
    i^n
    \int_{-\infty}^{t}d\tau_n
    \int_{-\infty}^{\tau_n}d\tau_{n-1}
    \cdots
    \int_{-\infty}^{\tau_2}d\tau_1
    \,
    E(\tau_n)\cdots E(\tau_1)
    \\
    &\times
    \mathcal{U}(t-\tau_n)\hat{\mu}^\times
    \mathcal{U}(\tau_n-\tau_{n-1})\hat{\mu}^\times
    \cdots
    \mathcal{U}(\tau_2-\tau_1)\hat{\mu}^\times
    \kket{\hat{\rho}_{\mathrm{eq}}}.
\end{split}
\label{eq:rho-n}
\end{equation}

At third order, with positive delays
$t_1=\tau_2-\tau_1$, $t_2=\tau_3-\tau_2$, and
$t_3=t-\tau_3$, the polarization is
\begin{equation}
\begin{split}
    P^{(3)}(t)
    ={}&
    \int_0^\infty dt_3
    \int_0^\infty dt_2
    \int_0^\infty dt_1\,
    S^{(3)}(t_3,t_2,t_1)
    \\
    &\times
    E(t-t_3)
    E(t-t_3-t_2)
    E(t-t_3-t_2-t_1),
\end{split}
\label{eq:P3}
\end{equation}
with the total matter response
\begin{equation}
    S^{(3)}(t_3,t_2,t_1)
    =
    i^3
    \langle\!\langle\hat{\mu}|
    \mathcal{U}(t_3)\hat{\mu}^\times
    \mathcal{U}(t_2)\hat{\mu}^\times
    \mathcal{U}(t_1)\hat{\mu}^\times
    |\hat{\rho}_{\mathrm{eq}}\rangle\!\rangle.
    \label{eq:S3}
\end{equation}
Here $S^{(n)}$ denotes a total or phase-selected response, whereas $R_p^{(n)}$ below denotes the time correlation function (with time variables) or the spectral response (with frequency variables) related to a Liouville pathway $p$. This follows the convention used in QuDPy and in the examples.

At fixed interaction order, expansion of the nested commutators produces ket/bra histories that can be represented by double-sided Feynman diagrams, each commutator acting from the left or from the right of the density operator \cite{tanimura1993liouville,mukamel1995principles}. The subsequent rotating-wave and phase-selection rules retain only the histories compatible with the chosen experimental channel. This separation is useful computationally: the physical model determines whether a requested transition block exists, while the protocol determines which histories are propagated and summed.

Each commutator separates into ket- and bra-side actions,
\begin{equation}
    \hat{\mu}^\times\hat{\rho}
    =
    \underbrace{\hat{\mu}\hat{\rho}}_{\text{ket action}}
    -
    \underbrace{\hat{\rho}\hat{\mu}}_{\text{bra action}}.
    \label{eq:commutator-ketbra}
\end{equation}
After splitting the transition operator into raising and lowering components according to the sign of the associated Bohr frequency, $\hat{\mu}=\hat{\mu}_++\hat{\mu}_-$, the elementary instructions are
\begin{align}
    \mathcal{V}_{K_u}[\hat{\rho}]&=\hat{\mu}_+\hat{\rho},
    &
    \mathcal{V}_{K_d}[\hat{\rho}]&=\hat{\mu}_-\hat{\rho},
    \nonumber\\
    \mathcal{V}_{B_u}[\hat{\rho}]&=\hat{\rho}\hat{\mu}_-,
    &
    \mathcal{V}_{B_d}[\hat{\rho}]&=\hat{\rho}\hat{\mu}_+.
    \label{eq:four-interaction-actions}
\end{align}
The symbols $K$ and $B$ identify ket- and bra-side actions, respectively. The subscripts $u$ and $d$ denote an upward or downward transition on that branch of the double-sided diagram. In the API-facing notation, the model-provided transition blocks are called $J_+$ and $J_-$; mathematically, these are the blocks of $\hat{\mu}_+$ and $\hat{\mu}_-$ used in Eq.~\eqref{eq:four-interaction-actions}.

The minus sign associated with every bra action is carried by the pathway coefficient. For a pathway $p$ with $N_{\mathrm{int}}=n$ interactions, $N_{\mathrm{bra}}$ of which act on the bra, the default coefficient is
\begin{equation}
    c_p
    =
    a_p\,i^{N_{\mathrm{int}}}(-1)^{N_{\mathrm{bra}}},
    \label{eq:pathway-prefactor}
\end{equation}
where $a_p=1$ by default and may be used for an explicitly supplied complex pathway amplitude. Dipole matrix elements are not included again in $c_p$: they are already contained in the interaction superoperators and the final observation operator. Pulse amplitudes and phases likewise remain associated with the applied fields or the protocol.

For excitation-manifold pathways, a dyad
$\ket{N_{\mathrm{ket}},a}\bra{N_{\mathrm{bra}},b}$
is assigned the coherence order
\begin{equation}
    q=N_{\mathrm{ket}}-N_{\mathrm{bra}}.
    \label{eq:coherence-order-convention}
\end{equation}
The four elementary changes are
\begin{equation}
    \Delta q(K_u)=+1,\qquad
    \Delta q(K_d)=-1,\qquad
    \Delta q(B_u)=-1,\qquad
    \Delta q(B_d)=+1.
    \label{eq:coherence-order-changes}
\end{equation}
These increments reproduce the UFSS instruction convention adopted by QuDPy and by the present solver \cite{rose2021ufss,shah2023qudpy}. The sign of $q$ is a bookkeeping convention; reversing the definition reverses all reported coherence-order signs without changing the underlying physical pathway.

\subsection{From a Liouville pathway to a chain of Green's functions}
\label{FromLiouville}

Expanding the commutators yields individual Liouville pathways. For a third-order pathway $p$, the time-domain correlation function is
\begin{equation}
    R_p^{(3)}(t_3,t_2,t_1)
    =
    c_p
    \langle\!\langle\hat{\mu}|
    \mathcal{U}(t_3)\mathcal{V}_{p,3}
    \mathcal{U}(t_2)\mathcal{V}_{p,2}
    \mathcal{U}(t_1)\mathcal{V}_{p,1}
    |\hat{\rho}_{\mathrm{eq}}\rangle\!\rangle,
    \label{eq:pathway-time-domain}
\end{equation}
where each $\mathcal{V}_{p,j}$ is selected from Eq.~\eqref{eq:four-interaction-actions}. Because the response is multilinear in the free-evolution intervals, any subset of the delays can be transformed independently by replacing the corresponding $\mathcal{U}(t_j)$ with $\mathcal{R}_\eta(\omega_j)$. For the common two-dimensional representation in which the first and third intervals are frequency resolved while $t_2$ remains a waiting time, the response can be conveniently rewritten as
\begin{equation}
\begin{split}
    R_p^{(3)}(\omega_3,t_2,\omega_1)
    ={}&
    c_p
    \langle\!\langle\hat{\mu}|
    \mathcal{R}_\eta(\omega_3)\mathcal{V}_{p,3}
    \mathcal{U}(t_2)\mathcal{V}_{p,2}
    \\
    &\times
    \mathcal{R}_\eta(\omega_1)\mathcal{V}_{p,1}
    |\hat{\rho}_{\mathrm{eq}}\rangle\!\rangle.
\end{split}
\label{eq:mixed-pathway-resolvent}
\end{equation}
No extra factor of $-i$ accompanies a frequency-resolved interval because Eq.~\eqref{eq:resolvent} defines the implemented resolvent directly. Equation~\eqref{eq:mixed-pathway-resolvent} is also a scalar product, $c_p\langle\!\langle G(\omega_3)|F(\omega_1)\rangle\!\rangle$, between a vector that depends only on $\omega_1$, $|F\rangle\!\rangle=\mathcal{V}_{p,3}\mathcal{U}(t_2)\mathcal{V}_{p,2}\mathcal{R}_\eta(\omega_1)\mathcal{V}_{p,1}|\hat{\rho}_{\mathrm{eq}}\rangle\!\rangle$, and one that depends only on $\omega_3$, $|G\rangle\!\rangle=\mathcal{R}_\eta(\omega_3)^\dagger|\hat{\mu}\rangle\!\rangle$. The implementation, which we call the factorized evaluation, cuts the chain between the two scanned frequency intervals (with $N_1$ and $N_2$ points respectively), evaluates each vector once per point of its own axis and obtains the whole map from one matrix product, which takes $N_1+N_2$ shifted solves for $N_1\times N_2$ points instead of $2N_1N_2$ (\ref{app:grid-evaluation}; Sec.~\SIsecgrid of the Supplementary Material). A phase-selected signal is assembled only after propagation,
\begin{equation}
    S_{\mathbf{s}}^{(n)}(\omega_1,t_2,\omega_3)
    =
    \sum_{p\in\mathcal{P}_{\mathbf{s}}}
    R_p^{(n)}(\boldsymbol{\xi}),
    \label{eq:phase-selected-signal}
\end{equation}
where $\boldsymbol{\xi}$ collects the variables scanned in the $n$ propagation intervals, a delay $t_j$ or a frequency $\omega_j$ for each (for instance $\boldsymbol{\xi}=(\omega_3,t_2,\omega_1)$ in Eq.~\eqref{eq:mixed-pathway-resolvent}), the sign signature $\mathbf{s}=(s_1,\ldots,s_n)$ of Sec.~\ref{sec:phase-matching} labels the phase-matching or phase-cycling channel and $\mathcal{P}_{\mathbf{s}}$ is the pathway set chosen by the protocol. The set may be supplied explicitly or enumerated from the signature, which gives the $2^n$ label sequences of $\mathbf{s}$ and lets the model decide which of them vanish; a set written by hand can miss sequences that relaxation or a thermal reference state make non-zero, as the third rephasing pathway of the two-level system of Sec.~\ref{sec:usage-model} shows. The solver lists the pathways that vanish on the whole grid and warns about those written by hand (companion API documentation). When an example deliberately studies a fixed subset, the solver does not enlarge that set merely because additional thermally initialized pathways are physically allowed.

\subsection{Observable functional and detection}
\label{observable_functional_and_detection}

The terminal contraction in Eqs.~\eqref{eq:S3} and \eqref{eq:mixed-pathway-resolvent} represents coherent polarization detection. More generally, let $\hat{\rho}_p^{(n)}(\boldsymbol{\xi})$ denote the state produced by the ordered propagation and interaction chain before multiplication by $c_p$. The contribution of a detected pathway to the final measurement in general is 
\begin{equation}
    R_{p,O}^{(n)}(\boldsymbol{\xi})
    =
    c_p\,
    \left\langle\!\left\langle \hat{O}|\hat{\rho}_p^{(n)}(\boldsymbol{\xi})\right\rangle\!\right\rangle = c_p\, \operatorname{Tr}\left[\hat{O}\hat{\rho}_p^{(n)}(\boldsymbol{\xi})\right],
    \label{eq:operator-observable}
\end{equation}

in which $\hat{O}$ is the observable operator. Coherent polarization follows from $\hat{O}=\hat{\mu}$, while population projectors, current operators, and mode-resolved observables share the same contraction, so a number of observables can be evaluated from a common propagated pathway state.

An action-detected measurement may require the stored pathway to end in a population or another detector-accessible sector. If it ends in a coherence, an additional phase-selected readout interaction can be applied after the final propagation interval,

\begin{equation}
    \hat{\rho}_{p,\mathrm{rd}}^{(n+1)}
    =
    \mathcal{V}_{\mathrm{rd}}\hat{\rho}_p^{(n)},
    \qquad
    R_{p,O}^{(n+1)}= 
    c_p c_{\mathrm{rd}}\operatorname{Tr} \!\left [\hat{O}\,\hat{\rho}_{p,\mathrm{rd}}^{(n+1)}\right], 
    \label{eq:readout-interaction}
\end{equation}

Here, $\mathcal{V}_{\mathrm{rd}}$ is one of the interaction of Eq~\eqref{eq:four-interaction-actions} and $c_{\mathrm{rd}}=+i$ for a ket-side and $-i$ for a bra-side readout interaction, the factor of Eq.~\eqref{eq:pathway-prefactor} for that interaction. This additional light--matter interaction converts the terminal coherence into a detector-accessible population or sector. It raises the physical perturbative order from $n$ to $n+1$ without introducing an additional propagation interval or spectral coordinate. If the stored pathway already ends in the sector required by the detector, the readout interaction can be omitted.

Dissipative events define a second class of action observables. For channel $\hat{L}_c$ with rate $\gamma_c$, the instantaneous detected jump rate is
\begin{equation}
    \operatorname{Tr}\!\left(\hat{O}_c^{\mathrm{rate}}\hat{\rho}\right), \quad \text{where} \quad \hat{O}_c^{\mathrm{rate}}=\varepsilon_{\mathrm{det},c}\,\gamma_c\,\hat{L}_c^\dagger\hat{L}_c, 
    \label{eq:detection_efficiency}    
\end{equation}
where $\varepsilon_{\mathrm{det},c}$ is the detection efficiency. The integrated mean count over a post-pathway window $[0,T]$ is

\begin{equation}
    N_{c,T}[\hat{\rho}] = \int_0^T ds\,\operatorname{Tr}\!\left[\hat{O}_c^{\mathrm{rate}}\,
    \mathcal{U}(s)\hat{\rho}\right]
    \label{eq:integrated-jump-observable}
\end{equation}

Because the trace is linear in the state, the window integral can be transferred to the detector,
\begin{equation}
    N_{c,T}[\hat{\rho}]=\langle\!\langle\hat{X}_{c,T}|\hat{\rho}\rangle\!\rangle, \qquad \hat{X}_{c,T}=\int_0^T ds\,\mathcal{U}^\dagger(s)\big[\hat{O}_c^{\mathrm{rate}}\big]
    \label{eq:integrated-detector}
\end{equation}


where $\mathcal{U}^\dagger(s)=\exp(\mathcal{L}^\dagger s)$ propagates operators in the Heisenberg picture. Since $\hat{O}_c^{\mathrm{rate}}$ is Hermitian and $\mathcal{U}^\dagger(s)$ preserves Hermiticity, $\hat{X}_{c,T}$ is Hermitian: it is the detection operator $\hat{O}$ of Eq.~\eqref{eq:operator-observable} for the integrated count,
$N_{c,T}=\operatorname{Tr}(\hat{X}_{c,T}\hat{\rho})$. QuDPy-FDGF computes $\hat{X}_{c,T}$ once per channel and window, without quadrature, from the exponential of an augmented generator,
\begin{equation}
    \begin{pmatrix}\kket{\hat{X}_{c,T}}\\ 1\end{pmatrix}
    =\exp\!\left[T
    \begin{pmatrix}
        \mathcal{L}^\dagger & \kket{\hat{O}_c^{\mathrm{rate}}}\\
        0 & 0
    \end{pmatrix}\right]
    \begin{pmatrix}0\\ 1\end{pmatrix},
    \label{eq:augmented-generator}
\end{equation}
applied without assembling $\mathcal{L}^\dagger$ in the sparse backend; each pathway state then requires a single contraction. The efficiency $\varepsilon_{\mathrm{det},c}$ is unrelated to the resolvent broadening $\eta$.

For an infinite window, $N_{c,T}$ diverges linearly unless the stationary count rate vanishes, since the rate of any state tends to $r_c\,\mathrm{Tr}\,\hat\rho$ with $r_c=\mathrm{Tr}(\hat O_c^{\mathrm{rate}}\hat\rho_{\mathrm{ss}})$. QuDPy-FDGF integrates the excess over this background, $\hat X_{c,[t_0,\infty)}=\int_{t_0}^{\infty}ds\,\allowbreak[\mathcal{U}^\dagger(s)-\mathcal{P}^\dagger]\allowbreak[\hat O_c^{\mathrm{rate}}]$, where $\mathcal{P}^\dagger[\hat A]=\mathrm{Tr}(\hat A\hat\rho_{\mathrm{ss}})\mathbb{1}$ projects on the stationary mode. It is $\exp(t_0\mathcal{L}^\dagger)\hat x$ with $\hat x$ the solution of one linear system with $\mathcal{L}^\dagger-|\mathbb{1}\rangle\!\rangle\langle\!\langle\hat\rho_{\mathrm{ss}}|$, which is invertible when the stationary state is unique. When the stationary state has no jumps (zero temperature, decay to the ground state) the background vanishes and the result is the total count of the state, its quantum yield; for a two-level emitter with a counted radiative channel and an uncounted non-radiative one it is $\gamma_r/(\gamma_r+\gamma_{nr})$. The preceding expressions separate pathway propagation, an optional readout interaction, and the final measurement. Their application to coherent and fluorescence-detected observables is illustrated in the listing of Sec.~\ref{sec:usage} and in Sec.~\SIsectwolevel of the Supplementary Material \cite{schroeter2018fluorescence,kuhn2020interpreting,oliver2018multidimensional}.

\subsection{Basis, transition decomposition, and excitation manifolds}
\label{sec:basis-decomposition}

All Hamiltonian, transition, detection, and collapse-operator blocks must be expressed in mutually consistent bases. When the model is provided in a site or orbital basis, a model adapter may diagonalize each independent sector,
\begin{equation}
    \hat{H}_{\mathrm{site}}^{(s)}U^{(s)}
    =
    U^{(s)}E^{(s)},
    \qquad
    \hat{A}_{\mathrm{eig}}^{(s)}
    =
    U^{(s)\dagger}\hat{A}_{\mathrm{site}}^{(s)}U^{(s)},
\end{equation}
and transform every operator with the same eigenvectors. The adapter may then separate the transition operator into $\hat{\mu}_+$ and $\hat{\mu}_-$ according to signed Bohr frequencies, a stated rotating-wave tolerance, and an optional window of Bohr frequencies that restricts the split to the band the pulses can drive. Alternatively, a model can supply the $J_+$ and $J_-$ transition blocks explicitly. The solver consumes this decomposition but does not infer its physical meaning.

Excitation manifolds and computational sectors are related but not identical concepts. A manifold is a physical ladder indexed by an excitation number and supports the coherence order in Eq.~\eqref{eq:coherence-order-convention}. A sector is a model-defined computational block and may instead label momentum, symmetry, particle number, or a combination of quantum numbers. Rectangular transition blocks between sectors are allowed, and formally valid pathways vanish when a required block is absent. The sector representation and basis transformation are detailed in \ref{app:sector-decomposition}.

\subsection{Phase matching, rephasing/nonrephasing selection, and signal assembly}
\label{sec:phase-matching}

A real pulse field contains positive- and negative-frequency components with opposite carrier wavevectors,
\begin{equation}
\begin{split}
    E_j(\boldsymbol{r},t)
    ={}&
    \mathcal{E}_j(t)
    \exp(i\boldsymbol{k}_j\cdot\boldsymbol{r}-i\omega_jt - i\phi_j)
    \\
    &+
    \mathcal{E}_j^*(t)
    \exp(-i\boldsymbol{k}_j\cdot\boldsymbol{r}+i\omega_jt _i\phi_j).
\end{split}
\label{eq:pulse_electric_field}
\end{equation}
Within the rotating-wave approximation, $\hat{\mu}_+$ pairs with the field component proportional to $\exp(-i\omega_jt)$, and $\hat{\mu}_-$ with its complex conjugate. Combining this pairing with Eq.~\eqref{eq:four-interaction-actions} gives
\begin{center}
\begin{tabular}{@{}c c c c@{}}
    \toprule
    Instruction & Action & $\Delta q_j$ & Wavevector sign $s_j$ \\
    \midrule
    $K_u$ & $\hat{\mu}_+\hat{\rho}$ & $+1$ & $+1$ \\
    $K_d$ & $\hat{\mu}_-\hat{\rho}$ & $-1$ & $-1$ \\
    $B_u$ & $\hat{\rho}\hat{\mu}_-$ & $-1$ & $-1$ \\
    $B_d$ & $\hat{\rho}\hat{\mu}_+$ & $+1$ & $+1$ \\
    \bottomrule
\end{tabular}
\end{center}
Thus $s_j=\Delta q_j$ in the adopted convention. The designation positive- or negative-frequency component refers to the transition or field decomposition; $\hat{\mu}_+$ and $\hat{\mu}_-$ should not themselves be identified generically as rotating and counter-rotating terms. The rotating-wave approximation is the rule that retains their resonant pairings.

Taking the combinations of wave vectors as an example, a phase-matched contribution accumulates in the direction
\begin{equation}
    \boldsymbol{k}_{\mathbf{s}}
    =
    \sum_{j=1}^{n}s_j\boldsymbol{k}_j.
    \label{eq:phase-matching-direction}
\end{equation}
For three pulses arriving in the order $(1,2,3)$, the standard third-order channels include
\begin{align}
    \boldsymbol{k}_{\mathrm{R}}
    &=-\boldsymbol{k}_1+\boldsymbol{k}_2+\boldsymbol{k}_3,
    \\
    \boldsymbol{k}_{\mathrm{NR}}
    &=+\boldsymbol{k}_1-\boldsymbol{k}_2+\boldsymbol{k}_3,
\end{align}
for rephasing and nonrephasing signals, respectively; in signature notation, $\mathrm{R}\equiv\mathbf{s}=(-1,+1,+1)$ and $\mathrm{NR}\equiv\mathbf{s}=(+1,-1,+1)$. In collinear geometries, the same pathway classes can be selected by phase cycling rather than spatial separation. Equation~\eqref{eq:phase-selected-signal} then assembles the selected pathways after their independent propagation.

\paragraph{Sign of the frequency axes}
For a phase-covariant generator (Sec.~\ref{sec:scope}), a frequency-resolved interval turns the coherence stored in that interval into a pole of Eq.~\eqref{eq:resolvent-spectral}: a dyad $\ketbra{x}{y}$ oscillates as $\exp[-i(E_x-E_y)t]$ and appears at $\omega_j=E_x-E_y$. The sign of $\omega_j$ is therefore the sign of the coherence order $q_j$ of the interval, positive when the ket lies in the higher manifold. In the rephasing channel the first interval carries $q_1=-1$, so that its resonances appear near $\omega_1=-\omega_{eg}$, whereas the nonrephasing channel ($q_1=+1$) and the detection interval ($q_n=+1$) resonate at positive frequencies; a two-quantum interval with $q=-2$ resonates near minus the two-excitation energies, as in Example~2. QuDPy-FDGF keeps these signed axes. The common experimental display of the rephasing signal against $|\omega_1|$ corresponds to the reflection $\omega_1\to-\omega_1$ of the computed map.

\paragraph{Detection operator}
Coherent detection contracts the final state with the transition operator, $\langle\!\langle\hat{\mu}|\hat{\rho}_p\rangle\!\rangle=\operatorname{Tr}(\hat{\mu}\hat{\rho}_p)$ for a Hermitian $\hat{\mu}$. For a phase-selected pathway the final state carries $q_n=+1$; it is a combination of dyads $\ketbra{x}{y}$ with $\ket{x}$ one manifold above $\ket{y}$, for which $\operatorname{Tr}(\hat{\mu}_+\ketbra{x}{y})=\bra{y}\hat{\mu}_+\ket{x}=0$. The full operator and its lowering part therefore give the same signal, $\operatorname{Tr}(\hat{\mu}\hat{\rho}_p)=\operatorname{Tr}(\hat{\mu}_-\hat{\rho}_p)$, which is the positive-frequency polarization radiated in the phase-matched direction; for the two-level system of Sec.~\ref{sec:usage-model} the two contractions agree exactly. This uses the excitation-number bookkeeping of the pathway generator: the default Bohr-frequency decomposition of \texttt{EigenbasisKModel} coincides with it only when the manifolds do not overlap, and the explicit $\hat{\mu}_\pm$ restore it otherwise (Sec.~\ref{sec:usage-beyond}). The distinction would matter only for states that mix several coherence orders, which the pathway construction does not produce for a phase-covariant generator (Sec.~\ref{sec:scope}).

\paragraph{From the computed response to the measured signal}
With $\hbar=1$, the pathway sum of Eq.~\eqref{eq:phase-selected-signal}, including the prefactors $c_p=i^n(-1)^{N_{\mathrm{bra}}}$, is the $n$th-order response function of the selected channel, Fourier--Laplace transformed in its frequency-resolved intervals. For impulsive pulses, the polarization radiated in the direction $\boldsymbol{k}_{\mathbf{s}}$ is proportional to $S_{\mathbf{s}}$ times the product of the field amplitudes, and the emitted field is proportional to $iP^{(n)}_{\mathbf{s}}$~\cite{mukamel1995principles}. A heterodyne measurement with a local oscillator of phase $\phi_{\mathrm{LO}}$ therefore records, up to a positive constant,
\begin{equation}
    S_{\mathrm{het}}
    \propto
    \operatorname{Re}\!\left[\exp(-i\phi_{\mathrm{LO}})\,i\,S_{\mathbf{s}}\right].
    \label{eq:heterodyne-signal}
\end{equation}
The plotting layer applies a phase factor $\exp(i\phi)S_{\mathbf{s}}$ with $\phi=$~\texttt{detection\_phase}, so that $\phi=\pi/2-\phi_{\mathrm{LO}}$; the choice $\phi=\pi/2$ displays $iS_{\mathbf{s}}$, proportional to the emitted field, whose real and imaginary parts are the absorptive and dispersive quadratures. With this choice, linear absorption appears as a negative real part, because the emitted field interferes destructively with the transmitted one, whereas in third-order spectra ground-state bleach and stimulated emission are positive and excited-state absorption is negative, as in transmission-change conventions; the two-level system of Sec.~\ref{sec:usage-model} and the two examples of Sec.~\ref{sec:examples} follow these signs. Purely absorptive two-dimensional line shapes follow from the real part of the sum of the rephasing signal, reflected in $\omega_1$, and the nonrephasing signal~\cite{khalil2003absorptive}. QuDPy-FDGF returns the two channels separately and leaves this combination to the analysis.

\subsection{Scope and approximations}
\label{sec:scope}

The formulation applies to a finite Hilbert space and assumes that both the Hamiltonian and the Markovian GKSL generator are independent of time. The optical response is perturbative and uses the dipole approximation (electric, or magnetic as in Example~1) and the rotating-wave approximation. The standard pathway protocol treats interactions as impulsive, although arbitrary time and frequency coordinates can be mixed between them. The model adapter must supply a physically consistent initial state, transition decomposition, sector structure, and dissipative channels. Non-Markovian bath memory, overlapping finite-duration pulses, explicitly time-dependent generators, ultrastrong driving, and transitions omitted by the supplied sector blocks require extensions beyond the present implementation.

The coherence order $q$ of Eq.~\eqref{eq:coherence-order-convention} is determined by
the interaction history of a pathway. It describes the coherence content of the
propagated state, a combination of dyads $\ket{x}\bra{y}$ with $N_x-N_y=q$, only when
three conditions hold. First, the generator has the U(1) symmetry generated by the
excitation-number operator $\hat N$,
$[\mathcal{L},\hat N^{\times}]=0$ with $\hat N^{\times}[\hat\rho]=[\hat N,\hat\rho]$
\cite{albert2014symmetries}, which block-diagonalizes $\mathcal{L}$ into sectors of
fixed coherence order; we call such a generator phase covariant, since it commutes with a global phase rotation $e^{i\varphi\hat N}$ of the excitations. This holds for the Hamiltonians and the dissipation channels of this paper. Second, the transition operators change the excitation number by one, $[\hat N,\hat\mu_\pm]=\pm\hat\mu_\pm$. Third, the reference state has coherence order zero. When all three hold, the signs of the frequency axes, the separation of the $+N$Q and $-N$Q ($N$-quantum) contributions, and
the equivalence $\operatorname{Tr}(\hat\mu\hat\rho_p)=\operatorname{Tr}(\hat\mu_-\hat\rho_p)$ of Sec.~\ref{sec:phase-matching} follow from $q$. A channel that breaks the symmetry, such as the Lindblad operator $\hat L=\hat\sigma_x$ on a two-level system, whose dissipator maps $\ket{g}\bra{e}$ to $\ket{e}\bra{g}-\ket{g}\bra{e}$, mixes coherence orders: the labels of the pathways remain exact, but the state is no longer an eigenvector of $\hat N^{\times}$, and the solver does not test this property.

\section{Software architecture}
\label{sec:architecture}

\begin{figure}[h]
    \centering
    \includegraphics[width=\linewidth]{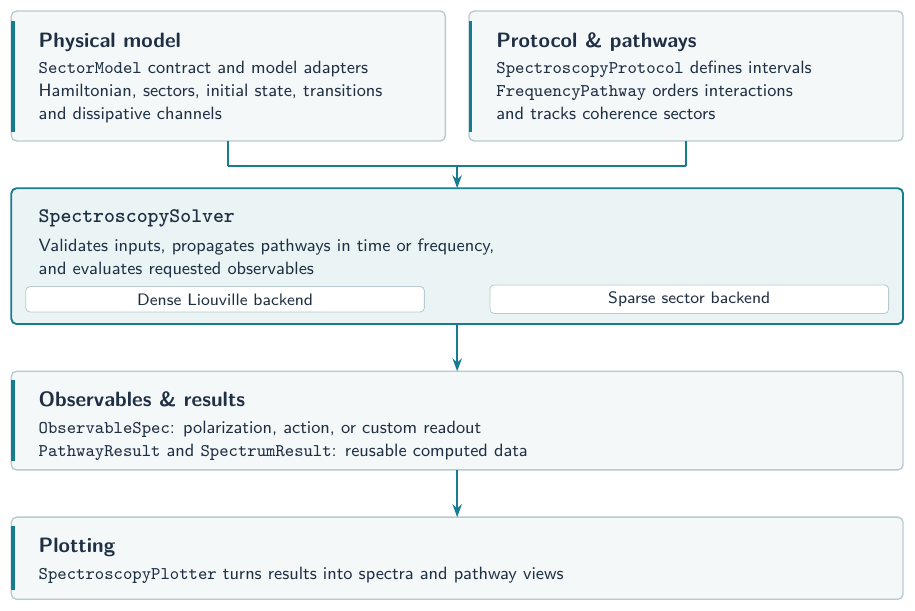}
    \caption{Functional organization of QuDPy-FDGF. A physical model and a spectroscopy protocol with selected pathways feed the solver, which propagates them through a dense Liouville or sparse sector backend. Requested observables produce reusable pathway and spectrum results for plotting.}
    \label{fig:Software_diagram}
\end{figure}

As summarized in Fig.~\ref{fig:Software_diagram}, QuDPy-FDGF separates five responsibilities. The \emph{model contract} provides the Hamiltonian, sector structure, state, transition operators, and dissipative channels. \emph{Protocols and pathways} specify the propagation intervals and ordered light-matter interactions. The \emph{solver core} validates these inputs and delegates time- and frequency-domain evolution to a numerical backend. \emph{Observables and results} define the terminal measurement independently of the propagated pathway, while the \emph{plotting} layer converts the resulting data containers into standard spectral representations. 

A calculation therefore proceeds from a physical model and a spectroscopy protocol to pathway propagation, observable evaluation, and visualization. This separation allows the same model and propagated pathways to be reused with different protocols, numerical backends, and detection operators. The mathematical sector representation and the equivalence of the numerical backends are described in \ref{app:sector-fdgf-realization}; constructors, accepted input formats, and the complete public interface are provided in the companion API documentation.

\section{Program usage}
\label{sec:usage}

This section walks through one complete calculation, the open two-level system of Sec.~\ref{sec:usage-model}, and relates each call to the formalism of Sec.~\ref{sec:theory}; the analysis of the model is in Sec.~\SIsectwolevel of the Supplementary Material. The constructors, optional arguments, and accepted input formats are listed in the companion API documentation.

\subsection{Installation}
QuDPy-FDGF~\cite{desmarais2026qudpyfdgf} requires Python~$\geq3.13$ with NumPy~\cite{harris2020numpy}, SciPy~\cite{virtanen2020scipy}, and Matplotlib~\cite{hunter2007matplotlib}; QuTiP~\cite{lambert2026qutip5,johansson2012qutip,johansson2013qutip2} and UFSS~\cite{rose2021ufss,rose2025ufsssoftware} are optional: they are used by the example notebooks, and the listing below builds its model with QuTiP. From a clone of the repository, \texttt{pip install -e .} installs the package under the import name \texttt{qudpy\_fdgf}. It was developed and tested on Windows~11 with Python~3.13.14, NumPy~2.5.0, SciPy~1.18.0, Matplotlib~3.11.0, and QuTiP~5.3.0, which are also the minimum versions; other platforms were not tested.

\subsection{The model of the listing}
\label{sec:usage-model}

The listing treats an open two-level system, whose dissipative dynamics and measured observables are known in closed form (\ref{app:two-level-analytical-benchmark}): a transition energy $\omega_{eg}=2$~eV, a dipole amplitude $d=0.5$, radiative relaxation $\hat{L}_{\mathrm{rad}}=\ketbra{g}{e}$ at rate $\gamma_1=0.04$~eV, and pure dephasing $\hat{L}_\phi=\hat{\sigma}_z$ at rate $\gamma_\phi/2$ where $\gamma_\phi=0.03$~eV, (Sec.~\SIsectwolevelmodel of the Supplementary Material). The optical coherence decays at $\Gamma_2=\gamma_1/2+\gamma_\phi=0.05$~eV, and the initial state is the ground state. The calculation targets the third-order rephasing signal at $t_2=10~\mathrm{eV}^{-1}$ with $\eta=0.002$~eV, from three pathways: the stimulated emission $R_1=(B_u,K_u,B_d)$, the ground-state bleaching $R_2=(B_u,B_d,K_u)$ \cite{mukamel1995principles,shah2023qudpy,hybl1998twodimensional}, and, excited-state absorption pathway $R_3=(B_u,K_u,K_u)$ is ruled out because of the limit of two-level system.

\subsection{Anatomy of a calculation}
Listing~\ref{lst:twolevel} computes the third-order rephasing response of the two-level system of Sec.~\ref{sec:usage-model} and evaluates three detection schemes from the same propagated pathways. It is organized in six steps, which follow the layers of Fig.~\ref{fig:Software_diagram}.

\begin{lstlisting}[language=Python, caption={Complete input for the two-level system of Sec.~\ref{sec:usage-model}: model, solver, pathways, protocol, and three observables evaluated from one propagation.}, label={lst:twolevel}]
import numpy as np
import qutip as qt

from qudpy_fdgf import (
    EigenbasisKModel,
    FrequencyPathway,
    ObservableSpec,
    SpectroscopySolver,
    standard_nq_protocol,
)

# 1. Model: QuTiP operators in the {|g>, |e>} basis (hbar = 1, energies in eV)
omega_eg = 2.0        # transition energy (eV)
dipole = 0.5          # transition dipole d
gamma_1 = 0.04        # radiative relaxation rate (eV)
gamma_phi = 0.03      # pure-dephasing rate (eV)

ground = qt.basis(2, 0)
excited = qt.basis(2, 1)

H = omega_eg * excited * excited.dag()
mu = dipole * (excited * ground.dag() + ground * excited.dag())

P_ground = ground * ground.dag()      # also the initial density matrix
P_excited = excited * excited.dag()
L_radiative = ground * excited.dag()  # |g><e|
L_dephasing = qt.sigmaz()

# 2. Solver: model with (operator, rate) channels, backend, broadening eta
model = EigenbasisKModel(
    H,
    mu,
    c_ops_raw=(
        (L_radiative, gamma_1),
        (L_dephasing, gamma_phi / 2),
    ),
    observable_op_arrays={
        "polarization": mu,
        "excited_population": P_excited,
    },
)

eta = 0.002          # resolvent broadening (eV)
solver = SpectroscopySolver(backend="dense", eta=eta)
solver.feed_model(model)          # default initial state: the ground state

# 3. Pathways: ordered interaction labels
pathway_r1 = FrequencyPathway(
    name="R1", interactions=("Bu", "Ku", "Bd"), component="rephasing", detection="polarization"
)
pathway_r2 = FrequencyPathway(
    name="R2", interactions=("Bu", "Bd", "Ku"), component="rephasing", detection="polarization"
)
pathway_r3 = FrequencyPathway(     # non-zero only through the radiative relaxation e -> g during t2
    name="R3", interactions=("Bu", "Ku", "Ku"), component="rephasing", detection="polarization"
)
pathways = (pathway_r1, pathway_r2, pathway_r3)

# 4. Protocol: t1 and t3 in frequency, the waiting time t2 in time
protocol = standard_nq_protocol(
    order=1,
    nq_interval=1,
    detection_interval=3,
    n_interactions=3,
    nq_axis="omega_1q",
    detection_axis="omega_emit",
)
t2 = 10.0                         # population waiting time (eV^-1)

# 5. Observables: one propagation, three detection schemes
radiative_channel = solver.jump_channel_names()[0]    # "c_op_0"
window = 5.0 / gamma_1

observables = {
    "polarization": "polarization",
    "action_population": ObservableSpec.action(
        "action_population", fourth_interaction="Bu", operator="excited_population"
    ),
    "action_fluorescence": ObservableSpec.mean_jump(
        "action_fluorescence", radiative_channel, time_window=(0.0, window), fourth_interaction="Bu"
    ),
}

omega_1q = np.linspace(-2.5, -1.5, 41)
omega_emit = np.linspace(1.5, 2.5, 41)
result = solver.generate_spectrum(
    protocol,
    axes={"omega_1q": omega_1q, "omega_emit": omega_emit},
    fixed_coordinates={"t2": t2},
    pathways=pathways,
    observables=observables,
)

# 6. Results: complex arrays indexed [omega_1q, omega_emit]
S_pol = result.observables["polarization"]["R1"]
S_pop = result.observables["action_population"]["R1"]
N_fl = result.observables["action_fluorescence"]["R1"]

i, j = np.unravel_index(np.argmax(np.abs(S_pol)), S_pol.shape)
print(f"peak = ({omega_1q[i]:.3f}, {omega_emit[j]:.3f}) eV")

identity_error = np.max(np.abs(S_pop + 1j * S_pol)) / np.max(np.abs(S_pol))
print(f"max|S_pop + i S_pol| / max|S_pol| = {identity_error:.1e}")

fluorescence_ratio = (N_fl[i, j] / S_pop[i, j]).real
print(f"N_fl / S_pop = {fluorescence_ratio:.6f}; 1 - exp(-5) = {1 - np.exp(-5):.6f}")
\end{lstlisting}

\begin{sloppypar}
\emph{Model.} The physical operators are built with QuTiP (NumPy arrays are accepted as well) in any basis. \texttt{EigenbasisKModel} diagonalizes the Hamiltonian, transforms every operator to the eigenbasis, and splits $\hat{\mu}$ into $\hat{\mu}_\pm$ according to the sign of the Bohr frequencies (Sec.~\ref{sec:basis-decomposition}). Each dissipative channel is an \texttt{(operator, rate)} pair: the rate is the Lindblad coefficient $\gamma_c$ of Eq.~\eqref{eq:gksl-generator} and is never folded into the operator, which is why the $\hat{\sigma}_z$ channel carries $\gamma_\phi/2$ (Sec.~\ref{Opensystemdynamics}). Named operators, here the dipole and the excited-state projector, are made available to the detection layer. Once the model is loaded, \texttt{solver.decay\_rates()} reports, in the eigenbasis of $\hat{H}$, the population and coherence decay rates implied by the declared channels; for this model it returns $\gamma_1$ for the excited-state population and $\Gamma_2=\gamma_1/2+\gamma_\phi$ for the optical coherence, which checks the normalization of every channel before a spectrum is computed.
\end{sloppypar}

\emph{Solver.} \texttt{SpectroscopySolver} selects the numerical backend and the regularization $\eta$ of Eq.~\eqref{eq:resolvent}. \texttt{feed\_model} validates the model against the backend and, unless another state is supplied, uses the ground state as the stationary reference state; another state, such as the stationary state of the generator (\texttt{initial\_state="steady"}), can be imposed.

\emph{Pathways.} A pathway is only an ordered list of interaction labels, $K_u$, $K_d$, $B_u$, $B_d$ (Eq.~\eqref{eq:four-interaction-actions}); the solver derives the coherence-order history and the prefactor $c_p$ of Eq.~\eqref{eq:pathway-prefactor}. The \texttt{component} field groups pathways into phase-matched sums, and \texttt{detection} names the operator used for coherent detection. The listing gives three rephasing pathways: $R_1$ and $R_2$ exist without population transfer, and $R_3$ acts on the population that the radiative channel brings from $\ket{e}$ to $\ket{g}$ during $t_2$ (Sec.~\SIsectwolevelpaths of the Supplementary Material).

\emph{Protocol.} The protocol states, interval by interval, whether the free evolution is represented by the propagator $\mathcal{U}(t)$ or by the resolvent $\mathcal{R}_\eta(\omega)$ (Sec.~\ref{twoequivalent}). Here the coherence intervals $t_1$ and $t_3$ are resolved in frequency on the axes \texttt{omega\_1q} and \texttt{omega\_emit}, while the waiting time $t_2$ stays in the time domain and is fixed through \texttt{fixed\_coordinates}. Every entry of the resulting array is one evaluation of the chain in Eq.~\eqref{eq:mixed-pathway-resolvent}; no time trace is propagated or Fourier transformed.

\emph{Observables.} The pathways are propagated once, and each observable is applied to the stored states (Sec.~\ref{observable_functional_and_detection}). The string \texttt{"polarization"} contracts the final coherence with the dipole, Eq.~\eqref{eq:operator-observable}. \texttt{ObservableSpec.action} applies the readout interaction of Eq.~\eqref{eq:readout-interaction}, here $B_u$, before projecting on $\hat{P}_e$. \texttt{ObservableSpec.mean\_jump} applies the same readout interaction and then integrates the radiative jump rate over the window $[0,5/\gamma_1]$, Eq.~\eqref{eq:integrated-jump-observable}; the integral is evaluated exactly through the integrated detector of Eq.~\eqref{eq:integrated-detector}. Jump channels are named in input order; \texttt{jump\_channel\_names()} returns these names.

\subsection{Correspondence with the formalism}
Table~\ref{tab:usage-correspondence} summarizes how the input of Listing~\ref{lst:twolevel} maps onto the objects of Sec.~\ref{sec:theory}.

\begin{table}[h]
\centering
\footnotesize
\begin{tabular}{@{}p{0.36\linewidth}p{0.30\linewidth}p{0.26\linewidth}@{}}
\toprule
Input & Object & Reference\\
\midrule
\texttt{H}, \texttt{mu}, \texttt{observable\_op\_arrays} & $\hat{H}$, $\hat{\mu}_\pm$, $\hat{O}$ & Secs.~\ref{Opensystemdynamics}, \ref{sec:basis-decomposition}\\
\texttt{c\_ops\_raw=((L, rate), ...)} & $\hat{L}_c$, $\gamma_c$, $\mathcal{D}_c$ & Eq.~\eqref{eq:gksl-generator}\\
\texttt{eta} & $\mathcal{R}_\eta(\omega)$ & Eq.~\eqref{eq:resolvent}\\
\texttt{interactions=("Bu", "Ku", "Bd")} & $\mathcal{V}_{p,j}$, $c_p$ & Eqs.~\eqref{eq:four-interaction-actions}, \eqref{eq:pathway-prefactor}\\
\texttt{standard\_nq\_protocol}, \texttt{axes}, \texttt{fixed\_coordinates} & $\mathcal{R}_\eta(\omega_1)$, $\mathcal{U}(t_2)$, $\mathcal{R}_\eta(\omega_3)$ & Eq.~\eqref{eq:mixed-pathway-resolvent}\\
\texttt{"polarization"} & $\langle\!\langle\hat{\mu}|$ & Eq.~\eqref{eq:operator-observable}\\
\texttt{ObservableSpec.action} & $\mathcal{V}_{\mathrm{rd}}$, then $\langle\!\langle\hat{P}_e|$ & Eq.~\eqref{eq:readout-interaction}\\
\texttt{ObservableSpec.mean\_jump} & $N_{c,T}$, $\hat{X}_{c,T}$ & Eqs.~\eqref{eq:integrated-jump-observable}, \eqref{eq:integrated-detector}\\
\bottomrule
\end{tabular}
\caption{Correspondence between the input of Listing~\ref{lst:twolevel} and the formalism.}
\label{tab:usage-correspondence}
\end{table}

\subsection{Output, units, and conventions}
\texttt{generate\_spectrum} returns a \texttt{SpectrumResult}. Its fields \texttt{axis\_names} and \texttt{axis\_values} follow the order of the keys of the mapping \texttt{axes}, and every array is indexed accordingly; in the listing this is also the order of the protocol, here \texttt{[omega\_1q, omega\_emit]}. The coherent signal of each pathway is stored in \texttt{pathways}, the phase-matched sums in \texttt{components}, and every requested observable in \texttt{observables[name][pathway]}, with the corresponding sums in \texttt{observable\_components}; solver diagnostics, such as linear-solver residuals, are collected in \texttt{diagnostics}. All quantities use $\hbar=1$: energies and frequencies share one unit (eV in the examples), and times are expressed in its inverse, so that $t_2=10~\mathrm{eV}^{-1}=6.58$~fs. In SI units, $1~\mathrm{eV}=1.602\,176\,634\times10^{-19}$~J and, with $\hbar=6.582\,119\,569\times10^{-16}$~eV\,s, $1~\mathrm{eV}^{-1}\simeq0.6582$~fs. Frequency axes are signed. In a rephasing pathway the first interval carries the coherence order $q=-1$, and its resonance therefore appears at negative $\omega_{1Q}$, as in the peak at $(-2,2)$~eV below (Sec.~\ref{sec:phase-matching}). The returned arrays are complex response functions; the plotting layer selects the absorptive and dispersive quadratures through a detection phase, \texttt{detection\_phase}$=\pi/2$ displaying $iS$, proportional to the emitted field (Eq.~\eqref{eq:heterodyne-signal}).

\subsection{Test run}
Run as a script, Listing~\ref{lst:twolevel} completes in a few seconds on a laptop (most of it the import of QuTiP) and prints
\begin{verbatim}
peak = (-2.000, 2.000) eV
max|S_pop + i S_pol| / max|S_pol| = 1.3e-16
N_fl / S_pop = 0.993262; 1 - exp(-5) = 0.993262
\end{verbatim}
Each line checks one layer against an analytical result. The peak lies at the signed resonance of the rephasing pathway. The action-detected population reproduces, to rounding error over the grid (the printed error is of the order of $10^{-16}$ and can differ between platforms),
\begin{equation}
  S^{(4)}_{P_e}(\omega_{1Q},t_2,\omega_{\rm emit})=-i\,S^{(3)}_{\rm pol}(\omega_{1Q},t_2,\omega_{\rm emit}),
  \label{eq:action_pol_identity}
\end{equation}
which follows from the selection rule of the readout interaction (Sec.~\SIsectwolevelreadout of the Supplementary Material). The ratio of the integrated fluorescence to the projected population equals $1-\exp(-\gamma_1T)$ (Sec.~\ref{sec:validation-checks}), since the window integral is evaluated exactly. The listing is extracted from the tagged cells of the notebook \texttt{examples/quickstart\_two\_level\_model.ipynb}, which adds the plots, the stationarity check, these identities and the sum of the three pathways against its closed form as assertions; the same checks are part of the test suite in \texttt{tests/}.

\subsection{Beyond the example}
\label{sec:usage-beyond}
\begin{sloppypar}
Other calculations change one element of Listing~\ref{lst:twolevel}. A thermal reference state is requested with a \texttt{ThermodynamicContext}, which the model interprets as the canonical state of its sectors, and any other mixed state is passed as a \texttt{DensityState}; \texttt{backend="sparse\_sector"} replaces the dense generator by its matrix-free sector action. The sparse sector backend is the default of the solver and the one that scales with the Hilbert-space dimension (Sec.~\ref{sec:perf-scaling}); the listing uses the dense backend because it is the faster one for a Hilbert space this small (Sec.~\SIsecresolved of the Supplementary Material), and \texttt{backend="auto"} selects the sparse one as well. Models with explicit excitation manifolds use \texttt{ExcitationSectorModel}, and the pathway set can be enumerated from the phase-matching signature with \texttt{pathways\_from\_signature}; these are the choices of Example~1. The pathways of a more complex model can also be generated by UFSS and translated into \texttt{FrequencyPathway} objects (Example~2). A new physical model only has to implement the \texttt{SectorModel} contract, which returns its sectors and the Hamiltonian, transition, observable, and collapse-operator blocks; the solver never imports a physical model.
\end{sloppypar}

\emph{Raising and lowering components.} By default \texttt{EigenbasisKModel} assigns every matrix element of the dipole to $\hat{\mu}_+$ or $\hat{\mu}_-$ according to the sign of its Bohr frequency. This rule coincides with the excitation-number bookkeeping assumed by the pathway generator when every state of the $(N+1)$-excitation manifold lies above every state of the $N$-excitation manifold, as in the two-level system. For sufficiently strong interactions the two conventions define different pathways. Supplying \texttt{j\_plus\_array} and \texttt{j\_minus\_array}, which store the physical operators $\hat{\mu}_+$ and $\hat{\mu}_-$, makes the choice explicit: the adapter transforms both operators into the eigenbasis of $\hat{H}$ without reclassifying their elements, and \texttt{transition\_decomposition()} returns \texttt{''explicit''} (Example~2).

\emph{Detection windows.} A finite window \texttt{time\_window=(t0, t1)} integrates the radiative count exactly. The window \texttt{(t0, np.inf)} counts the emission until the system has relaxed to its stationary state, above the stationary rate; it requires a unique stationary state (Sec.~\ref{sec:validation-checks}).

\section{Example calculations}
\label{sec:examples}

The open two-level system of Sec.~\ref{sec:usage-model} is the tutorial of Sec.~\ref{sec:usage}, and its closed-form response is the analytic reference of Sec.~\ref{sec:validation-checks}. Two examples follow, in which no closed form is available: a closed many-body system with a thermal initial state and unequal excitation sectors (Example~1), and a small open system whose spectrum depends on the detection, which is also taken to fifth order in its closed limit (Example~2).

\subsection{Example 1: Thermal XXZ spin chain driven by THz pulses}

This example applies the sector-resolved formulation to a closed many-body system: a field-polarized spin-$\frac{1}{2}$ XXZ chain with interacting magnon manifolds, prepared in a thermal state. It tests what the two-level system of Sec.~\ref{sec:usage-model} does not: excitation sectors of unequal dimension, a canonical initial state spread over several sectors, rectangular magnetic-transition blocks between successive magnon manifolds, and the enumeration of every rephasing sequence with \texttt{pathways\_from\_signature}. The third-order rephasing response is evaluated with its eight sequences. The size of the model also serves as the reference for the cost measurements of Sec.~\ref{sec:performance}. Example~1 tests the solver on a thermally populated many-body model; it is not a simulation of a particular magnet. The impulsive rotating-wave pulses are an assumption whose validity at THz bandwidths is not assessed here (Sec.~\ref{sec:scope}), and the line widths are set by $\eta$ alone.

\subsubsection{Model and thermal state}

We consider a periodic chain of $L$ spin-$\frac{1}{2}$ sites described by
\begin{equation}
\begin{split}
    \hat{H} ={}& J_{xy}\sum_j\left(\hat{S}_j^x \hat{S}_{j+1}^x + \hat{S}_j^y \hat{S}_{j+1}^y\right)\\
    &+ J_z\sum_j \hat{S}_j^z \hat{S}_{j+1}^z - g\mu_B B_0\sum_j \hat{S}_j^z ,
\end{split}
\label{eq:ex1_Hamiltonian}
\end{equation}
where the site index is taken modulo $L$. The transverse exchange $J_{xy}$ propagates spin deviations, whereas the longitudinal exchange $J_z$ produces a nearest-neighbor magnon interaction. For the convention used here, $J_z<0$ favors adjacent magnons and can support a bound bimagnon \cite{wortis1963bound,wu2022fewmagnon}. The mapping to hard-core magnons and the physical interpretation of the model are developed in Sec.~\SIsecxxz of the Supplementary Material. We use the representative parameters
\begin{equation}
    \begin{gathered}
    L=6,\qquad J_{xy}=-0.30~\mathrm{meV},\qquad J_z=-0.55~\mathrm{meV},\\
    g=2.10,\qquad B_0=7.1773~\mathrm{T},
    \end{gathered}
    \label{eq:ex1_parameters}
\end{equation}
with $\mu_B=0.0578838~\mathrm{meV/T}$. These parameters define a generic easy-axis ferromagnetic chain in the THz energy range and are not fitted to a specific material.

The Hamiltonian conserves the magnon number
\begin{equation}
    \hat{N}=\sum_j\left(\frac{1}{2}-\hat{S}_j^z\right),
    \qquad [\hat{H},\hat{N}]=0,
\label{eq:ex1_magnon_number}
\end{equation}
so the Hilbert space decomposes into excitation manifolds $\mathcal{H}_N$ of dimension $\binom{L}{N}$. These fixed-$N$ manifolds are the sectors supplied to the solver. They should not be confused with crystal-momentum sectors: momentum remains an additional quantum number within each manifold.

The magnetic component of a spatially uniform THz pulse couples to the total transverse magnetization, $\hat{V}(t)=-\hat{\mathbf{m}}\cdot\boldsymbol{B}(t)$: in the notation of Sec.~\ref{sec:theory}, the transition operator is the magnetization and the field is the magnetic field of the pulse, and the pathway algebra is unchanged. After absorbing the overall magnetic-dipole scale into the signal amplitude, its excitation-raising and excitation-lowering components are
\begin{equation}
    \hat{M}_{+}=\frac{1}{\sqrt{L}}\sum_j\hat{S}_j^{-},
    \qquad
    \hat{M}_{-}=\hat{M}_{+}^{\dagger}.
\label{eq:ex1_transition_operator}
\end{equation}
Here, raising refers to magnon number: $\hat{M}_+$ increases $N$ even though $\hat{S}_j^-$ lowers the microscopic spin projection. The operator is represented by rectangular blocks connecting $\mathcal{H}_N$ to $\mathcal{H}_{N+1}$, and the lowering blocks are their adjoints. A uniform field is the $k=0$ Fourier component of the magnetization, so that $\hat{M}_\pm$ conserves the crystal momentum.

The initial state is the canonical density operator
\begin{equation}
    \hat{\rho}_\beta=\frac{1}{Z} \sum_{N,a} \exp\!\left(-\frac{E_{N,a}}{k_BT}\right) \ket{N,a}\bra{N,a},
    \qquad
    Z= \sum_{N,a} \exp\!\left(-\frac{E_{N,a}}{k_BT}\right),
\label{eq:ex1_thermal_state}
\end{equation}
where $\ket{N,a}$ is eigenstate $a$ in the $N$-magnon manifold. At $T=4~\mathrm{K}$, for which $k_BT=0.3447~\mathrm{meV}$, the three lowest sector populations are
\begin{equation}
    P_0=0.885331,\qquad
    P_1=0.102734,\qquad
    P_2=0.010908.
    \label{eq:ex1_thermal_populations}
\end{equation}
The one- and two-magnon populations are therefore too large to justify a pure ground-state initial condition, and the Hilbert space is truncated at $N_{\max}=4$:
\begin{equation}
    \mathcal{H}_{\mathrm{ret}} =\bigoplus_{N=0}^{4}\mathcal{H}_N,\qquad D=1+6+15+20+15=57.
        \label{eq:ex1_thermal_truncation}
\end{equation}
Only $4.34\times10^{-6}$ of the equilibrium population lies above this space. A third-order pathway that starts in the sector $N$ can reach $N+2$, so that the sectors from which a pathway can leave the retained space are $N\geq3$; their population is $1.03\times10^{-3}$. No collapse channels are included, so the evolution is purely Hamiltonian, and the parameter $\eta=0.03~\mathrm{meV}$ supplies the causal regularization and a Lorentzian broadening of the resolvents; it is not a microscopic relaxation or dephasing rate.

\subsubsection{Bright magnon and bimagnon}

Within the one-magnon manifold, translational invariance gives the dispersion
\begin{equation}
    \omega_1(k) =  g\mu_BB_0-J_z+J_{xy}\cos k .
    \label{eq:ex1_energies}
\end{equation}
For the six-site chain, the one-magnon band extends from $1.1224$ to $1.7224~\mathrm{meV}$ ($0.2714$ to $0.4165$~THz). Because $J_{xy}<0$, the $k=0$ state lies at the bottom of the band. Acting on the polarized ground state, the uniform magnetization operator gives
\begin{equation}
    \hat{M}_+\ket{G}
    =
    \frac{1}{\sqrt L}\sum_j\ket{j}
    =
    \ket{k=0},
\end{equation}
so a spatially uniform THz field selects this bright state from the polarized ground state \cite{lu2017thz2d}.

The two-magnon manifold is organized by the total momentum $K=2\pi m/L$, which the uniform field conserves. Starting from the bright magnon, which has $K=0$, the field reaches only the three two-magnon levels of $K=0$, and the lowest of them,
\begin{equation}
    E_{\min}^{(2)}=2.1108~\mathrm{meV}
\end{equation}
above the polarized ground state, is the bound bimagnon. Its probability of finding the two magnons on neighboring sites, $P_{\mathrm{nn}}=\langle\sum_j\hat{n}_j\hat{n}_{j+1}\rangle=0.6656$, is to be compared with the uncorrelated value $6/\binom{L}{2}=0.4$. The next level of the same sector, at $2.5583~\mathrm{meV}$, has $P_{\mathrm{nn}}=0.2889$ and is the first scattering level. Their distance, $0.4475~\mathrm{meV}$, is a finite-size gap of this six-site ring and not a binding energy: on a ring the relative momenta of the pair are quantized, and the lowest level of the free hard-core pair of the same total momentum is $2\omega_1(\pi/L)=2.3253~\mathrm{meV}$ (Eq.~\eqref{eq:ex1_energies} evaluated at $k=\pi/L$), $0.2145~\mathrm{meV}$ above the bound level. For the infinite chain the bound state at $K=0$ lies at $2(g\mu_BB_0-J_z)+J_z+J_{xy}^2/J_z=2.1313~\mathrm{meV}$, valid for $|J_z|>|J_{xy}|$, that is $0.1136~\mathrm{meV}$ below the threshold $2\omega_1(0)$ of the $K=0$ continuum.

\begin{figure}[htbp]
    \centering
    \includegraphics[width=\linewidth]{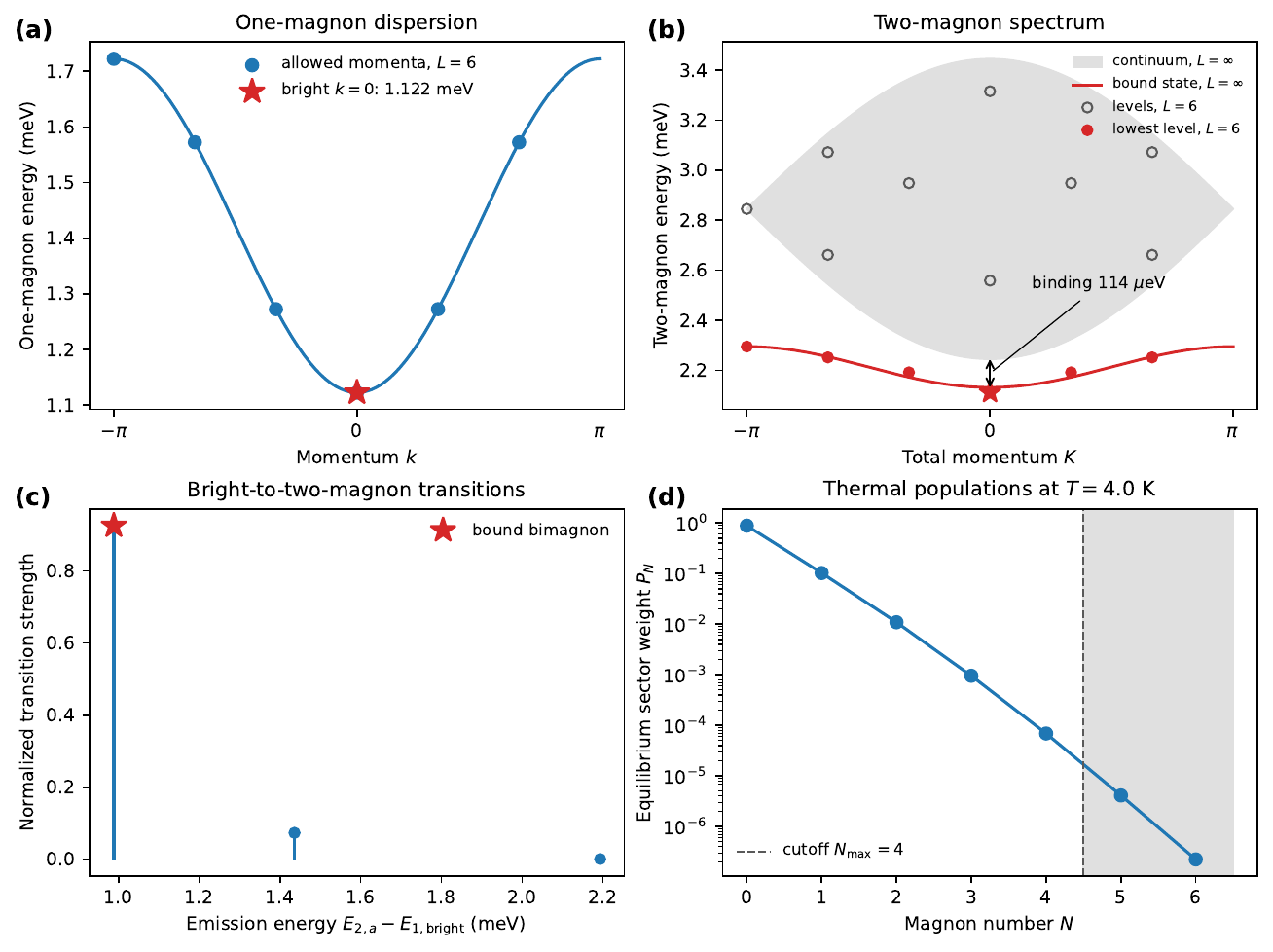}
    \caption{Excitation structure and thermal preparation of the six-site XXZ chain. (a) One-magnon dispersion: the analytic curve $\omega_1(k)$ and the six momenta allowed on the ring. The star marks the $k=0$ state selected by the uniform THz field. (b) Two-magnon spectrum as a function of the total momentum $K$: the levels of the six-site ring (the lowest of each $K$ in red), the continuum of the infinite chain (grey band) and its bound state $E(K)=2\varepsilon+V+4t^2\cos^2(K/2)/V$ (red curve), with $\varepsilon=g\mu_BB_0-J_z$, $t=J_{xy}/2$, and $V=J_z$. The arrow is the binding energy of the infinite chain at $K=0$, $0.114~\mathrm{meV}$, measured from the threshold $2\omega_1(0)$; the star is the lowest level of the ring at $K=0$, which lies $0.134~\mathrm{meV}$ below it. (c) One-to-two-magnon transition strengths from the bright state, which reach the three two-magnon levels of total momentum $K=0$. The lowest is the bound bimagnon, at $0.9884~\mathrm{meV}$ from the bright state, with $92.55\%$ of the accessible strength; the gap to the next line, $0.4475~\mathrm{meV}$, is a finite-size gap (see text). (d) Canonical magnon-sector populations at $T=4~\mathrm{K}$. The dashed boundary after $N=4$ marks the retained thermal Hilbert space; the total population beyond it is $4.34\times10^{-6}$.}
    \label{fig:ex1_manifolds_thermal}
\end{figure}

Starting from the bright $k=0$ state, the bimagnon carries $92.55\%$ of the transition strength generated by $\hat{M}_+$, and its one-to-two-magnon transition energy is
\begin{equation}
    E_{\min}^{(2)}-\omega_1(0)
    =
    0.9884~\mathrm{meV}
    \simeq 0.2390~\mathrm{THz}.
    \label{eq:ex1_bimagnon_transition}
\end{equation}
This transition is the excited-state-absorption feature of the nonlinear calculation. It lies $E_{\min}^{(2)}-2\omega_1(0)=-0.1341~\mathrm{meV}$ from the bright line: that displacement is what the spectrum shows. It is the finite-size value of the binding energy of the bimagnon relative to the threshold $2\omega_1(0)$ of the $K=0$ continuum, which is $0.1136~\mathrm{meV}$ for the infinite chain. Figure~\ref{fig:ex1_manifolds_thermal} shows the dispersion of the magnon, the two-magnon spectrum as a function of the total momentum, the transition strengths from the bright state, and the thermal sector populations.

\subsubsection{Third-order rephasing response}

The rephasing signal has the signature $\mathbf{s}=(-1,+1,+1)$, with the coherence-order history $q=(-1,0,+1)$: the first interaction lowers $q$ by one ($B_u$ or $K_d$) and the other two raise it ($K_u$ or $B_d$). This makes $2\times2\times2=8$ instruction sequences, which the function \texttt{pathways\_\allowbreak from\_\allowbreak signature} enumerates for the signature \texttt{"-++"}, so that none is left out by hand. They are listed in Table~\ref{tab:ex1_sequences}. For the polarized ground state, three of them are non-zero: the ground-state bleach $R_{\mathrm{GSB}}=(B_u,B_d,K_u)$, the stimulated emission $R_{\mathrm{SE}}=(B_u,K_u,B_d)$, and the excited-state absorption $R_{\mathrm{ESA}}=(B_u,K_u,K_u)$. The five others start with a ket lowering or end with two bra lowerings, which the ground state cannot sustain, and are non-zero only when populated sectors exist: at $4~\mathrm{K}$ they are activated by the thermal state. The labels GSB, SE, and ESA therefore name pathway topologies, not processes that start in the ground state: the same sequences act on every populated diagonal block of $\hat{\rho}_\beta$. The sum of the eight is the complete rephasing response of the impulsive rotating-wave model of Sec.~\ref{sec:theory}; the non-rephasing signal has another signature, $\mathbf{s}=(+1,-1,+1)$.

\begin{table}[h]
\centering
\footnotesize
\setlength{\tabcolsep}{6pt}
\begin{tabular}{@{}llccc@{}}
\toprule
sequence & label & \shortstack{population after\\the second interaction} & final coherence & \shortstack{non-zero\\for $N=0$}\\
\midrule
$(B_u,B_d,K_u)$ & GSB & $N$ & $|N{+}1\rangle\langle N|$ & yes \\
$(B_u,K_u,B_d)$ & SE & $N{+}1$ & $|N{+}1\rangle\langle N|$ & yes \\
$(B_u,K_u,K_u)$ & ESA & $N{+}1$ & $|N{+}2\rangle\langle N{+}1|$ & yes \\
$(B_u,B_d,B_d)$ & -- & $N$ & $|N\rangle\langle N{-}1|$ & no \\
$(K_d,K_u,K_u)$ & -- & $N$ & $|N{+}1\rangle\langle N|$ & no \\
$(K_d,K_u,B_d)$ & -- & $N$ & $|N\rangle\langle N{-}1|$ & no \\
$(K_d,B_d,K_u)$ & -- & $N{-}1$ & $|N\rangle\langle N{-}1|$ & no \\
$(K_d,B_d,B_d)$ & -- & $N{-}1$ & $|N{-}1\rangle\langle N{-}2|$ & no \\
\bottomrule
\end{tabular}
\caption{The eight rephasing sequences, $\mathbf{s}=(-1,+1,+1)$, for an initial state in the magnon sector $N$ ($B$ and $K$: bra and ket side; $u$ and $d$: raising and lowering of the magnon number). The state after the second interaction is a population of the sector listed, and the final coherence $|N_{\mathrm{ket}}\rangle\langle N_{\mathrm{bra}}|$ is the one that emits. The last column says whether the sequence survives for the polarized ground state, $N=0$.}
\label{tab:ex1_sequences}
\end{table}

The excitation and emission intervals are evaluated directly in the frequency domain, while the waiting time is fixed at $t_2=0$. The spectral windows are
\begin{equation}
    -1.55\leq\omega_{1Q}\leq-0.55~\mathrm{meV},
    \qquad
    0.55\leq\omega_{\mathrm{emit}}\leq1.55~\mathrm{meV},
    \label{eq:ex1_windows}
\end{equation}
each sampled on a $60\times60$ grid ($\Delta\omega=0.017~\mathrm{meV}$), and the calculation uses the sparse backend with $N_{\max}=4$ ($D=57$). The generator of this closed model is diagonal in the eigenbasis, so that the solver solves each shifted system by division (a sum over states); its agreement with the dense backend is given in Sec.~\SIsecbackends of the Supplementary Material. For branches that start in the zero-magnon sector, all three sequences of Table~\ref{tab:ex1_sequences} that survive share the excitation pole $\omega_{1Q}=-\omega_1(0)=-1.1224~\mathrm{meV}$. The GSB and SE contributions emit at the bright one-magnon energy, $\omega_{\mathrm{emit}}=1.1224~\mathrm{meV}$, whereas the ESA contribution emits at the bright-to-bimagnon transition, $\omega_{\mathrm{emit}}=0.9884~\mathrm{meV}$. Their separation, $0.134~\mathrm{meV}$, exceeds twice the regularization, $2\eta=0.060~\mathrm{meV}$, so the bimagnon feature is resolved on this grid, and it would be for any $\eta<0.067~\mathrm{meV}$. The widths of all the lines of this closed model are set by $\eta$ alone: they are those of the regularization and not of a physical relaxation.

\begin{figure}
\centering
\includegraphics[width=\linewidth]{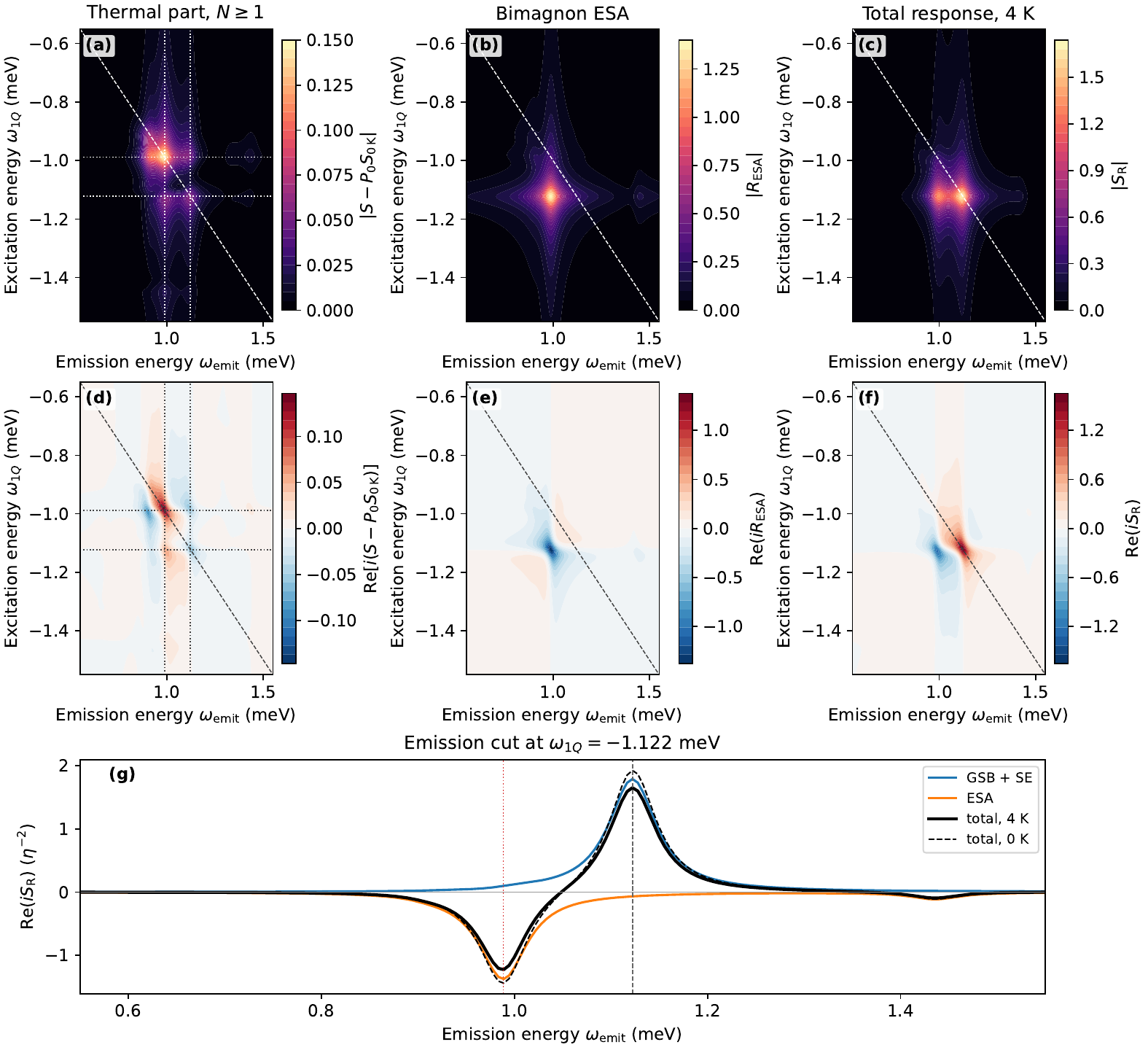}
\caption{Third-order rephasing response of the six-site XXZ chain at $t_2=0$, with its eight sequences, in units of $\eta^{-2}$. Top row, modulus of (a) the thermal part of the response at $4~\mathrm{K}$, $S-P_0S_{0\,\mathrm{K}}$, that is, the contribution of the initial sectors $N\geq1$ (the dotted lines mark the bright and bimagnon energies), (b) the ESA sequence, and (c) the complete rephasing response at $4~\mathrm{K}$. Second row, (d)--(f), real part of $iS$ (the display convention of Sec.~\ref{sec:phase-matching}; red positive, blue negative) of the same three quantities. The dashed line of the maps is the antidiagonal $\omega_{\mathrm{emit}}=-\omega_{1Q}$. (g) Emission cut at the grid point nearest the excitation pole, $\omega_{1Q}=-1.122~\mathrm{meV}$, showing, with their sign, $R_{\mathrm{GSB}}+R_{\mathrm{SE}}$ (positive), $R_{\mathrm{ESA}}$ (negative), the complete response at $4~\mathrm{K}$, and the complete response of the pure ground state (dashed). Vertical markers identify the bright one-magnon emission at $1.1224~\mathrm{meV}$ and the bound-bimagnon ESA pole at $0.9884~\mathrm{meV}$. The figure is drawn on a $181\times181$ grid ($\Delta\omega=0.0056~\mathrm{meV}$); the numbers of the text are those of the $60\times60$ grid of Eq.~\eqref{eq:ex1_windows}.}
\label{fig:ex1_thermal_spectroscopy}
\end{figure}

Thermally populated sectors add hot-band poles to both axes. They remain compatible with the uniform-field selection rule because $\hat{M}_\pm$ conserves the crystal momentum. The thermal part of the response, $S(4~\mathrm{K})-P_0S(0~\mathrm{K})$, is exactly the contribution of the initial sectors $N\geq1$, since $N=0$ holds only the polarized ground state, and is shown in Fig.~\ref{fig:ex1_thermal_spectroscopy}(a) and~(d): its strongest feature sits where both axes carry the one-to-two-magnon energy $0.9884~\mathrm{meV}$, the bimagnon line seen from the thermally populated one-magnon sector. The five sequences that vanish for the ground state add $4.4\%$ of the maximum of the complete response at $4~\mathrm{K}$.

Panels~(d)--(f) show the real part of $iS$ and panel~(g) the emission cut, hence the sign of the lines. In the display convention of Sec.~\ref{sec:phase-matching}, the real part of $iS$, the ground-state bleach and the stimulated emission are positive and the excited-state absorption is negative, as for the two-level system and for Example~2, so that the bimagnon appears as a negative line $0.134~\mathrm{meV}$ below the positive bright line. At the excitation pole of the ground state and at $\omega_{\mathrm{emit}}=\omega_1(0)$ the bleach and the emission each equal $1/\eta^2$ exactly, since the matrix element of the uniform probe from the polarized state is one; the absorption at $\omega_{\mathrm{emit}}=0.9884~\mathrm{meV}$ is $-1.54/\eta^2$, larger than each of them because the probe takes the bright magnon to the bimagnon with a transition strength $1.54$ times that of the first step (the sum of the two-magnon strengths is $2-2/L$ times it, of which the bimagnon carries $92.55\%$).

\subsubsection{Pulses and approximations}
\label{sec:ex1-pulses}

The bright magnon lies at $1.1224~\mathrm{meV}=0.2714~\mathrm{THz}$ and the bimagnon line at $0.2390~\mathrm{THz}$ ($1~\mathrm{meV}=0.2418~\mathrm{THz}$), so that the window of Eq.~\eqref{eq:ex1_windows} is $1~\mathrm{meV}$, or $0.24~\mathrm{THz}$, wide. The signal is the response of impulsive pulses in the rotating-wave approximation: each interaction is an instant whose spectrum is flat across the window, and the counter-rotating terms are dropped. The impulsive rotating-wave response is the limit of a calculation with the pulses themselves, which is not part of the solver: the propagation of three Gaussian pulses of duration $0.02~\mathrm{meV}^{-1}$ in the rotating-wave field, which uses neither resolvents nor pathways, reproduces the $60\times60$ map to $1.0\times10^{-3}$ (Table~\ref{tab:checks}), and thereby checks the signs, prefactors, and kernel of the eight sequences. This check holds within the impulsive rotating-wave model. In the experiment that motivates this example \cite{lu2017thz2d}, signals of second order in the field are also observed (second-harmonic and difference-frequency generation); they are excluded here by construction, since only the third-order rephasing signal is computed.

\subsubsection{Checks}

Three numerical checks validate the model construction and the propagation without relying on the appearance of the spectra.

First, the numerical one-magnon Hamiltonian reproduces the analytical dispersion of Eq.~\eqref{eq:ex1_energies} to machine precision, $\max_k|E_{k,\mathrm{num}}^{(1)}-E_{k,\mathrm{analytic}}^{(1)}|=4.4\times10^{-16}~\mathrm{meV}$.

Second, the canonical density operator of the full $2^6=64$-dimensional spin space is normalized and positive, $|\mathrm{Tr}\,\hat{\rho}_\beta-1|=2.2\times10^{-16}$ and $\min\,\mathrm{eig}(\hat{\rho}_\beta)=2.2\times10^{-7}>0$.

Third, the spectrum converges with the magnon cutoff. Against $N_{\max}=5$ ($D=63$), on every fourth point of the grid, the largest deviation of the eight-sequence response relative to its maximum is $6.4\times10^{-2}$, $4.3\times10^{-3}$, and $2.4\times10^{-4}$ for $N_{\max}=2$, 3, and 4. It follows the population of the initial sectors from which the pathways reach the boundary, $1.15\times10^{-1}$, $1.19\times10^{-2}$, and $1.03\times10^{-3}$, to within a factor of a few; this population is a diagnostic, not a bound on each point of the map. The calculation uses $N_{\max}=4$.

The agreement of the dense and sparse backends and the independent time-domain propagation of the previous subsection are in Sec.~\ref{sec:performance}.

\subsubsection{Scope}

The model assumes exact magnon-number conservation, periodic translation symmetry, a spatially uniform field, and a six-site chain without a dissipative bath. The canonical density operator specifies the initial populations but does not thermalize the system dynamically during the pulse sequence, and $\eta$ is a causal regularization that broadens the lines, not a microscopic temperature-dependent linewidth. The model omits longer-range and interchain exchange, Dzyaloshinskii--Moriya interactions, disorder, and phonons. A static transverse field would couple the magnon sectors and leave no sector structure; the cost of such a model is measured in Sec.~\SIsecscaling of the Supplementary Material, but its spectrum is not computed here. The finite chain replaces the continua of an extended material by discrete levels. Within these limits, the example shows how thermal populations, unequal excitation sectors, a bright one-magnon transition, and an attractive bimagnon enter the third-order rephasing response.

\subsection{Example 2: Two coupled dissipative modes, detection-resolved response}
\label{sec:example-modes}

The second example tests what a closed system cannot: a small open system whose spectrum depends on the observable that reads it out. The model is the coupled-mode model of the Perspective of Paiva-Ortega \emph{et al.}~\cite{paivaortega2026detection}, whose apparent linewidths depend on the readout; here this dependence is computed with several readouts (the polarization, projectors on the one-quantum states, and the excitation number). The example exercises collapse channels whose rates are stored separately from their operators, decay rates read from the generator, pathways written by hand and generated with UFSS, and several readouts evaluated from one propagation. The same model is taken to fifth order in its closed limit in Sec.~\ref{sec:example-modes-chi5}; the comparison with the data deposited with the Perspective is in Sec.~\ref{sec:perf-published}.

\subsubsection{Model}

Each mode $m\in\{a,b\}$ is associated with the annihilation (creation) operators $\hat{b}_m$ ($\hat{b}_m^\dagger$), 
and is considered up to three levels per mode, which gives $D=9$ states in total. The Hamiltonian and the transition operator are
\begin{equation}
    \hat{H}=\sum_{m}\left[\omega_m \hat{b}_m^\dagger \hat{b}_m + \frac{\Delta}{2}\,\hat{b}_m^\dagger \hat{b}_m^\dagger \hat{b}_m \hat{b}_m \right]
    +J\big(\hat{b}_a^\dagger\hat{b}_b+\hat{b}_b^\dagger\hat{b}_a\big),
    \qquad
    \hat{\mu}=\sum_m d_m\big(\hat{b}_m+\hat{b}_m^\dagger\big),
    \label{eq:cm_hamiltonian}
\end{equation}
with $\omega_a=1.0$~eV, $\omega_b=1.2$~eV, $J=1$~meV, $\Delta=10$~meV, and $d_a=d_b=1$, the parameters of the script deposited with the Perspective. The anharmonicity $\Delta$ lifts the two-quantum level of each mode and displaces the excited-state absorption from the stimulated emission. The ground state is $\ket{g}=\ket{0,0}$, and the two one-quantum eigenstates, $\ket{e_a}\approx\ket{1,0}$ and $\ket{e_b}\approx\ket{0,1}$, lie near $1.0$ and $1.2$~eV.

Dissipation is described by seven channels of GKSL form, of four kinds (Table~\ref{tab:cm_channels}). The rate $\gamma_c$ of each channel is the Lindblad coefficient of Eq.~\eqref{eq:gksl-generator} and is stored separately from the operator $\hat{L}_c$ (Sec.~\ref{sec:usage}). The model is built with QuTiP and passed to \texttt{EigenbasisKModel}, which receives the channels as \texttt{(operator, rate)} pairs and the raising part of the transition operator, $\hat{\mu}_+=\sum_m d_m\hat{b}_m^\dagger$, explicitly. The solver uses the dense backend ($81\times81$ Liouville matrix) with $\eta=10~\mu$eV, small against the physical widths of several meV, so that the lines of the maps are those of the generator.

\begin{table}[h]
\centering
\footnotesize
\begin{tabular}{@{}>{\raggedright\arraybackslash}p{0.22\linewidth}lr>{\raggedright\arraybackslash}p{0.36\linewidth}@{}}
\toprule
Channel & $\hat{L}_c$ & coefficient (meV) & Effect\\
\midrule
pure dephasing (each mode) & $\hat{n}_m$ & $\gamma_n=10$ & the coherence between the levels $n$ and $n'$ decays at $\gamma_n(n-n')^2/2$, so $\ket{0}\bra{1}$ at $\gamma_n/2$\\
relaxation (each mode) & $\hat{b}_m$ & $k=1$ & level $n$ decays to $n-1$ at the rate $nk$\\
transfer, upper to lower mode & $\ket{1_a0_b}\bra{0_a1_b}$ & $k_{ba}=5$ & downhill transfer within the one-quantum manifold, no uphill return\\
radiative decay (each mode) & $\ket{0}\bra{1}_m$ & $k_{\mathrm{rad}}=0.1$ & emission from the one-quantum level of the mode\\
\bottomrule
\end{tabular}
\caption{Dissipative channels of Example~2: operator $\hat{L}_c$, Lindblad coefficient $\gamma_c$ (named in the third column), and effect. The transfer coefficient $k_{ba}$ is 5~meV in the baseline case and 10~meV in the second case of Table~\ref{tab:cm_widths}.}
\label{tab:cm_channels}
\end{table}

Besides the transition operator, the readouts below use three detector operators: the projector $\hat{A}_{\mathrm{PL}}=\ket{e_a}\bra{e_a}+\ket{e_b}\bra{e_b}$ on the two one-quantum eigenstates, a photoluminescence-type readout that gives no weight to the ground state or to the two-quantum states; its unbalanced version, $2\ket{e_a}\bra{e_a}+\ket{e_b}\bra{e_b}$; and the excitation number $\hat{N}=\hat{n}_a+\hat{n}_b$.

\subsubsection{Initial state and decay rates}

The calculation starts from $\hat{\rho}_0=\ket{g}\bra{g}$, which has to be stationary: no channel moves population out of $\ket{0,0}$, and $\max|\mathcal{L}\hat{\rho}_0|=0$ for the generator that QuTiP builds from the same channels. The stationary state of the generator, computed by \texttt{solver.steady\_state()}, is unique and equal to this state. \texttt{solver.decay\_rates()} reports, in the eigenbasis of $\hat{H}$, the decay rate of every coherence that the declared channels imply. The coherences $\ket{g}\bra{e_a}$ and $\ket{g}\bra{e_b}$, which set the widths of the two diagonal peaks, decay at
\begin{equation}
    \Gamma_a=\frac{\gamma_n+k+k_{\mathrm{rad}}}{2}=5.55~\mathrm{meV},
    \qquad
    \Gamma_b=\Gamma_a+\frac{k_{ba}}{2}=8.05~\mathrm{meV},
    \label{eq:cm_rates}
\end{equation}
where the coefficients are those of Table~\ref{tab:cm_channels} and the pure-dephasing rate of an optical coherence is $\gamma_n/2$ (Sec.~\ref{Opensystemdynamics}); the second coherence loses in addition the population that the transfer channel takes from $e_b$. These rates agree with the eigenvalues of the QuTiP Liouvillian whose eigenvectors have the largest overlap on each coherence, to better than $10^{-10}$~eV, which checks the normalization of every channel before a spectrum is computed.

\subsubsection{Pathways and coherent response}

The rephasing third-order response of a model with a two-quantum manifold has three pathways, the ground-state bleach $R_{\mathrm{GSB}}=(B_u,B_d,K_u)$, the stimulated emission $R_{\mathrm{SE}}=(B_u,K_u,B_d)$, and the excited-state absorption $R_{\mathrm{ESA}}=(B_u,K_u,K_u)$, all with the coherence-order history $q=(-1,0,+1)$. They are written as \texttt{FrequencyPathway} objects, or generated by UFSS from the phase-discrimination pattern \texttt{[(0,1),(1,0),(1,0)]} with \texttt{maximum\_manifold=2} and translated by \texttt{translate\_ufss\_diagrams}; the generated set is the one written by hand. The excited-state absorption is displaced from the stimulated emission by $\Delta$ and has the opposite sign. The protocol is that of Sec.~\ref{sec:usage}: $t_1$ and $t_3$ are resolved in frequency, on the axes $-1.30\leq\omega_{1Q}\leq-0.90$~eV and $0.90\leq\omega_{\rm emit}\leq1.30$~eV with 201 points each, 2~meV apart, to resolve lines about 6~meV wide, and the waiting time $t_2=10~\mathrm{eV}^{-1}$ stays in the time domain. The rephasing peaks lie at $\omega_{1Q}\approx-\omega_m$ and $\omega_{\rm emit}\approx\omega_{m'}$: two diagonal and two cross peaks.

\subsubsection{Readouts and detector weights}

\begin{figure}[t]
    \centering
    \includegraphics[width=\linewidth]{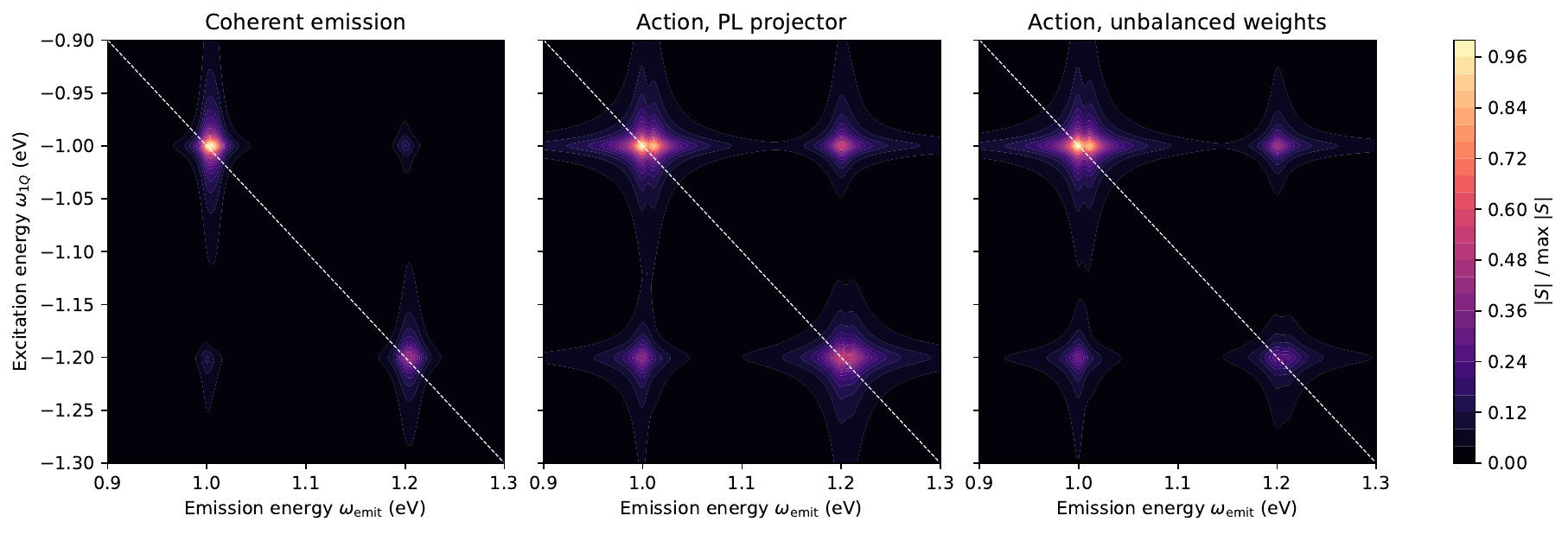}
    \caption{Total rephasing response $|S|$ of the coupled modes at $t_2=10$~eV$^{-1}$ ($201\times201$ points, $\eta=10~\mu$eV), normalized to the maximum of each panel: coherent emission, and two action readouts evaluated from the same propagation, with the projector $\hat{A}_{\mathrm{PL}}$ and with its unbalanced version. The dashed line is the antidiagonal $\omega_{\rm emit}=-\omega_{1Q}$ on which the diagonal peaks lie. The readout with the excitation number gives the coherent map, Eq.~\eqref{eq:cm_weights}, and is not drawn.}
    \label{fig:cm_maps}
\end{figure}

The three pathways are propagated once, and four observables read the same states: the polarization, which is the emitted field, and three action readouts evaluated with \texttt{ObservableSpec.action} and \texttt{fourth\_interaction=("Bu","Kd")}, the sum of the two alternatives that bring the final coherence ($q=+1$) to a population ($q=0$) (Sec.~\SIsectwolevelreadout of the Supplementary Material): the projector $\hat{A}_{\rm PL}$, its unbalanced version, and the excitation number $\hat{N}$. Figure~\ref{fig:cm_maps} shows the total rephasing response of three of them. The coherent map has the two diagonal peaks and, weaker, the two cross peaks. The projector gives the same positions with other relative weights and broader lines; doubling the weight of $e_a$ raises the peaks at $\omega_{\rm emit}=\omega_a$ against those at $\omega_b$.

\paragraph{Weights} Summed over the two alternatives, the readout acts on the final state as the commutator with the lowering part $\hat{\mu}_-$ of the transition operator, up to the phase of the interaction, so that the action signal of a detector $\hat{D}$ is proportional to $\operatorname{Tr}(\hat{D}[\hat{\mu}_-,\hat{\rho}^{(3)}])=\operatorname{Tr}([\hat{D},\hat{\mu}_-]\hat{\rho}^{(3)})$. For a detector that is diagonal in the eigenbasis of $\hat{H}$, with the values $D_x$ on the levels, every final coherence $\ket{x}\bra{y}$, with $x$ one manifold above $y$, enters with the weight $D_x-D_y$:
\begin{equation}
    S^{(4)}_{D}=-i\sum_{x>y}\,(D_x-D_y)\,P_{xy},
    \qquad
    S^{(3)}_{\mathrm{pol}}=\sum_{x>y}P_{xy},
    \label{eq:cm_weights}
\end{equation}
where $P_{xy}$ is the contribution of that coherence to the polarization. The excitation number counts one quantum per manifold: $D_x-D_y=1$ for every final coherence, so that $S^{(4)}_N=-i\,S^{(3)}_{\mathrm{pol}}$ pathway by pathway, the many-level counterpart of Eq.~\eqref{eq:action_pol_identity}. The projector $\hat{A}_{\rm PL}$ counts only the one-quantum states. Its weight is $1$ for the final coherence $\ket{e}\bra{g}$ of the ground-state bleach and of the stimulated emission, as for $\hat{N}$, but $-1$ for the final coherence of the excited-state absorption, which lies between a two-quantum and a one-quantum state ($0-1$). The excited-state absorption, which has the opposite sign in the coherent sum, then enters with the sign of the other pathways, and the apparent lines are broader.

\begin{figure}[t]
    \centering
    \includegraphics[width=\linewidth]{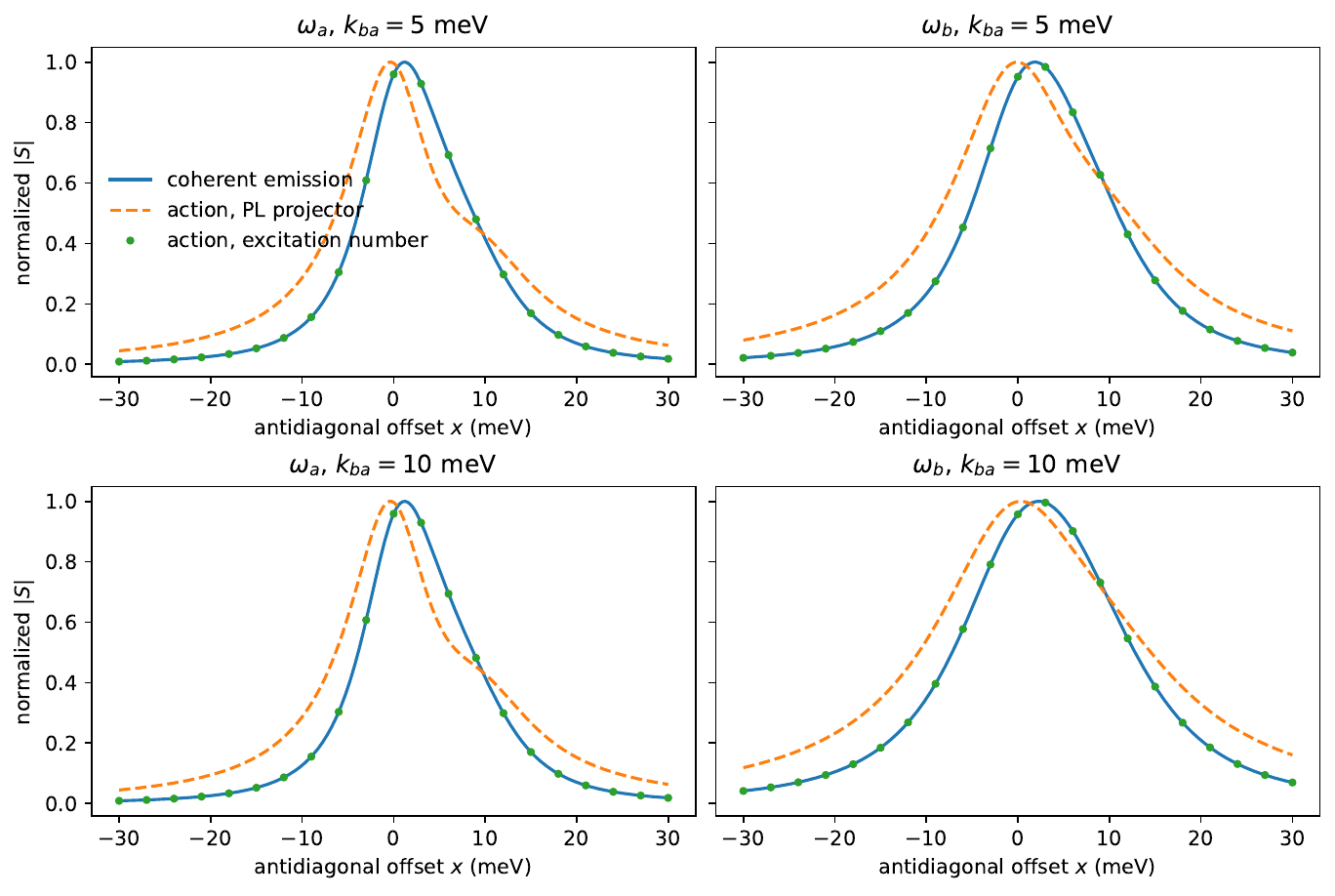}
    \caption{Antidiagonal cuts of the diagonal peaks, normalized to their maximum: coherent emission (solid), action with the projector $\hat{A}_{\mathrm{PL}}$ (dashed), and action with the excitation number (dots, on the coherent curve). Rows: $k_{ba}=5$ and $10$~meV; columns: the peaks at $\omega_a$ and $\omega_b$.}
    \label{fig:cm_cuts}
\end{figure}

\paragraph{Linewidths} To compare the lines, each diagonal peak is cut along the antidiagonal, $(\omega_{1Q},\omega_{\rm emit})=(-\omega_m+x,\ \omega_m+x)$, along which a single pathway with decay rate $\Gamma$ gives $|S|\propto1/(\Gamma^2+x^2)$, a Lorentzian of half width $\Gamma$. The solver evaluates any list of frequencies, so the cut is evaluated at the requested frequencies, without interpolation, on a square patch of $241\times241$ points, 0.25~meV apart, centred on $(-\omega_m,\omega_m)$, whose diagonal is the cut. The half width at half maximum of a Lorentzian with an offset, fitted to $|S|$ along the cut, is the linewidth of Table~\ref{tab:cm_widths}, and Fig.~\ref{fig:cm_cuts} shows the cuts. The second case doubles the transfer rate: only the generator changes, and the pathways and the readouts are the same.

\begin{table}[h]
\centering
\footnotesize
\begin{tabular}{@{}lrrrr@{}}
\toprule
& \multicolumn{2}{c}{$k_{ba}=5$~meV} & \multicolumn{2}{c}{$k_{ba}=10$~meV}\\
\cmidrule(lr){2-3}\cmidrule(lr){4-5}
readout & $\omega_a$ & $\omega_b$ & $\omega_a$ & $\omega_b$\\
\midrule
coherent emission & 5.65 & 7.30 & 5.65 & 9.08 \\
action, $\hat{A}_{\mathrm{PL}}$ & 7.22 & 10.28 & 7.24 & 12.51 \\
action, unbalanced projector & 7.23 & 10.18 & 7.24 & 12.24 \\
action, $\hat{N}$ & 5.65 & 7.30 & 5.65 & 9.08 \\
\bottomrule
\end{tabular}
\caption{Half width at half maximum (meV) of the two diagonal peaks along the antidiagonal, from a Lorentzian with an offset fitted to $|S|$, for the coherent emission and three action readouts of one propagation, at two transfer rates.}
\label{tab:cm_widths}
\end{table}

Three features stand out. The excitation-number readout has exactly the width of the coherent emission, as Eq.~\eqref{eq:cm_weights} requires; the projector is broader than both at each peak; and doubling the transfer rate widens the peak at $\omega_b$, whose population leaves, while the peak at $\omega_a$ barely changes.

\subsubsection{Checks and scope}

\paragraph{Excitation number} With the two readout alternatives summed, $\max|S_N+i\,S_{\mathrm{pol}}|/\max|S_{\mathrm{pol}}|=3\times10^{-16}$ for the sum of the pathways, and $1.3\times10^{-16}$, $1.7\times10^{-16}$, and $3.5\times10^{-16}$ for $R_{\mathrm{GSB}}$, $R_{\mathrm{SE}}$, and $R_{\mathrm{ESA}}$ taken one by one, with dissipation present. Each alternative alone breaks the identity: the relative deviation is 0.91 with $B_u$ alone and 1.71 with $K_d$ alone, so that the identity tests the readout step itself.

\paragraph{Width of one pathway} For the ground-state bleach alone, the antidiagonal cut of each diagonal peak is a Lorentzian whose half width is the decay rate of the coherence plus $\eta$: 5.53 and 7.98~meV, against 5.56 and 8.06~meV, within 1.1\,\%; the tail of the other peak, 0.2~eV away, accounts for the difference. In the sum of the three pathways, the excited-state absorption, displaced by $\Delta$ and of opposite sign, changes the apparent width: the half widths of the total response in Table~\ref{tab:cm_widths} are not the decay rates of Eq.~\eqref{eq:cm_rates}.

\paragraph{Scope} The model is the zero-temperature coupled-mode model of the Perspective: local Markovian channels, the ground state as initial state, impulsive interactions in the rotating-wave approximation, and no inhomogeneous broadening or vibrational structure. The readout operators are modeling choices that the solver applies as defined. The projector $\hat{A}_{\rm PL}$ stands for a photoluminescence-type readout of the one-quantum populations; it does not simulate a particular photoluminescence or photocurrent experiment, whose detection chain defines its own operator. The linewidths of Table~\ref{tab:cm_widths} are those of the total response along the antidiagonal, as an experiment would extract them, not the decay rates of the generator. The deposit of the Perspective uses the excitation number as its action detector; the projector $\hat A_{\rm PL}$ is a readout chosen here, and its linewidths in Table~\ref{tab:cm_widths} are not compared with published values.

\subsubsection{Closed limit: fifth-order two-quantum response}
\label{sec:example-modes-chi5}

Without dissipation, the Hamiltonian of Eq.~\eqref{eq:cm_hamiltonian} is a closed anharmonic system, and its fifth-order double-quantum ($2Q$) response tests the solver on another kind of calculation: five interactions, pathways generated by UFSS, and a pathway set whose completeness depends on the Hilbert space. A fifth-order pathway can reach three excitations before the detection \cite{roslyak2013multiexcitons}, and the truncation to three levels per mode used above contains neither $\ket{3,0}$ nor $\ket{0,3}$. The Hamiltonian is therefore taken on the ten states with $n_a+n_b\leq3$, ordered by manifold. Since $\hat{H}$ conserves $n_a+n_b$, this does not change the energies of the retained manifolds, which lie at $1.000$ and $1.200$~eV (one excitation), at $E_1=2.010$, $E_2=2.200$, and $E_3=2.410$~eV (two), and at $3.03$, $3.21$, $3.41$, and $3.63$~eV (three). The third-order response of the previous subsections, which never leaves two excitations, is identical on the states with $n_a+n_b\leq2$, on those with $n_a+n_b\leq3$, and on three levels per mode. The raising operator $\hat{\mu}_+$ is supplied explicitly, as above, and the model has no collapse channel: the resolvent broadening $\eta=2$~meV, a fifth of the anharmonicity, is the only width of the lines.

\paragraph{Pathways and protocol} The $2Q$ signal is selected by the phase-discrimination signature $--+++$, written in UFSS notation as \texttt{[(0,1),\allowbreak(0,1),\allowbreak(1,0),\allowbreak(1,0),\allowbreak(1,0)]} and passed to \texttt{ufss.DiagramGenerator} with \texttt{maximum\_manifold=3} \cite{rose2021ufss,rose2025ufsssoftware}. \texttt{translate\_ufss\_diagrams} converts the diagrams into \texttt{FrequencyPathway} objects, and \texttt{pathways\_from\_signature("--+++", max\_manifold=3, min\_manifold=0)} returns the same seven sequences without UFSS. They all have the coherence-order history
\begin{equation}
    q=(-1,-2,-1,0,+1).
\end{equation}
The first two interactions are $B_u$ and build the double-quantum coherence $q=-2$; the last three raise $q$ by one unit each. Since $K_u$ and $B_d$ both carry $\Delta q=+1$, eight orderings of the last three interactions are compatible with this history, and the generator returns seven. The all-$B_d$ ordering is missing: after two bra excitations a third bra interaction would take the bra below the ground state, a constraint that does not depend on the truncation. The all-$K_u$ ordering stacks three ket excitations and reaches the three-excitation manifold, so it vanishes at $N_{\max}=2$.

The protocol is \texttt{standard\_nq\_protocol(order=2, nq\_interval=2, detection\_interval=5, n\_interactions=5)}. The double-quantum and emission intervals are resolved in frequency, on $-2.46\leq\omega_{2Q}\leq-1.96$~eV (251 points) and $0.96\leq\omega_{\rm emit}\leq1.26$~eV (301 points), and the three remaining delays are fixed at $t_1=t_3=t_4=2~\mathrm{eV}^{-1}$. The spectrum is generated with \texttt{generate\_nq\_spectrum(2, \ldots)}, which separates the $\pm2Q$ contributions; every selected pathway carries $q=-2$ in the second interval, so the $+2Q$ component is identically zero, which checks the coherence-order bookkeeping but is not an independent validation of the sign convention. The $251\times301$ map takes about 0.1~s on one core with the dense backend (Liouville dimension 100).

\begin{figure}[t]
    \centering
    \includegraphics[width=\linewidth]{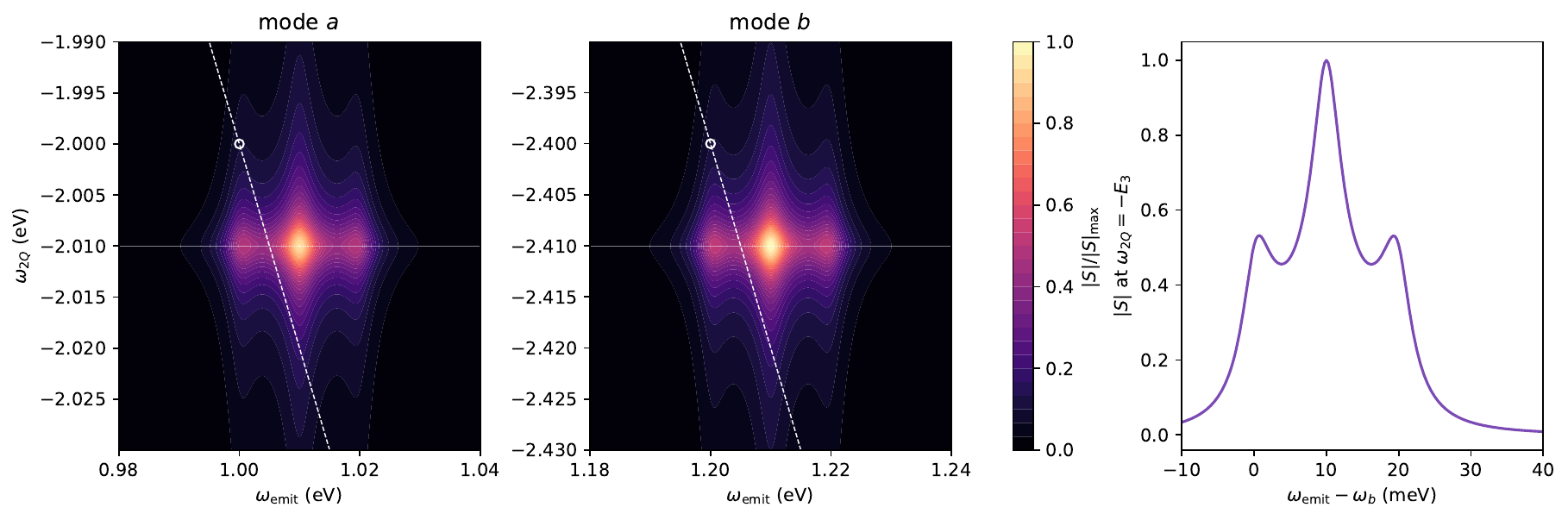}
    \caption{Fifth-order $2Q$ response of the closed coupled-mode model on the ten states with $n_a+n_b\leq3$ ($\eta=2$~meV, $t_1=t_3=t_4=2$~eV$^{-1}$). Left and centre: the groups of lines of the lower and of the upper mode, $|S|$ normalized to the maximum of the map and drawn on a grid four times finer than the $251\times301$ grid of the calculation ($0.25$~meV in both directions); the solid white line marks $\omega_{2Q}=-(2\omega_m+\Delta)$, and the dashed line $\omega_{2Q}=-2\omega_{\mathrm{emit}}$, on which the three lines of a harmonic mode ($\Delta=0$) would merge at the open circle, $(\omega_m,-2\omega_m)$. Right: emission profile at $\omega_{2Q}=-E_3$, with the lines at $\omega_b$, $\omega_b+\Delta$, and $\omega_b+2\Delta$.}
    \label{fig:cm_chi5_map}
\end{figure}

\paragraph{Spectrum} Figure~\ref{fig:cm_chi5_map} shows the two groups of lines of the map. Each mode shows the double-quantum coherence at $\omega_{2Q}=-(2\omega_m+\Delta)$, that is $-2.010$ and $-2.410$~eV, and three emission lines, from the final coherences $\ket{1}\bra{g}$, $\ket{2}\bra{1}$, and $\ket{3}\bra{2}$ of the mode, at $\omega_m$, $\omega_m+\Delta$, and $\omega_m+2\Delta$. The third family exists only because the three-excitation manifold is retained. The strongest line is the central one of the upper mode, at $(-2.410,\,1.210)$~eV; the central line of the lower mode has $0.94$ of its height, and the outer lines about $0.5$. The outer lines are pulled by about 1~meV towards the central one, whose tail adds to them. The state $\ket{1,1}$ at $\omega_a+\omega_b$ does not appear: for independent modes its contribution cancels between the pathways, and it is excited only through the coupling, with a weight, near $\omega_{2Q}=-2.2$~eV, of $1.3\times10^{-3}$ of the maximum at $J=1$~meV and $3.6\times10^{-2}$ at $J=8$~meV.

\begin{figure}[t]
    \centering
    \includegraphics[width=\linewidth]{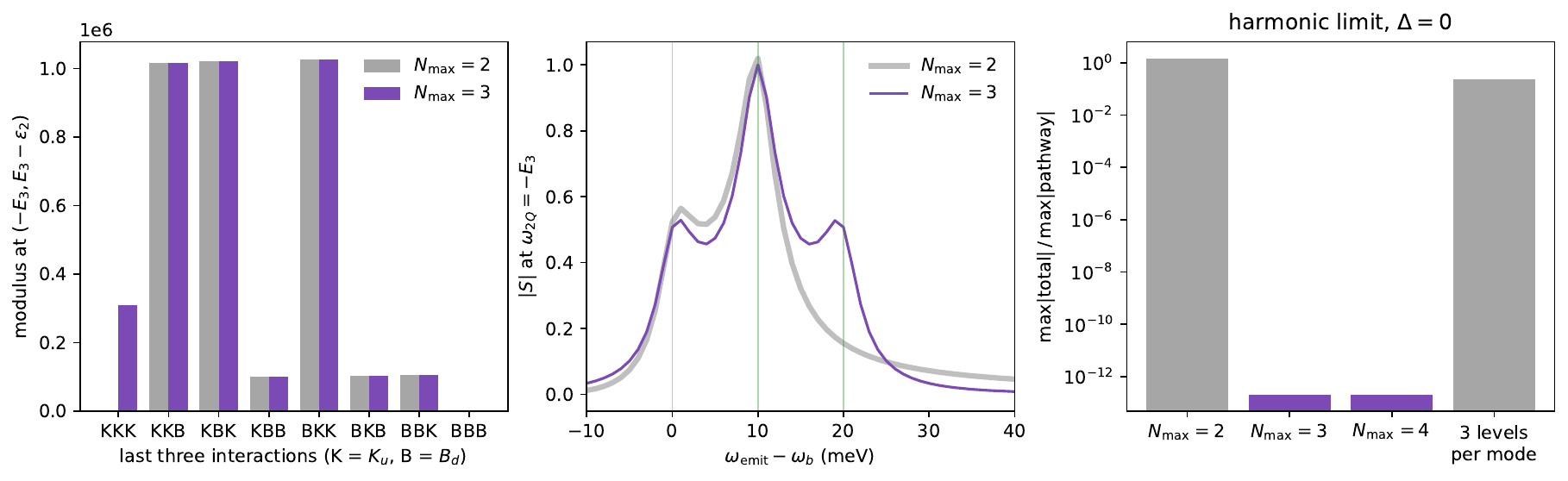}
    \caption{Checks of the fifth-order calculation. (a) Modulus of the pathway response at the strongest line, $(-E_3,\,E_3-\varepsilon_2)$, for the eight orderings of the last three interactions (K stands for $K_u$ and B for $B_d$; the common leading pair $B_u,B_u$ is omitted), at $N_{\max}=2$ and $N_{\max}=3$. (b) Emission profile at $\omega_{2Q}=-E_3$ for the two truncations, normalized to the $N_{\max}=3$ maximum; green lines mark the transitions $E^{(3)}_m-E_k$. (c) Harmonic limit, $\Delta=0$: largest modulus of the sum of the pathways over the largest modulus of a single pathway, for $N_{\max}=2$, $3$, and $4$ and for three levels per mode.}
    \label{fig:cm_chi5_validation}
\end{figure}

\paragraph{Validation} Figure~\ref{fig:cm_chi5_validation} collects three checks. Panel~(a) resolves the response at the strongest line into the eight orderings of the last three interactions. Six of the seven generated orderings are identical at $N_{\max}=2$ and $N_{\max}=3$; the all-$K_u$ ordering rises from zero to $3.1\times10^{5}$ when the three-excitation manifold is kept, about 30\,\% of the largest orderings, and the all-$B_d$ ordering is zero at both truncations. The two vanishing orderings therefore have different origins, and the solver distinguishes them without being told. Panel~(b) shows the emission profile at $\omega_{2Q}=-E_3$ for the two truncations: $N_{\max}=2$ omits the line at $\omega_b+2\Delta$, whose final coherence is $\ket{3}\bra{2}$, and, through interference, changes the others. The maximum changes by only $+1.9\,\%$, but the complex spectra differ by $94.5\,\%$ of the maximum, because the missing pathway is large.

For a harmonic system driven linearly there is no nonlinear response \cite{roslyak2013multiexcitons}, and with $\Delta=0$ the pathways must cancel (panel~(c)). At $N_{\max}=3$ the largest modulus of their sum is $2\times10^{-13}$ of the largest single pathway, and $N_{\max}=4$ gives the same value: no pathway of the selected signal needs four excitations. At $N_{\max}=2$ the ratio is $1.4$: the seven terms sum to zero, so the six that survive sum to exactly minus the omitted one, and the residual harmonic signal is the missing pathway itself. With three levels per mode, which keep $\ket{2,1}$, $\ket{1,2}$, and $\ket{2,2}$ but neither $\ket{3,0}$ nor $\ket{0,3}$, the ratio is $0.23$. The complete three-excitation manifold is therefore the smallest truncation that recovers the cancellation.

A second exact property concerns the coupling. For $J=0$ the Hamiltonian and the transition operator are sums of commuting one-mode terms, and the response of independent systems is the sum of their responses at every order. The two-mode map at $J=0$ equals the sum of the maps of two one-mode models of four levels, built separately from the two-mode model, to $2\times10^{-13}$ of its maximum.

\paragraph{Scope} Truncating by excitation number leaves the resonance positions exact, since $\hat{H}$ conserves $n_a+n_b$, but not the amplitudes. The closed model is not a prediction for the dissipative model of the previous subsections: its widths are set by $\Gamma_a$ and $\Gamma_b$ of Eq.~\eqref{eq:cm_rates} there, and here $\eta$ is a numerical broadening.

\section{Validation and performance}
\label{sec:performance}

The numbers of this section are produced by the scripts of the \texttt{validation/\allowbreak benchmarks} directory of the repository, which build the two-level system and the two examples through a common module that reproduces their published values; the results are stored in \texttt{validation/\allowbreak results/\allowbreak data}. Timings were measured on one core of a laptop processor (BLAS restricted to one thread), with Python~3.13, NumPy~2.5, SciPy~1.18, and QuTiP~5.3, and the memory is the peak working set of the process above its working set after the imports; they indicate scaling rather than optimized performance. Two questions are kept apart. The first is the cost of evaluating a response at chosen frequencies, which the direct evaluation answers at each of them up to $\eta$ and the solver tolerance, with no truncation or leakage from a finite window (Sec.~\ref{sec:perf-scaling}, and Secs.~\SIsecgrid and~\SIsecscaling of the Supplementary Material). The second is the cost of a resolved map on a regular grid, whose number of points grows as the lines narrow and for which a time-domain route is a serious competitor; it is compared at equal resolution and accuracy in Sec.~\ref{sec:perf-resolved}. The comparison with the released QuDPy is in Sec.~\ref{sec:perf-published}; the comparison with a factorized time route, which gives context only, is in Sec.~\ref{sec:perf-resolved}.

\subsection{Analytic and internal checks}
\label{sec:validation-checks}
Table~\ref{tab:checks} lists the checks that have a closed-form or independent reference, with the agreement obtained. The analysis behind the first five rows, for the model of Listing~\ref{lst:twolevel}, is in Sec.~\SIsectwolevel of the Supplementary Material, and the test suite (247 tests, about 30~s) repeats them.

\begin{table}[!htbp]
\centering
\footnotesize
\begin{tabular}{@{}p{0.41\linewidth}p{0.33\linewidth}p{0.19\linewidth}@{}}
\toprule
Check & Reference & Deviation\\
\midrule
\multicolumn{3}{@{}l}{\emph{Open two-level system, closed forms (Listing~\ref{lst:twolevel})}}\\
Amplitudes of $R_1$, $R_2$, $R_3$ versus $t_2$ & $e^{-\gamma_1t_2}$, $1$, $-(1-e^{-\gamma_1t_2})$ & machine precision\\
Emission pole of $R_1$ & $\omega_{eg}-i(\Gamma_2+\eta)$ & $2.000-0.052\,i$~eV\\
Action-detected population & $-i\,S^{(3)}_{\mathrm{pol}}$, Eq.~\eqref{eq:action_pol_identity} & $1.3\times10^{-16}$\\
Integrated fluorescence, $T=5/\gamma_1$ & $P_e(1-e^{-\gamma_1T})$ & machine precision\\
Fluorescence on an infinite window & quantum yield 1 & $10^{-10}$\\
\midrule
\multicolumn{3}{@{}l}{\emph{Identities and limits of Example~2}}\\
Excitation-number readout & $-i\,S^{(3)}_{\mathrm{pol}}$, Eq.~\eqref{eq:cm_weights} & $3\times10^{-16}$\\
Closed limit, harmonic modes & fifth-order signal zero & $2\times10^{-13}$\\
Closed limit, $J=0$ & sum of the one-mode responses & $2\times10^{-13}$\\
\midrule
\multicolumn{3}{@{}l}{\emph{Independent propagation (Example~1)}}\\
Rephasing map ($60\times60$) & three Gaussian pulses of duration $0.02$~meV$^{-1}$ in the rotating-wave field, without resolvents or pathways & $1.0\times10^{-3}$\\
\midrule
\multicolumn{3}{@{}l}{\emph{Dense and sparse backends (Sec.~\SIsecbackends of the Supplementary Material)}}\\
Sparse backend (division or GMRES) against the dense backend, four calculations & dense assembly of $\mathcal{L}$ & $3\times10^{-16}$ to $6\times10^{-11}$\\
\bottomrule
\end{tabular}
\caption{Checks against closed-form results, limits and independent calculations. The deviation is the largest deviation relative to the largest value of the map. The deviation of the trapezoidal option of the integrated fluorescence ($2.1\times10^{-4}$ at 101 steps, $\propto n^{-2}$) is given in Sec.~\SIsectwolevel of the Supplementary Material, and those of each readout alone and of the truncation at two excitations (the harmonic-limit signal is then 1.4 of a single pathway) in Sec.~\ref{sec:example-modes}.}
\label{tab:checks}
\end{table}

\subsection{Comparison with published data and with QuDPy}
\label{sec:perf-published}
The Perspective on detection-defined dephasing in two-dimensional spectroscopy~\cite{paivaortega2026detection} was computed with QuDPy~\cite{shah2023qudpy,shah2022qudpysoftware}, and its data and scripts are deposited~\cite{paivaortega2026data}. They allow two checks on a model that is not one of ours, the two coupled modes of Example~2 with the parameters of the deposited script: against the published spectra, and against the original code run on the same diagrams.

\paragraph{Published spectra}
For each of the six scenarios of the deposit, the spectra of the individual Liouville diagrams are stored on a $382\times382$ frequency grid obtained by transforming a time window of $801$~eV$^{-1}$ sampled every $2.1$~eV$^{-1}$, which is a spacing of $7.85$~meV in frequency, and the half widths of the two diagonal peaks, $a$ and $b$, are obtained from a three-parameter Lorentzian fitted to ten samples along the anti-diagonal through the peak. We evaluate QuDPy-FDGF at the same ten frequencies, with the frequency axis of the transform of the stored time grid, and apply the same estimator to our values. The diagrams are compared one by one, which does not depend on the sign convention used to combine them: the three coherent diagrams $R_1$, $R_2$, $R_3$ and their sum $R_1+R_2-R_3$, and the four fluorescence diagrams $Q_{3a}$, $Q_{4a}$, $Q_{2b}$, $Q_{8b}$ of the action-detected signal, whose detection operator is the excitation number. Our evaluation involves no time window and no time step.

\begin{table}[!htbp]
\centering
\footnotesize
\setlength{\tabcolsep}{3.5pt}
\begin{tabular}{@{}lrrrrrrrr@{}}
\toprule
& \multicolumn{4}{c}{peak $a$} & \multicolumn{4}{c}{peak $b$}\\
\cmidrule(lr){2-5}\cmidrule(lr){6-9}
& \multicolumn{2}{c}{HWHM (meV)} & ampl. & shape & \multicolumn{2}{c}{HWHM (meV)} & ampl. & shape\\
\cmidrule(lr){2-3}\cmidrule(lr){6-7}
diagram & stored & FDGF & ratio & (\%) & stored & FDGF & ratio & (\%)\\
\midrule
$R_1$ (SE) & 5.56 & 5.60 & 0.999 & 0.6 & 8.11 & 8.17 & 1.002 & 2.5 \\
$R_2$ (GSB) & 5.32 & 5.32 & 0.982 & 0.2 & 7.88 & 7.88 & 0.954 & 1.0 \\
$R_3$ (ESA) & 7.53 & 7.51 & 0.988 & 0.6 & 10.33 & 10.35 & 0.990 & 1.7 \\
$R_1+R_2-R_3$ & 5.16 & 5.16 & 1.009 & 0.2 & 7.10 & 7.10 & 0.983 & 0.4 \\
\midrule
$Q_{3a}$ (SE, $B_u$) & 5.61 & 5.65 & 0.999 & 0.6 & 8.16 & 8.22 & 1.002 & 2.5 \\
$Q_{4a}$ (GSB, $B_u$) & 5.37 & 5.37 & 0.982 & 0.2 & 7.93 & 7.93 & 0.954 & 1.0 \\
$Q_{2b}$ (ESA, $B_u$) & 7.83 & 7.80 & 0.987 & 0.7 & 10.53 & 10.56 & 0.989 & 1.7 \\
$Q_{8b}$ (ESA, $K_d$) & 8.10 & 8.07 & 0.985 & 0.8 & 10.71 & 10.73 & 0.989 & 1.6 \\
\bottomrule
\end{tabular}
\caption{Reference scenario of the Perspective ($k_{ba}=5$~meV), diagram by diagram, for the two peaks of the map. Half width at half maximum (meV) from the deposited data (stored) and from QuDPy-FDGF at the same ten frequencies, both through the same three-parameter Lorentzian; ampl.: mean ratio of the two amplitudes in units of the scale of the transform of the deposited time window ($1$ is a perfect agreement); shape: largest difference between the ten samples once this scale is removed, in percent of the largest sample.}
\label{tab:perspective-diagrams}
\end{table}

In the reference scenario the half widths of the eight rows agree within 0.8\,\%, the amplitudes within 5\,\% and the shapes within 2.5\,\% of the peak. Table~\SItabscenarios of the Supplementary Material gives the same comparison for the six scenarios, where the rate of transfer between the two modes varies by a factor of two hundred and the relaxation rate is changed in the last one. Over the six scenarios, the 96 half widths (eight rows, two peaks) agree within 1.4\,\% (0.4\,\% on average), the amplitudes within 8\,\% and the shapes within 3.2\,\%. These differences are of the order of the finite window of the stored run, over which the slowest coherence has decayed by a factor $e^{-4.4}$, of its time step of $2.1$~eV$^{-1}$, which gives a spacing of $7.85$~meV against widths of 5 to 15~meV, and of the three-parameter fit of ten samples; the original QuDPy converges to the present values as its window grows (Table~\ref{tab:perspective-qudpy}). We did not isolate the origin of the amplitude differences, which reach 8\,\%. The half widths of the individual diagrams differ by a factor of 1.5 among themselves (5.3 to 8.1~meV for peak $a$ in the reference scenario), so the agreement is not that of one broad line: each diagram, and the coherent sum whose width is the smallest of the eight, is reproduced.

\paragraph{What this comparison establishes}
For every diagram, QuDPy-FDGF reproduces the stored spectrum within the tolerances above, so that it evaluates the same chain of interactions and propagators for the same model. The reference is a published study and a code of the same group, for the same model: the agreement is a consistency check between two implementations of the same equations, not an independent confirmation of the model. Two things are not compared. First, the combination of the four action-detected diagrams into one signal, which depends on the relative sign of each diagram; here the signs are those of Eq.~\eqref{eq:pathway-prefactor} and of the readout interaction of Eq.~\eqref{eq:readout-interaction}, and with them the excitation-number readout gives exactly the coherent widths (Eq.~\eqref{eq:cm_weights}, Table~\ref{tab:cm_widths}); the combination used for the action-detected widths of Ref.~\cite{paivaortega2026detection} is corrected in Ref.~\cite{paivaortega2026erratum}. Second, the readout of Example~2 with the projector $\hat A_{\mathrm{PL}}$, which is not a detector of the deposit.

\paragraph{Original QuDPy}
The deposit holds results, not the cost of obtaining them, and its window is that of the Perspective. We therefore also ran the original code (version~1.1.0, the public \texttt{coherence2d} with the integrator \texttt{mesolve} of QuTiP at its default tolerances, unmodified) on the same model, with the same diagrams, the delay $t_2=10$~eV$^{-1}$, and the ground state as initial state: the three coherent diagrams, and the four fluorescence diagrams, which the public version handles because it accepts four interactions. QuDPy returns the signal $S(t_1,t_3)$ of each diagram on $N$ samples per time axis, a window $T=N\Delta t$. We transform it to the frequency grid with the kernel $\exp[(i\omega-\eta)t]$ and compare it with QuDPy-FDGF at the same frequencies, after the constant factor of the sign convention of each diagram (a power of $i$, identical at all frequencies and checked on the maps; it is not fitted). The error is the largest deviation relative to the largest value of the QuDPy-FDGF map, the largest over the diagrams. Two grids are used: that of the Perspective ($T=801$~eV$^{-1}$, which puts $57\times57$ frequencies in the band of the map at 7.8~meV), and a grid resolved at 2~meV, a quarter of the width of the narrowest line ($T=1500$~eV$^{-1}$, $N=716$, $226\times226$ frequencies). Table~\ref{tab:perspective-qudpy} gives the results.

\begin{table}[!htbp]
\centering
\footnotesize
\setlength{\tabcolsep}{2.5pt}
\begin{tabular}{@{}lllrrrrrr@{}}
\toprule
& & & \multicolumn{4}{c}{QuDPy} & \multicolumn{2}{c}{QuDPy-FDGF}\\
\cmidrule(lr){4-7}\cmidrule(lr){8-9}
diagrams & grid & step (meV) & $N$ & error & MB & $t$ (s) & MB & $t$ (s)\\
\midrule
coherent, 3 diagrams & $57\times57$ & 7.8 & 382 & $1.7\times10^{-2}$ & 263 & 115 & 1.4 & 0.025 \\
action, 4 diagrams &  &  & 382 & $1.7\times10^{-2}$ & 430 & 2193 & 1.9 & 0.037 \\
\midrule
coherent, 3 diagrams & $226\times226$ & 2.0 & 716 & $1.2\times10^{-3}$ & 877 & 321 & 14 & 0.055 \\
action, 4 diagrams &  &  & 716 & $1.2\times10^{-3}$ & 1242 & 9133 & 21 & 0.10 \\
\bottomrule
\end{tabular}
\caption{Original QuDPy (time propagation with \texttt{mesolve}, $N$ samples per time axis) against QuDPy-FDGF (direct frequency-domain evaluation, dense backend) on the two coupled modes of Example~2, for the three coherent diagrams and the four action diagrams, on the grid of the Perspective and on a grid resolved at 2~meV. Error: largest deviation of QuDPy relative to the largest value of the QuDPy-FDGF map, largest over the diagrams. MB: peak working set above that of the process after the imports; $t$: total time of the diagrams of the row, on one core.}
\label{tab:perspective-qudpy}
\end{table}

The error of QuDPy is that of its window. The signal of a shorter window is the prefix of the signal of a longer one, so that one run gives the error as a function of $T$: on the resolved grid, truncating the coherent signals at $T=300$, 500, 801, 1000, 1250, and 1500~eV$^{-1}$ gives errors of 0.36, 0.13, $2.4\times10^{-2}$, $8.2\times10^{-3}$, $2.5\times10^{-3}$, and $1.2\times10^{-3}$, a factor of 300 for a window five times longer, as the slowest coherence ($\Gamma=5.6$~meV) decays; the action diagrams behave in the same way. The $1.7\times10^{-2}$ of the window of the Perspective is the order of the deviations of Table~\ref{tab:perspective-diagrams}, and an agreement of $10^{-3}$ requires a window almost twice as long, which QuDPy-FDGF does not need: it has no window and no time step, and its result at a frequency is the response with the apodization $\eta$, here $10^{-5}$~eV. It also needs between 0.025 and 0.10~s and between 1.4 and 21~MB, where QuDPy needs between 2~min and 2.5~h and between 0.26 and 1.2~GB, since it stores the $N^2$ propagated states. These ratios, of $5\times10^3$ to $9\times10^4$, are those of the released QuDPy used as is, and they are not a measure of the time-domain approach in general, for two reasons. QuDPy integrates the equation of motion separately for each $(t_1,t_3)$ pair, which a code written for these maps would not do; the time route of Sec.~\ref{sec:perf-resolved} samples the chain once per axis, is closer to what such a code would do, and is beaten by the direct evaluation by a smaller factor (it is itself not the most optimized route possible, see the end of that section). And a time-domain map covers the whole band allowed by its time step, whereas the direct evaluation is made only at the frequencies of interest, which is the use that matters here but not an equal task.

\subsection{Cost versus Hilbert-space dimension}
\label{sec:perf-scaling}
For the spin chain of Example~1 enlarged from $L=4$ to 10 sites ($D=16$ to 386, on a $9\times9$ subgrid of the published window), the dense backend needs 0.7~GB and 29~s at $D=57$ and was not run beyond; the sparse backend with its defaults reaches $D=386$ in 1.6~s with 238~MB. For this closed chain the defaults solve each shifted system by division (a sum over states), so this comparison does not measure an iterative solver; with plain GMRES the number of iterations grows with $D$ and as $\eta$ decreases. The measurements, with the figure, the agreement of the dense and sparse backends, the evaluation of a grid and a model without sector structure, are in Secs.~\SIsecscaling, \SIsecbackends{} and~\SIsecgrid{} of the Supplementary Material.

\subsection{A resolved map at equal resolution}
\label{sec:perf-resolved}
The comparison with a time-domain route must be made at the same resolution and the same accuracy, and the resolution depends on the linewidth. We compare the direct evaluation of the library (dense backend), the sum over states (the generator diagonalized once, each resolvent a division in its eigenbasis) and a factorized time route (the chain sampled once per axis with the one-step propagator and transformed with the kernel $\exp[(i\omega-\eta)t]$, the cheapest step and length being searched for each target error), on the two-level system and on Example~2 in its third-order and closed fifth-order forms; the definitions and Table~\SItabresolved{} are in Sec.~\SIsecresolved{} of the Supplementary Material. For a strongly damped line the direct evaluation and the time route are comparable at a tolerance of $10^{-2}$; the direct evaluation is faster as soon as the tolerance is tighter or the line narrower, by a factor of 2.4 to 190 in the cases tested, and the time route does not reach $10^{-4}$ within 9\,000 samples per axis for lines 0.56~meV wide. The number of points of a resolved map grows as the lines narrow, and the cost of the direct evaluation grows with it; what is gained is that no time step and no window have to be chosen. The sum over states is faster than both for these small models; it is the method of choice when the generator can be diagonalized (this is the idea of the Liouvillian-eigenbasis route of UF$^2$ \cite{rose2019uf2,rose2021uf2open}, here without finite pulses), and the direct evaluation is the one that remains available in the sparse backend beyond that limit.

\paragraph{Scope of the comparison}
Neither time-domain route of this section is tuned. The original QuDPy of Sec.~\ref{sec:perf-published} is used as released, and the factorized time route was written for this comparison: it forms the full $N_t\times N_t$ matrix of samples before transforming it, and transforming the two families of propagated vectors first and contracting them afterwards gives the same maps to $3\times10^{-13}$ of their maximum and, in our tests, a time shorter by a factor of 3 to 12. The speed ratios of Table~\ref{tab:perspective-qudpy} and of Table~\SItabresolved{} of the Supplementary Material are therefore those of these implementations. They are not a bound on what a time-domain method can reach, and no general advantage of the frequency domain over the time domain is claimed. What does not depend on the implementation is that the direct evaluation has no window and no time step to choose.

\section{Discussion}
\label{sec:discussion}

\paragraph{What the framework separates}
QuDPy-FDGF evaluates a nonlinear signal as an ordered chain of interaction superoperators $\mathcal{V}_{p,j}$, time propagators $\mathcal{U}(t)=\exp(\mathcal{L}t)$, frequency-domain resolvents $\mathcal{R}_\eta(\omega)$, and a detection functional. Each element is specified independently: the physical model supplies the Hamiltonian, transition, detection, and dissipative blocks; the protocol fixes which intervals are resolved in time or in frequency; the pathway set fixes the phase-matching or phase-cycling channel; and the observable layer fixes the measured quantity. The examples show that this separation is not only an organizational choice. In the two-level system the same third-order pathway states yield the coherent polarization, the projected excited-state population, and the integrated radiative count, and the identity $S^{(4)}_{P_e}=-i\,S^{(3)}_{\mathrm{pol}}$ is recovered to machine precision. In Example~2 the same three pathways are read by the polarization and by three population-type detectors, and the weight that a detector gives to each final coherence, Eq.~\eqref{eq:cm_weights}, shows why the excitation number reproduces the coherent signal to $3\times10^{-16}$, with dissipation present, whereas a projector on the one-quantum states does not. Detection-dependent quantities can therefore be compared at fixed dynamics, which is the situation encountered when coherently detected and action-detected spectra of the same sample are interpreted together.

\paragraph{References for the checks}
Each part of the chain is checked against a reference; for the coupled modes it comes from the same group. The open two-level system has a closed-form response, which fixes the amplitudes, the poles, and the integrated fluorescence yield (Sec.~\ref{sec:validation-checks}). The coupled modes of Example~2 come from a published study of the same group: their Liouville diagrams, compared one by one, reproduce the deposited data to 1.4\,\% in half width over six scenarios, a difference of the order of the finite window, the time step and the fit of the stored run, and the original QuDPy, run on the same diagrams, approaches the same maps as its window grows, with an error of 0.36 at $T=300~\mathrm{eV}^{-1}$ and $1.2\times10^{-3}$ at $1500~\mathrm{eV}^{-1}$ (Sec.~\ref{sec:perf-published}). This is a consistency check between implementations of the same model; the combination of the action-detected diagrams is not part of it (Sec.~\ref{sec:perf-published}). The propagation of three Gaussian pulses, which uses neither resolvents nor pathways, reproduces the rephasing map of Example~1 to $10^{-3}$ in the impulsive limit, and the harmonic limit of Example~2 cancels its fifth-order signal to $2\times10^{-13}$ once the three-excitation manifold is retained. These checks establish that the chains of Sec.~\ref{sec:theory} are evaluated correctly for the models and conventions tested, not that a model is valid for a given experiment.

\paragraph{Direct frequency-domain evaluation}
Because every frequency-resolved interval is a shifted linear system, the spectral grid is chosen independently of any propagation window, and no discrete Fourier transform, window truncation, or zero padding enters the result; the response at each requested frequency depends only on $\eta$ and on the tolerance of the linear solve. The regularization $\eta$ plays two distinct roles that should be kept apart. As a numerical parameter it moves the shifted system away from the poles of $\mathcal{L}$, which affects the cost and the conditioning of the linear solve; as a physical parameter it adds a uniform Lorentzian broadening to every frequency-resolved interval. In the closed limit of Example~2, $\eta=2$~meV is the only width of the lines; it resolves the three emission lines of each mode, 10~meV apart, and changing it requires no re-propagation. When microscopic dephasing and relaxation are present, as in the two-level system and in the third-order Example~2, where $\eta=10~\mu\mathrm{eV}$ stays far below line widths of 5.6 and 8.1~meV, $\eta$ can be taken much smaller than the physical widths, and the extracted poles coincide with those of the generator. In a closed spin chain $\eta$ alone keeps the shifted systems away from singularity, and the number of plain GMRES iterations per solve grows as it decreases (from about 110 to about 1\,100 between 0.3 and 0.03~meV for Example~1), whereas, in the cases we measured, physical decay rates limit this growth (Sec.~\SIsecscaling of the Supplementary Material). For a closed model given in its eigenbasis the generator is diagonal and the sparse backend solves each shifted system by a division, the sum over states, with no iteration at any $\eta$ measured; this should not be mistaken for a speed-up of the iterative solver on a general generator.

\paragraph{Numerical representation and cost}
The dense Liouville backend constructs the matrix of $\mathcal{L}$ explicitly and serves as a reference for small systems. The sector-resolved backend applies $\mathcal{L}$ through its blocks and never forms the $D_{\mathrm{H}}^2\times D_{\mathrm{H}}^2$ matrix; for pure initial states and unitary time-domain intervals it propagates ket and bra branches separately. Its memory requirement is nevertheless still set by one density vector of size $D_{\mathrm{H}}^2$ as soon as a frequency-domain interval, a mixed state, or a dissipative generator is involved: the present implementation extends the accessible Hilbert-space dimension by avoiding the dense generator, not by avoiding the density vector (Sec.~\SIsecscaling of the Supplementary Material). The sector structure is what keeps $D_{\mathrm{H}}$ small. A static transverse field couples the magnon sectors, the chain becomes a single sector of $2^L$ states, and the calculation, which still involves only $\eta$ and the linear solve, is limited by the diagonalization and the density vectors, not by the resolvent (Sec.~\SIsecscaling of the Supplementary Material). The factorized evaluation of a grid reduces the number of shifted solves from $N^2$ to $N$ per pathway, at the price of storing the vectors of one side of the cut, and the grid can be processed in slabs when this memory is limited. The comparison with the released QuDPy (Sec.~\ref{sec:perf-published}) reflects, in part, that the released code integrates the equation of motion separately for each pair of delays; no general speed advantage of the frequency domain over optimized time-domain codes is claimed. For the smallest models ($D_{\mathrm{H}}\leq10$) the dense backend is faster than the sparse one, whose cost per solve dominates, and the sum over states is faster than both. The two backends agree to within the GMRES tolerance for dissipative nonlinear responses and to rounding when the diagonal solve is exact (Sec.~\SIsecbackends of the Supplementary Material).

\paragraph{Scope and limitations}
The assumptions of the formulation are listed in Sec.~\ref{sec:scope}. Not covered by the present release are non-Markovian bath memory, hierarchical equations of motion, Redfield generators that are not of GKSL form, pulses of finite duration and overlap, counter-rotating terms, signals of other orders in the field, and strong-field corrections. The diagonal preconditioner is exact only when the generator is diagonal in the supplied basis; with collapse channels that couple coherences it reduces the number of iterations but does not make it independent of $\eta$. The package does not construct dissipative channels that satisfy the balance conditions required by a thermal reference state, and the combination of a thermal reference state, dissipation and several excitation sectors is not demonstrated by the examples of this paper: Example~1 is closed and Example~2 starts from the ground state. Finally, a calculation is only as complete as the pathway set supplied: the solver does not add pathways that the user has not requested, and Example~1 obtains its eight rephasing sequences, including those that only a thermal state activates, from the signature with \texttt{pathways\_from\_signature}.

\paragraph{Outlook}
Several extensions follow directly from the present structure. Low-rank and tensor-network representations of the density vector, already explored in the experimental modules, would remove the remaining $D_{\mathrm{H}}^2$ bottleneck for weakly mixed states, and are a possible route to models without sector structure beyond a dozen sites. Because every spectrum is an explicit, differentiable function of the model parameters, the same chains of propagators and resolvents can be differentiated to obtain parameter Jacobians, for example for sensitivity and identifiability analyses of multidimensional spectra across detection schemes, or for fitting models to measured data. Finite-pulse envelopes and time-dependent generators would require replacing single resolvents by integrals over the pulse spectra, which fits the same interval-by-interval design; the validation folder already holds a time-domain engine for Gaussian pulses of finite duration, used here only as an independent check of the impulsive limit.

\section{Conclusions}
\label{sec:conclusions}

QuDPy-FDGF computes pathway- and observable-resolved nonlinear optical responses of finite quantum systems with unitary or Markovian GKSL dynamics. Each Liouville pathway is evaluated as a chain of light--matter interactions, time propagators, and frequency-domain resolvents, so that frequency axes are obtained directly from shifted linear systems while waiting times remain in the time domain. A model-agnostic block contract separates the physical model from pathway construction, numerical propagation, and detection, and the same propagated pathway states serve coherent polarization, projected populations, readout-assisted action detection, and instantaneous or integrated jump observables. Dense and sector-resolved backends provide, respectively, a reference implementation and a matrix-free realization for sector-structured models.

The two-level system and the two examples validate complementary parts of the framework. The open two-level system reproduces the analytical third-order response and its action-detected counterparts. A thermally initialized XXZ chain (Example~1) exercises unequal excitation sectors, rectangular transition blocks, mixed initial states, and the complete set of eight rephasing sequences, among them the five that only a thermal state activates, and it locates the bimagnon as the excited-state-absorption line. Two coupled dissipative modes taken from a published study (Example~2) reproduce the published Liouville diagrams one by one, show that the readout changes the lines, and, in their closed limit, reproduce the harmonic-limit cancellation of the fifth-order two-quantum signal once the three-excitation manifold is retained; the original QuDPy, run on the same diagrams, approaches the same maps as its time window grows. The matrix-free sparse backend reproduces the dense one to within the linear-solver tolerance, including for dissipative nonlinear responses. The sparse backend keeps the memory at the level of density vectors rather than Liouville matrices and, for closed models in their eigenbasis, solves its shifted systems by division, a sum over states, when the generator is diagonal. The factorized evaluation of a grid takes a number of solves proportional to the number of points per axis. The cost of the evaluation is compared with the released QuDPy, and no general speed advantage of the frequency domain over the time domain is claimed. The code is released under the GPLv3 license, together with the examples and a separate reference manual of its programming interface.

\newpage
\bibliographystyle{elsarticle-num}
\bibliography{references}
\newpage

\appendix
\section{Sector-resolved numerical realization of the FDGF formalism}
\label{app:sector-fdgf-realization}

This appendix connects the frequency-domain Green's-function formulation in the main text to its sector-resolved numerical realization. It specifies the mathematical objects exchanged between a physical model and the solver, the action of a Liouville-space pathway on sector blocks, and the relation between the dense and matrix-free backends. Software signatures, accepted input formats, aliases, and plotting options are provided in the separate API documentation.

\subsection{Sector decomposition and basis transformation}
\label{app:sector-decomposition}

Let the truncated Hilbert space be written as a direct sum
\begin{equation}
    \mathcal{H}
    =
    \bigoplus_{s\in\mathcal{S}}\mathcal{H}_s,
    \qquad
    D_{\mathrm{H}}
    =
    \sum_{s\in\mathcal{S}} d_s,
    \qquad
    d_s=\dim(\mathcal{H}_s).
    \label{eq:app-sector-direct-sum}
\end{equation}
The labels in \(\mathcal S\) may denote excitation manifolds, momentum points, symmetry sectors, or another model-defined partition. For an operator \(\hat{A}\), the block that maps a source sector \(s\) to a target sector \(t\) is
\begin{equation}
    \hat{A}_{ts}
    =
    \hat{P}_t \hat{A} \hat{P}_s
    \in
    \mathbb{C}^{d_t\times d_s},
    \label{eq:app-operator-block}
\end{equation}
where \(\hat{P}_s\) projects onto \(\mathcal{H}_s\). This notation does not require the Hamiltonian to be block diagonal: inter-sector blocks \(\hat{H}_{ts}\), with \(t\neq s\), may be retained when they are part of the physical model.

When a model is supplied in a site or orbital basis, every operator must be transformed with the same eigenvectors used to diagonalize the Hamiltonian. For an independent momentum sector \(k\),
\begin{equation}
    \hat{H}_{\mathrm{site}}^{(k)}U^{(k)}
    =
    U^{(k)}E^{(k)},
    \qquad
    \hat{A}_{\mathrm{eig}}^{(k)}
    =
    U^{(k)\dagger}
    \hat{A}_{\mathrm{site}}^{(k)}
    U^{(k)}.
    \label{eq:app-basis-transformation}
\end{equation}
The transformation applies equally to light--matter operators, detection operators, and dissipative jump operators. The raising and lowering components \(\hat{\mu}_+\) and \(\hat{\mu}_-\) (stored as \texttt{j\_plus\_array} and \texttt{j\_minus\_array} in the software interface) may then be supplied directly by the model or separated by the model adapter using signed energy differences and a stated rotating-wave tolerance. The provided independent-\(k\) adapter introduces no coupling between different momentum points; a general sector model may nevertheless contain inter-sector couplings through Eq.~\eqref{eq:app-operator-block}.

A density operator is stored through sector-pair blocks
\begin{equation}
    \rho_{ab}
    =
    \hat{P}_a\hat{\rho} \hat{P}_b
    \in
    \mathbb{C}^{d_a\times d_b}.
    \label{eq:app-density-block}
\end{equation}
Diagonal blocks describe populations within a sector, while \(a\neq b\) describes a coherence between sectors. This block representation is the common starting point for pure-state, low-rank, and full-density calculations.

\subsection{Blockwise realization of Liouville pathways}
\label{app:pathway-block-action}

For a sector block \(\rho_{ab}\), left and right multiplication by an operator \(\hat{A}\), written \(\hat{A}_{\mathrm{L}}\) and \(\hat{A}_{\mathrm{R}}\) in Mukamel's superoperator notation, are assembled from compatible blocks:
\begin{align}
    \hat{A}_{\mathrm{L}}\rho_{ab}
    &=
    \hat{A}\rho_{ab}
    =
    \sum_t \hat{A}_{ta}\rho_{ab},
    &
    (a,b)&\longrightarrow(t,b),
    \label{eq:app-left-block-action}
    \\
    \hat{A}_{\mathrm{R}}\rho_{ab}
    &=
    \rho_{ab}\hat{A}
    =
    \sum_t \rho_{ab}\hat{A}_{bt},
    &
    (a,b)&\longrightarrow(a,t).
    \label{eq:app-right-block-action}
\end{align}
Only dimensionally and sector-compatible terms are retained. A formally allowed interaction therefore contributes zero when the required operator block or target sector is absent from the chosen physical truncation. This is the mechanism by which the sector structure prunes inaccessible pathways without modifying the pathway algebra.

For excitation-manifold sectors, a dyad
\(\ket{N_{\mathrm{ket}},a}\bra{N_{\mathrm{bra}},b}\)
has coherence order
\begin{equation}
    q=N_{\mathrm{ket}}-N_{\mathrm{bra}}.
    \label{eq:app-coherence-definition}
\end{equation}
The elementary increments follow the canonical interaction convention defined in the main text. They determine the requested sequence of sector pairs, whereas Eqs.~\eqref{eq:app-left-block-action} and \eqref{eq:app-right-block-action} determine whether that sequence exists in the supplied model. The coherence-order history is consequently a pathway label, not a substitute for the sector-compatibility test.

For a pathway \(p\) containing \(N_{\mathrm{int}}\) interactions, of which \(N_{\mathrm{bra}}\) act from the right, the default response prefactor is
\begin{equation}
    c_p
    =
    a_p\,i^{N_{\mathrm{int}}}(-1)^{N_{\mathrm{bra}}},
    \label{eq:app-general-pathway-prefactor}
\end{equation}
where \(a_p\) is an optional complex pathway amplitude. An explicitly supplied prefactor replaces Eq.~\eqref{eq:app-general-pathway-prefactor}. Field phases or alternative diagrammatic sign conventions can therefore be included without changing the sector propagation.

\subsection{Common generator and mixed time--frequency propagation}
\label{app:common-generator}

Both numerical backends implement the time-independent GKSL generator $\mathcal{L}$ defined in Eq.~\eqref{eq:gksl-generator}. A time-domain interval applies
\begin{equation}
    \operatorname{vec}(\hat{\rho}(t))
    =
    \exp(\mathcal{L}t)\operatorname{vec}(\hat{\rho}(0)),
    \label{eq:app-time-action}
\end{equation}
whereas a frequency-domain interval evaluates the causal resolvent of Eq.~\eqref{eq:resolvent} through the shifted linear system
\begin{equation}
    \left[
        (\eta-i\omega)\mathbb{1}-\mathcal{L}
    \right]x
    =
    \operatorname{vec}(\hat{\rho}).
    \label{eq:app-frequency-action}
\end{equation}
Consequently, the mixed time--frequency chain in Eq.~\eqref{eq:mixed-pathway-resolvent} is assembled interval by interval: each declared time coordinate uses Eq.~\eqref{eq:app-time-action}, and each declared frequency coordinate uses Eq.~\eqref{eq:app-frequency-action}. No discrete Fourier transform of a propagated time trace is required for a frequency-domain interval, and no additional $-i$ prefactor is introduced by that interval.

The broadening \(\eta>0\) shifts the resolvent away from its poles and may also represent phenomenological homogeneous broadening. It is distinct from the microscopic rates \(\gamma_c\) entering the generator and from any efficiency multiplying an action-detected observable.

\subsection{Evaluation of a grid}
\label{app:grid-evaluation}

A pathway response is the functional $\langle\!\langle d|\,\mathcal{O}_K\cdots\mathcal{O}_1|\hat{\rho}_0\rangle\!\rangle$ of an ordered chain of operations $\mathcal{O}_k$ (interaction superoperators, propagators~\eqref{eq:app-time-action} and resolvents~\eqref{eq:app-frequency-action}) followed by a detection vector $d$. Evaluating it independently at each point of a grid repeats the same solves: for two frequency axes of $N_1$ and $N_2$ points, the pointwise evaluation solves $2N_1N_2$ shifted systems per pathway, although only $N_1$ distinct resolvents enter the first interval and $N_2$ the last. Cutting the chain after the operation $c-1$, between the two scanned intervals, gives
\begin{equation}
    \begin{gathered}
    S_p(\boldsymbol{\xi}_L,\boldsymbol{\xi}_R)
    = c_p\,\langle\!\langle G(\boldsymbol{\xi}_R)\,|\,F(\boldsymbol{\xi}_L)\rangle\!\rangle,\\
    |F\rangle\!\rangle=\mathcal{O}_{c-1}\cdots\mathcal{O}_1|\hat{\rho}_0\rangle\!\rangle,
    \qquad
    |G\rangle\!\rangle=(\mathcal{O}_K\cdots\mathcal{O}_c)^\dagger|d\rangle\!\rangle,
    \end{gathered}
    \label{eq:app-factorized}
\end{equation}
where $\boldsymbol{\xi}_L$ and $\boldsymbol{\xi}_R$ are the points of the axes before and after the cut. The adjoint of a resolvent is the resolvent of the adjoint generator, $\mathcal{R}_\eta(\omega)^\dagger=[(\eta+i\omega)\mathbb{1}-\mathcal{L}^\dagger]^{-1}$, with $\mathcal{L}^\dagger$ the Heisenberg-picture generator; that of a propagator is $\exp(\mathcal{L}^\dagger t)$; the adjoint of a ket interaction $\hat{\rho}\mapsto T\hat{\rho}$ is $\hat{Y}\mapsto T^\dagger\hat{Y}$ and that of a bra interaction $\hat{\rho}\mapsto\hat{\rho}T^\dagger$ is $\hat{Y}\mapsto\hat{Y}T$, with the blocks of the transition operator $T$ acting through their adjoints. The detection vector is $\operatorname{vec}(\hat{O}^\dagger)$ for an operator observable. A readout interaction of an action-detected observable is absorbed into it, $d\to\sum_b\bar{\epsilon}_b\,\mathcal{V}_b^\dagger d$, and an integrated jump count uses its integrated detector $X$ (\ref{app:jump-observables}). Each side is evaluated once per point of its own axes and the grid follows from one matrix product, $N_1+N_2$ solves per pathway and observable. For three or more scanned axes the cut is placed where the number of solves is smallest.

Three properties reduce the work further. Pathways that begin with the same operations share them: the three rephasing pathways of the open two-level system start with the same bra interaction and the same resolvent $\mathcal{R}_\eta(\omega_1)$, and end in the same resolvent $\mathcal{R}_\eta(\omega_3)^\dagger$ applied to the detection vector, so that the whole map takes $N_1+N_2$ solves for all of them. Several observables differ only in $d$, so they share the forward part, and their detection vectors are the columns of one adjoint chain. Operations at fixed coordinates between the two scanned intervals, such as the waiting-time propagator, are applied on the side that has fewer vectors.

The factorized evaluation stores the vectors of one side for the whole grid, $D_{\mathrm L}=D_{\mathrm H}^2$ complex numbers each, and streams the other side in slabs of its first axis, so that the memory held by the Liouville vectors can be bounded. The number of solves is that of the unslabbed evaluation when the grid fits in memory; with several slabs the chains of the streamed side are rebuilt for each slab, the prefixes shared between pathways are lost and the number of solves grows (Sec.~\ref{sec:perf-scaling}). It gives the same numbers as the pointwise evaluation up to the accuracy of the linear solves, which is the only place where the two differ. Both are available (the pointwise one for validation), and the diagnostics of a result report the strategy used and the number of solves and propagations performed. A closed model with a pure initial state and time intervals only is the exception: the sparse backend evaluates it as a rank-one dyad of Hilbert-space vectors, which is cheaper than any Liouville vector, and does not use the factorization.

\subsection{Equivalence and scaling of the numerical backends}
\label{app:numerical-backends}

The dense and matrix-free backends differ only in how they represent and apply Eqs.~\eqref{eq:app-time-action} and \eqref{eq:app-frequency-action}. They do not define different Hamiltonians, dissipators, pathways, or observables. Table~\ref{tab:app-backend-comparison} summarizes the two realizations.

\begin{table}[ht]
    \centering
    \caption{Numerical realization of the common sector-resolved problem.}
    \label{tab:app-backend-comparison}
    \begin{tabular}{@{}l p{0.34\linewidth}p{0.34\linewidth}@{}}
        \toprule
        & Dense Liouville representation & Matrix-free sector representation \\
        \midrule
        Generator
        & Explicit \(D_{\mathrm L}\times D_{\mathrm L}\) matrix
        & Linear action assembled from sector blocks
        \\
        Time interval
        & Dense matrix exponential
        & Krylov exponential action
        \\
        Frequency interval
        & Dense shifted linear solve
        & Division when the generator is diagonal, otherwise GMRES with an optional diagonal preconditioner
        \\
        Primary allocation
        & \(O(D_{\mathrm{H}}^{4})\) for the Liouville matrix
        & Sector operators and, when required, an \(O(D_{\mathrm{H}}^{2})\) density vector
        \\
        Principal role
        & Reference calculations and small systems
        & Sector-resolved calculations and larger sparse models
        \\
        \bottomrule
    \end{tabular}
\end{table}

Here \(D_{\mathrm L}=D_{\mathrm{H}}^{2}\). Materializing the Liouville generator therefore requires \(D_{\mathrm L}^{2}=D_{\mathrm{H}}^{4}\) complex entries. The matrix-free representation avoids this allocation by evaluating the Hamiltonian and dissipative actions directly from their sector blocks, as products of the blocks with the whole matrix $\hat\rho$ ($2+5n_c$ products for $n_c$ dissipative channels). It does not, however, remove the \(D_{\mathrm{H}}^{2}\) state-vector requirement of a general mixed-state, dissipative, or frequency-domain calculation.

The iterative solves report their accuracy. For the shifted system $Sx=b$, $S=(\eta-i\omega)\mathbb{1}-\mathcal{L}$, the code records the number of GMRES iterations and the true relative residual $\|b-Sx\|/\|b\|$, a backward error, and it reports an indicator of the size of the condition number of $S$. For a normal generator the eigenvalues of $S$ are $\eta+\kappa_\nu-i(\omega-\Omega_\nu)$ with $\kappa_\nu\geq0$, hence $\|S^{-1}\|\leq1/\eta$ and $\kappa(S)\leq(\eta+|\omega|+\|\mathcal{L}\|)/\eta$: the relative error of a solve is at most this number times its relative residual, and the conditioning is set by $\|\mathcal{L}\|/\eta$ whatever the dimension. For a dissipative generator the resolvent has norm at most $1/\eta$ in the trace norm, since $\exp(\mathcal{L}t)$ is completely positive and trace preserving, and at most $\sqrt{D_{\mathrm H}}/\eta$ in the Hilbert--Schmidt norm of the solver, so that the indicator can be exceeded by up to a factor $\sqrt{D_{\mathrm H}}$. It is therefore not a bound for a dissipative generator, and $\|\mathcal{L}\|$ is itself obtained by power iteration. We have not measured how close the examples come to this worst case; the error of their solves is controlled by the residual and by the agreement with the dense backend (Sec.~\SIsecbackends of the Supplementary Material).

For a pure initial state undergoing unitary time-domain evolution, the ket and bra branches may instead be propagated separately and recombined as a rank-one dyad. This representation is mathematically equivalent to propagating the corresponding density matrix, up to the numerical accuracy of the Krylov action. A full Liouville-space state is required once the calculation involves a mixed state, a dissipative generator, or the general shifted solve of Eq.~\eqref{eq:app-frequency-action}.

\subsection{Applicability and numerical controls}
\label{app:applicability-controls}

The sector partition fixes the physical truncation before propagation. Increasing the number or dimension of sectors tests truncation convergence, while changing the linear-algebra backend tests the numerical realization at fixed physics. These two convergence questions should not be conflated.

Dense calculations provide a direct reference whenever the full Liouville matrix is affordable. Matrix-free frequency-domain calculations are additionally controlled by the residual of the shifted iterative solve. Failure to satisfy the requested tolerance is treated as a convergence failure rather than as a valid spectrum. Near a resolvent pole, the conditioning depends on both the physical generator and \(\eta\); convergence with respect to \(\eta\), truncation, and iterative tolerance must therefore be assessed separately when peak positions or amplitudes are reported quantitatively.

The present realization assumes a finite-dimensional truncation and a time-independent Markovian generator. Sectorization reduces storage and exposes selection rules, but it does not by itself approximate non-Markovian memory, finite pulse overlap, time-dependent generators, or transitions missing from the model-provided operator blocks.

\section{Analytical benchmark for the dissipative two-level model}

\label{app:two-level-analytical-benchmark}

This appendix collects the closed-form results against which the two-level system of Sec.~\ref{sec:usage-model} is validated. We use the ordered basis \(\{\ket{g},\ket{e}\}\), set \(\hbar=1\), and take
\begin{equation}
    \hat{H}=\omega_{eg}\ket{e}\bra{e},
    \qquad
    \hat{\mu}=d\left(\ket{e}\bra{g}+\ket{g}\bra{e}\right),
    \label{eq:app-two-level-H-J}
\end{equation}
with radiative relaxation \(\hat{L}_{\mathrm{rad}}=\ket{g}\bra{e}\) at rate \(\gamma_1\) and pure dephasing \(\hat{L}_\phi=\hat{\sigma}_z\) at rate \(\gamma_\phi/2\). Evaluating the two dissipators element by element (Supplementary Material, Sec.~\SIsecderivation), or equivalently applying the normalization rule of Sec.~\ref{Opensystemdynamics}, gives the equations of motion
\begin{equation}
\begin{gathered}
    \dot{\rho}_{ee}=-\gamma_1\rho_{ee},
    \qquad
    \dot{\rho}_{gg}=+\gamma_1\rho_{ee},
    \\
    \dot{\rho}_{eg}=-\left(i\omega_{eg}+\Gamma_2\right)\rho_{eg},
    \qquad
    \Gamma_2=\frac{\gamma_1}{2}+\gamma_\phi,
\end{gathered}
    \label{eq:app-two-level-eom}
\end{equation}
with solutions
\begin{equation}
\begin{gathered}
    \rho_{eg}(t)=\rho_{eg}(0)\,\exp[-(i\omega_{eg}+\Gamma_2)t],
    \\
    \rho_{ee}(t)=p_0\exp(-\gamma_1t),
    \qquad
    \rho_{gg}(t)=p_0\left(1-\exp(-\gamma_1t)\right)
\end{gathered}
    \label{eq:app-population-solution}
\end{equation}
for an excited contribution of weight \(p_0\). For the parameters of Sec.~\ref{sec:usage-model}, \(\Gamma_2=0.04/2+0.03=0.05\)~eV.

\subsection{Analytical population pathway}
\label{app:population-pathway}

The population benchmark uses the ordered interactions \((K_u,B_u)\).  With
\(\hat{\mu}_+=d\ket{e}\bra{g}\) and \(\hat{\mu}_-=\hat{\mu}_+^\dagger\), the first interaction
acts on the ket and the second on the bra:
\begin{align}
    \hat{\rho}_g &\xrightarrow{K_u}\hat{\mu}_+\hat{\rho}_g
    =d\ket{e}\bra{g},\\
    d\ket{e}\bra{g}&\xrightarrow{B_u}
    d\ket{e}\bra{g}\hat{\mu}_-
    =d^2\ket{e}\bra{e}.
    \label{eq:app-population-preparation}
\end{align}
Thus, before the waiting-time propagation, the pathway carries the
unnormalized excited population
\begin{equation}
    p_0=d^2.
    \label{eq:app-population-weight}
\end{equation}
Equations~\eqref{eq:app-population-solution} then give
\begin{align}
    P_e(t)&=d^2\exp(-\gamma_1t),\\
    P_g(t)&=d^2\left(1-\exp(-\gamma_1t)\right).
    \label{eq:app-pathway-populations}
\end{align}
For \(d=0.5\), the initial contribution is
\begin{equation}
    P_e(0)=d^2=0.25.
    \label{eq:app-initial-population}
\end{equation}

\subsection{Jump-rate and integrated-fluorescence observables}
\label{app:jump-observables}

Since \(\hat{L}_{\mathrm{rad}}^\dagger\hat{L}_{\mathrm{rad}}=\hat{P}_e\), the instantaneous radiative jump rate and the integrated count over a post-pathway window \([0,T]\) are
\begin{equation}
\begin{gathered}
    I_{\mathrm{rad}}(t)=\gamma_1P_e(t)=\gamma_1d^2\exp(-\gamma_1t),
    \\
    N_{\mathrm{rad}}(T)=\int_0^T I_{\mathrm{rad}}(s)\,ds=P_e(0)\left(1-\exp(-\gamma_1T)\right).
\end{gathered}
    \label{eq:app-integrated-jump-solution}
\end{equation}
The count is the expected number of radiative events in the post-pathway window, not the emission generated before the final pathway state. For \(T=5/\gamma_1\) and \(d=0.5\), \(N_{\mathrm{rad}}=0.25\,(1-\exp(-5))=0.2483155\), which the calculation of Sec.~\ref{sec:validation-checks} reproduces to machine precision with the exact window integral of Eq.~\eqref{eq:integrated-detector}.

\subsection{Analytical third-order rephasing response}
\label{app:chi3-rephasing}

The two-level rephasing calculation uses
\begin{align}
    R_1&=(B_u,K_u,B_d),\\
    R_2&=(B_u,B_d,K_u),\\
    R_3&=(B_u,K_u,K_u).
    \label{eq:app-rephasing-pathways}
\end{align}
The first interaction creates \(\rho_{ge}\), which evolves during the
first interval as
\begin{equation}
    \rho_{ge}(t_1)
    =\rho_{ge}(0)\exp[(i\omega_{eg}-\Gamma_2)t_1].
    \label{eq:app-negative-coherence-propagation}
\end{equation}
Pathway \(R_1\) creates an excited population during \(t_2\), whereas
\(R_2\) creates a ground-state population.  The corresponding population
factors are therefore \(\exp(-\gamma_1t_2)\) and \(1\), respectively.
Pathway \(R_3\) starts from the same excited population as \(R_1\), and its
third interaction \(K_u\) acts on the ground-state population that the
radiative channel has produced from it; its population factor is
\(1-\exp(-\gamma_1t_2)\) (Eq.~\eqref{eq:app-population-solution}), so that it vanishes
at \(t_2=0\) and, without the radiative channel, at all \(t_2\).  After
the third interaction, the three pathways end in \(\rho_{eg}\), which evolves as
\begin{equation}
    \rho_{eg}(t_3)
    =\rho_{eg}(0)\exp[-(i\omega_{eg}+\Gamma_2)t_3].
    \label{eq:app-positive-coherence-propagation}
\end{equation}

With the default pathway prefactor
\begin{equation}
    c_p=i^3(-1)^{N_{\mathrm{bra}}}=
    \begin{cases}
        -i & (p=R_1,R_2;\ N_{\mathrm{bra}}=2),\\
        +i & (p=R_3;\ N_{\mathrm{bra}}=1),
    \end{cases}
    \label{eq:app-pathway-prefactor}
\end{equation}
and four dipole factors, the time-domain rephasing response is
\begin{equation}
    \begin{split}
    R_{\mathrm{R}}^{(3)}(t_3,t_2,t_1)
    ={}&-id^4\exp[-\Gamma_2(t_1+t_3)]
    \exp(+i\omega_{eg}t_1-i\omega_{eg}t_3)\\
    &\times\left[\exp(-\gamma_1t_2)+1-\left(1-\exp(-\gamma_1t_2)\right)\right],
    \end{split}
    \label{eq:app-time-rephasing-response}
\end{equation}
where the three terms of the bracket are \(R_1\), \(R_2\), and \(R_3\), and the bracket equals \(2\exp(-\gamma_1t_2)\).  The pair \(R_1+R_2\) alone would give \(1+\exp(-\gamma_1t_2)\), which differs from the complete response at every \(t_2>0\).

With the implemented causal resolvent $\mathcal{R}_{\eta}$, the corresponding frequency-domain expression is
\begin{equation}
    S_{\mathrm{R}}^{(3)}(\omega_{1Q},\omega_{\mathrm{emit}};t_2)
    =
    \frac{-2id^4\exp(-\gamma_1t_2)}
    {\left[\Gamma_2+\eta-i(\omega_{1Q}+\omega_{eg})\right]
     \left[\Gamma_2+\eta-i(\omega_{\mathrm{emit}}-\omega_{eg})\right]}.
    \label{eq:app-frequency-rephasing-response}
\end{equation}
The factor $-i$ in the numerator is the third-order pathway prefactor inherited from Eq.~\eqref{eq:app-time-rephasing-response}; it is not introduced by either frequency-domain interval. Each denominator follows directly from the modal form in Eq.~\eqref{eq:resolvent-spectral}.
The first denominator is resonant at \(\omega_{1Q}=-\omega_{eg}\), while
the emission denominator is resonant at
\(\omega_{\mathrm{emit}}=+\omega_{eg}\).  This gives the expected rephasing
peak at \((-\omega_{eg},+\omega_{eg})\).  Equation
\eqref{eq:app-frequency-rephasing-response} is the analytical benchmark for
the numerical spectrum generated in Sec.~\ref{sec:validation-checks}, which sums the three pathways
\(R_1\), \(R_2\), and \(R_3\).  The pair \(R_1+R_2\) alone gives the same
expression with \(2\exp(-\gamma_1t_2)\) replaced by \(1+\exp(-\gamma_1t_2)\).

\end{document}


\maketitle

This document collects background material and additional results for the examples of the main text. Section~\ref{app:spin-half-xxz-chain} develops the physics of the spin-$\frac{1}{2}$ XXZ chain of Example~1, with the binding energy of its bimagnon; Section~\ref{supp:twolevel-analysis} analyses the open two-level model of the quickstart, whose response is known in closed form: pathways, choice of the readout interaction, and radiative observables; Section~\ref{supp:two-level-derivation} details the element-by-element derivation behind its analytical benchmark (\ref{M-app:two-level-analytical-benchmark} of the main text); Section~\ref{supp:perspective-scenarios} gives the comparison with the data of the Perspective for the six scenarios of its deposit (Example~2); and Section~\ref{supp:scaling} gives the cost of the spin chain of Example~1 as a function of the Hilbert-space dimension, with its figure and numbers; Section~\ref{supp:backends} the agreement of the dense and sparse backends; Section~\ref{supp:grid} the evaluation of a grid; and Section~\ref{supp:resolved} the comparison of the direct evaluation with a time-domain route at equal resolution. Sections~\ref{supp:scaling} to~\ref{supp:resolved} contain the measurements behind the performance statements of Sec.~\ref{M-sec:performance}; Sections~\ref{app:spin-half-xxz-chain} to~\ref{supp:perspective-scenarios} are not needed to reproduce or verify the calculations of the main text.

\section{Physical interpretation of the spin-$\frac{1}{2}$ XXZ chain}
\label{app:spin-half-xxz-chain}

This section develops the physical model used in Example~1 of the main text. The model is a
minimal description of interacting magnetic excitations in a field-polarized,
quasi-one-dimensional magnet. It is not fitted to a specific compound. Its
purpose is to retain the ingredients needed to distinguish a propagating
one-magnon excitation from an interacting two-magnon state while preserving a
transparent connection between the microscopic Hamiltonian, the excitation
sectors, and the magnetic transitions supplied to the solver.

\subsection{Local moments and exchange interactions}

Each lattice site carries an effective spin-$\frac{1}{2}$ degree of freedom.
The two local basis states, $\ket{\uparrow}_j$ and $\ket{\downarrow}_j$, are
eigenstates of $\hat{S}_j^z$ and represent the two orientations of a localized
magnetic doublet along the quantization axis. They need not correspond to an
isolated electron spin: in a magnetic material they may represent a
low-energy effective doublet obtained after higher-energy electronic and
crystal-field states have been eliminated. The periodic boundary condition
$\hat{S}_{L+1}^{\alpha}=\hat{S}_1^{\alpha}$ removes edge effects from the finite
calculation; it should not be interpreted as requiring a material arranged in
a microscopic ring.

The unperturbed Hamiltonian is
\begin{equation}
\hat{H} =
J_{xy}\sum_j
\left(\hat{S}_j^x\hat{S}_{j+1}^x+\hat{S}_j^y\hat{S}_{j+1}^y\right)
+J_z\sum_j \hat{S}_j^z\hat{S}_{j+1}^z
-g\mu_BB_0\sum_j\hat{S}_j^z.
\label{eq:app-ex1-xxz-hamiltonian}
\end{equation}
The constants $J_{xy}$ and $J_z$ are effective nearest-neighbor magnetic
exchange energies. They absorb the microscopic electronic processes that
produce the magnetic coupling. Equation~\eqref{eq:app-ex1-xxz-hamiltonian}
does not distinguish among direct exchange, superexchange, or other
material-dependent mechanisms. Using $\hat{S}_j^{\pm}=\hat{S}_j^x\pm i\hat{S}_j^y$, the
transverse part can be written as
\begin{equation}
\hat{H}_{xy}
=\frac{J_{xy}}{2}\sum_j
\left(\hat{S}_j^+\hat{S}_{j+1}^-+\hat{S}_j^-\hat{S}_{j+1}^+\right).
\label{eq:app-ex1-transverse-exchange}
\end{equation}
It exchanges the configurations
$\ket{\downarrow_j\uparrow_{j+1}}$ and
$\ket{\uparrow_j\downarrow_{j+1}}$, allowing a local spin deviation to move
without changing the total number of deviations. The $J_z$ term is diagonal
in this basis and assigns an energy according to whether neighboring spins
are parallel or antiparallel. With the sign convention of
Eq.~\eqref{eq:app-ex1-xxz-hamiltonian}, negative $J_{xy}$ and $J_z$ describe
ferromagnetic exchange, while $|J_z|>|J_{xy}|$ defines an easy-axis regime.
The Zeeman term lowers the energy of spins aligned with the static field
$B_0\hat{\mathbf z}$ and stabilizes the polarized reference state
\begin{equation}
\ket{G}=\ket{\uparrow\uparrow\cdots\uparrow}.
\label{eq:app-ex1-polarized-state}
\end{equation}

\subsection{Hard-core magnons}

A magnon is a mobile spin deviation above the polarized reference. Defining
\begin{equation}
\hat{n}_j=\frac{1}{2}-\hat{S}_j^z,
\qquad
\hat{m}_j^{\dagger}=\hat{S}_j^-,
\qquad
\hat{m}_j=\hat{S}_j^+,
\label{eq:app-ex1-magnon-definitions}
\end{equation}
gives $n_j=0$ for an up spin and $n_j=1$ for a down spin. Substituting
$\hat{S}_j^z=1/2-\hat{n}_j$ into Eq.~\eqref{eq:app-ex1-xxz-hamiltonian}, rewriting the
transverse exchange with Eq.~\eqref{eq:app-ex1-transverse-exchange}, and
subtracting the energy $E_G$ of Eq.~\eqref{eq:app-ex1-polarized-state} yields
\begin{equation}
\begin{split}
\hat{H}-E_G={}&
\left(g\mu_BB_0-J_z\right)\sum_j\hat{n}_j
+\frac{J_{xy}}{2}\sum_j
\left(\hat{m}_j^{\dagger}m_{j+1}+m_{j+1}^{\dagger}\hat{m}_j\right)
\\
&+J_z\sum_j\hat{n}_j\hat{n}_{j+1}.
\end{split}
\label{eq:app-ex1-magnon-hamiltonian}
\end{equation}
This form separates three physical effects. The first term is the local energy
cost of creating a spin deviation. The second is a hopping term that turns a
local deviation into a delocalized spin wave. The third is a nearest-neighbor
magnon interaction. For $J_z<0$, two deviations on adjacent sites have a lower
energy than two well-separated deviations, providing the attraction that can
produce a bimagnon bound state \cite{wortis1963bound,wu2022fewmagnon}.

The operators $\hat{m}_j$ are not unrestricted bosonic operators. Since a
spin-$\frac{1}{2}$ can be flipped only once,
\begin{equation}
\left(\hat{m}_j^{\dagger}\right)^2=0,
\qquad
n_j\in\{0,1\}.
\label{eq:app-ex1-hard-core-constraint}
\end{equation}
Magnons in this representation are therefore hard-core particles. At low
density they retain the intuitive language of bosonic spin waves, but the
on-site exclusion remains exact and becomes important in multi-magnon
manifolds.

\subsection{Excitation number, momentum, and solver sectors}

The total magnon number is
\begin{equation}
\hat{N}=\sum_j \hat{n}_j=\frac{L}{2}-\hat{S}_{\mathrm{tot}}^z,
\qquad
[\hat{H},\hat{N}]=0.
\label{eq:app-ex1-conserved-number}
\end{equation}
The spin Hilbert space consequently decomposes as
\begin{equation}
\mathcal{H}=\bigoplus_{N=0}^{L}\mathcal{H}_N,
\qquad
\dim\mathcal{H}_N=\binom{L}{N}.
\label{eq:app-ex1-sector-decomposition}
\end{equation}
These fixed-$N$ spaces are the excitation sectors passed to the solver. They
should not be confused with crystal-momentum sectors. Translation invariance
provides an additional quantum number $k$ within each $\mathcal{H}_N$, but a
given excitation manifold generally contains several momenta. The numerical
sector label answers how many spins are reversed, whereas momentum describes
how a state transforms under translation around the chain.

This distinction also explains the rectangular transition blocks used in
Example~1. The operator $\hat{M}_+$ increases $N$ by one, so its block from
$\mathcal{H}_N$ to $\mathcal{H}_{N+1}$ has dimensions
$\binom{L}{N+1}\times\binom{L}{N}$. The Hamiltonian is diagonalized inside
each fixed-$N$ block, and the same eigenvectors transform every corresponding
transition block. Population blocks lie on $(N,N)$ in Liouville space, while
coherences generated by the THz interactions lie on sector pairs such as
$(N+1,N)$ or $(N+2,N)$.

\subsection{One-magnon band}

Let $\ket{j}=\hat{m}_j^{\dagger}\ket{G}$ denote a spin deviation localized at site
$j$. Translation eigenstates in the one-magnon manifold are
\begin{equation}
\ket{k}=\frac{1}{\sqrt L}\sum_{j=0}^{L-1}\exp(ikj)\ket{j},
\qquad
k=\frac{2\pi q}{L},
\label{eq:app-ex1-one-magnon-state}
\end{equation}
where $q$ is an integer modulo $L$. Applying
Eq.~\eqref{eq:app-ex1-magnon-hamiltonian} gives the dispersion
\begin{equation}
\omega_1(k)=g\mu_BB_0-J_z+J_{xy}\cos k.
\label{eq:app-ex1-one-magnon-dispersion}
\end{equation}
The Zeeman and longitudinal-exchange terms set the overall magnon gap, whereas
$J_{xy}$ produces a bandwidth $2|J_{xy}|$. For the ferromagnetic convention
$J_{xy}<0$, the band minimum occurs at $k=0$. Thus the one-magnon eigenstates
are collective waves extending over the chain, even though they were
constructed from local spin flips.

The meV energy scale of Example~1 is directly in the THz range through
$E=hf$: $1~\mathrm{meV}$ corresponds to approximately
$0.2418~\mathrm{THz}$. The THz frequency characterizes the transition energy
being driven and detected. It does not mean that the exchange interaction is
itself a propagating THz electromagnetic interaction.

\subsection{Two-magnon continuum and bound bimagnon}

The two-magnon basis consists of configurations
\begin{equation}
\ket{j,\ell}=\hat{m}_j^{\dagger}m_{\ell}^{\dagger}\ket{G},
\qquad j<\ell.
\label{eq:app-ex1-two-magnon-basis}
\end{equation}
The transverse exchange moves either spin deviation, while the last term of
Eq.~\eqref{eq:app-ex1-magnon-hamiltonian} changes the energy when the two
deviations occupy neighboring sites. Without an attractive interaction, the
two-particle spectrum is primarily scattering-like and is related to sums of
one-magnon energies. Negative $J_z$ enhances adjacent configurations and can
separate a bound branch from this scattering continuum, as in the established
few-magnon spectrum of ferromagnetic XXZ chains
\cite{wortis1963bound,wu2022fewmagnon}.

A bimagnon is bound in the relative coordinate, not fixed at one position. Its two reversed spins preferentially remain near one another, while the center of mass can remain delocalized and carry a well-defined total momentum $K=2\pi m/L$. The binding energy is defined at fixed $K$, in the sector reached by the transitions of interest: the uniform THz field creates the bright magnon at $k=0$ and the transitions that follow stay in the sector $K=0$. On a ring of $L$ sites the hard-core exclusion shifts the momenta of the free pair, so that the lowest free level of this sector is $2\omega_1(\pi/L)$ and not $2\omega_1(0)$; as $L$ grows the free levels become dense and approach $2\omega_1(0)$, the threshold of the $K=0$ continuum of the infinite chain. Three quantities must not be confused: the gap $E_b=E^{(2)}_{s}(K=0)-E^{(2)}_{\min}$ between the lowest level of the sector and its first scattering level, a finite-size quantity; the distance from the lowest free hard-core level, $2\omega_1(\pi/L)-E^{(2)}_{\min}$; and the binding energy relative to the threshold, $2\omega_1(0)-E^{(2)}_{\min}$, whose infinite-chain value, $2\omega_1(0)-\big[2(g\mu_BB_0-J_z)+J_z+J_{xy}^2/J_z\big]$, is $0.1136$~meV for these parameters. The diagnostics used in the main text are $E_b$ and the probability $P_{\mathrm{nn}}$ that two magnons are on neighboring sites,
\begin{equation}
P_{\mathrm{nn}}=\left\langle\sum_j\hat{n}_j\hat{n}_{j+1}\right\rangle ,
\label{eq:app-ex1-bimagnon-diagnostics}
\end{equation}
which is $L/\binom{L}{2}=0.4$ for an uncorrelated pair on a ring of $L=6$ sites. At $K=0$ the two-magnon sector has three levels, at $2.1108$, $2.5583$, and $3.3155$~meV, with $P_{\mathrm{nn}}=0.67$, $0.29$, and $0.05$: the lowest level is the bound bimagnon, and the two others are scattering levels (their pair probability is below the uncorrelated value, as for hard-core magnons). The gap is $E_b=0.447$~meV; against the lowest free hard-core level, $2\omega_1(\pi/L)=2.325$~meV, the bound level is $0.214$~meV lower; and against the threshold $2\omega_1(0)$ it is $0.134$~meV lower, the finite-size value of the binding energy, which is also the displacement of the excited-state-absorption line from the bright line in the third-order spectrum of the main text. On a six-site ring the scattering continuum is a set of discrete levels, so ``continuum'' refers to the branch that becomes dense as the chain length increases.

\subsection{Coupling to a uniform THz magnetic field}

The oscillating magnetic field of the THz pulse couples to the transverse
magnetization. After absorbing the overall magnetic-dipole scale into the
signal amplitude, the excitation-raising component used in Example~1 is
\begin{equation}
\hat{M}_+=\frac{1}{\sqrt L}\sum_j\hat{S}_j^-,
\qquad
\hat{M}_-=\hat{M}_+^{\dagger}.
\label{eq:app-ex1-uniform-magnetization}
\end{equation}
Here ``raising'' refers to the magnon number: $\hat{S}_j^-$ lowers the microscopic
spin projection but creates one excitation. A far-field THz wavelength is
much larger than an atomic lattice spacing, so the field is effectively
uniform across the model chain and transfers negligible crystal momentum
\cite{lu2017thz2d}. Consequently,
\begin{equation}
\hat{M}_+\ket{G}
=\frac{1}{\sqrt L}\sum_j\ket{j}
=\ket{k=0},
\qquad
\langle k|\hat{M}_+|G\rangle=\delta_{k,0}.
\label{eq:app-ex1-uniform-selection-rule}
\end{equation}
The complete one-magnon band remains part of the Hamiltonian, but only the
$k=0$ state is directly bright from the polarized ground state in this ideal
translation-invariant geometry.

At finite temperature, the same operator connects thermally occupied states
according to $\Delta N=+1$ and, for a uniform probe, $\Delta k=0$. The
third-order pathway can therefore pass from a populated $N$-magnon state to
the $(N+1)$- and $(N+2)$-magnon manifolds. In particular, the
$N=1\rightarrow2$ transition reached by an excited-state-absorption-like
branch provides access to the bound bimagnon. Thermal occupation creates
additional hot-band transitions rather than changing the Hamiltonian itself
\cite{kubo1957response,bibikov2023magnon}. The canonical density operator used
by the solver is an initial statistical mixture; because no thermal bath is
included, the subsequent calculation does not describe dynamical
thermalization.

\subsection{Physical scope of the example}

The model represents the coherent magnetic response of a short easy-axis spin
chain with translation invariance and a longitudinal static field. It retains
anisotropic nearest-neighbor exchange, magnetic coupling to a THz field,
hard-core one- and two-magnon physics, and a controlled thermal initial state.
These ingredients are sufficient to demonstrate how a collective one-magnon
resonance and an interaction-shifted bimagnon transition enter a
sector-resolved nonlinear spectrum.

Several effects needed for a material-specific prediction are intentionally
absent: longer-range and interchain exchange, Dzyaloshinskii--Moriya
interactions, disorder, phonons, crystal-field levels outside the effective
doublet, and microscopic relaxation or dephasing channels. The six-site
periodic chain also produces discrete finite-size levels rather than the
dense continua of a macroscopic sample. The parameter $\eta$ therefore acts
only as a causal resolvent broadening. Within these limits, Example~1 should
be read as a physically interpretable solver demonstration for generic
one-dimensional THz magnets, not as a quantitative simulation of a named
material.

\newpage
\section{The open two-level system of the quickstart}
\label{supp:twolevel-analysis}

This section collects the analysis of the model of the listing of the main text (Sec.~\ref{M-sec:usage}, Listing~\ref{M-lst:twolevel}), whose response is known in closed form: its rates, the pathways of the rephasing signal and the pathway that radiative relaxation makes necessary, the choice of the readout interaction, and the radiative observables. The main text quotes the numbers (Sec.~\ref{M-sec:validation-checks}); the derivation of the closed forms is in \ref{M-app:two-level-analytical-benchmark} of the main text and, element by element, in Sec.~S3.

\subsection{Model and decay rates}
\label{supp:twolevel-model}

The listing of the main text treats an open two-level system, for which the dissipative dynamics and the measured observables are available in closed form. The field-free Hamiltonian and light-matter interaction operator are

\begin{equation}
    \hat{H}= \begin{pmatrix}
        0&0\\ 0 & \omega_{eg}
    \end{pmatrix}, \hspace{3 cm} 
    \hat{\mu} = d 
    \begin{pmatrix}
        0 & 1 \\ 1 & 0
    \end{pmatrix},
\end{equation}

with a transition energy $\omega_{eg}=2$~eV and dipole amplitude $d=0.5$. 
Radiative relaxation is described by
\begin{equation}
    \hat{L}_{\mathrm{rad}} = \ket{g}\bra{e}, \hspace{2 cm} \gamma_1 = 0.04~\mathrm{eV},
\end{equation}
and pure dephasing by
\begin{equation}
    \hat{L}_\phi = \hat{\sigma}_z, \hspace{3.0 cm} \gamma_{\phi} = 0.03~\mathrm{eV}.
\end{equation}

Because $(\mathcal{D}_\phi[\hat{\rho}])_{eg} = (\hat{\sigma}_z\hat{\rho}\hat{\sigma}_z - \hat{\rho})_{eg} = -2\rho_{eg}$, the rate associated with $\hat{L}_\phi$ in the Lindblad generator is set to $\gamma_\phi/2 $. The resulting decay rate of the optical coherence is therefore 
\begin{equation}
    \Gamma_2 = \frac{\gamma_1}{2} + \gamma_\phi = 0.05 \text{ eV}.
    \label{eq:supp-coherence-decay}
\end{equation}

The initial state is the ground-state density operator $\hat{\rho}_g = \ket{g} \bra{g}$, which is stationary under the field-free generator. The closed-form expressions used as references are collected in \ref{M-app:two-level-analytical-benchmark} and compared with the calculation in Sec.~\ref{M-sec:validation-checks} of the main text.

\subsection{Rephasing pathways}
\label{supp:twolevel-pathways}

The complete rephasing signal is the sum over three pathways, which the radiative channel makes necessary (without population transfer $R_1$ and $R_2$ suffice \cite{mukamel1995principles,shah2023qudpy,hybl1998twodimensional}):
\begin{align}
R_1 &= (B_u,K_u,B_d),\\
R_2 &= (B_u,B_d,K_u),\\
R_3 &= (B_u,K_u,K_u).
\end{align}

All three pathways have the coherence-order history
\begin{equation}
    q = (-1,0,+1).
\end{equation}

The first interaction creates a negative one-quantum coherence during the excitation interval. The second converts this coherence into a population, and the third interaction produces the positive one-quantum coherence detected through the polarization. In $R_1$, the system occupies the excited-state population during the waiting time and gives a stimulated-emission contribution. $R_2$ occupies the ground-state population and gives a ground-state-bleach contribution, following the standard third-order pathway interpretation \cite{hybl1998twodimensional,mukamel1995principles}. The sequence $R_3$ is the excited-state-absorption pathway of a three-level system, and it vanishes here when no population is transferred, because the Hilbert space contains no doubly excited manifold. With the radiative channel, however, the excited population created by the first two interactions partly relaxes to $\ket{g}$ during $t_2$, and the third interaction $K_u$ then excites this population. $R_3$ therefore carries the weight $1-\exp(-\gamma_1 t_2)$ and the sign opposite to $R_2$: it describes the refilling of the ground-state bleach. It vanishes at $t_2=0$, and $R_2+R_3$ decays as $\exp(-\gamma_1 t_2)$, so that the complete rephasing signal decays with the population lifetime, as in the optical Bloch model \cite{hybl1998twodimensional}. A diagram generator must keep this sequence when relaxation couples the manifolds \cite{rose2021ufss}.

The rephasing response is evaluated as

\begin{equation}
    \begin{aligned}
        S^{(3)}_{\rm R}(\omega_{1Q},t_2,\omega_{\rm emit})
        ={}& \sum_{p\in\{R_1,R_2,R_3\}} c_p\,
        \langle\langle \hat{\mu} |\,
        \mathcal{R}_{\eta}(\omega_{\rm emit})\,\mathcal{V}_{p,3}\,
        \mathcal{U}(t_2)\,\mathcal{V}_{p,2} \\
        &\times
        \mathcal{R}_{\eta}(\omega_{1Q})\,\mathcal{V}_{p,1}\,
        | \hat{\rho}_g \rangle\!\rangle .
    \end{aligned}
\end{equation}
where $c_p$ is the pathway response prefactor defined in \ref{M-app:pathway-block-action}. The closed form of this sum is given in \ref{M-app:chi3-rephasing}.

The excitation and emission intervals are treated directly in the frequency domain, while the population interval is propagated in time. We use $t_2 = 10~\mathrm{eV}^{-1}$ and a resolvent broadening $\eta=0.002$ eV. The excitation axis spans $-2.5$ to $-1.5$~eV and the emission axis spans $1.5$ to $2.5$~eV. The calculated spectrum is finite over the complete grid and has its maximum at 
\begin{equation}
    (\omega_{\text{1Q}}, \omega_{\text{emit}}) = (-2.000, 2.000) \text{ eV},
\end{equation}

 as expected from the signed-frequency convention for a rephasing pathway. Along the emission axis the response is a simple pole at $\omega_{eg}- i(\Gamma_2 + \eta)$, combining the physical coherence decay of Eq.~\eqref{eq:supp-coherence-decay} with the phenomenological broadening of the frequency-domain resolvent; the numerically extracted pole is $2.000-0.052\,i$ eV. Note that a map of $|S|$ appears $\sqrt{3}$ times broader than this pole width, since the modulus of a Lorentzian amplitude reaches half maximum at $\pm \sqrt{3} \Gamma$.

\subsection{Choosing the readout interaction}
\label{supp:twolevel-readout}

The population and fluorescence observables of the listing apply a readout interaction, \texttt{fourth\_\allowbreak interaction="Bu"}, after the last interval of the pathway (Sec.~\ref{M-observable_functional_and_detection}). The stored pathway therefore remains the third-order sequence ($q=-1,0,+1$), whereas the complete action-detected interaction history is ($q=-1,0,+1,0$): the fourth interaction converts the terminal coherence into a fourth-order population, introduces neither an additional delay nor a new frequency axis, and lets several observables reuse the same propagated pathway.

The projection label is a physical choice rather than a bookkeeping one. The third-order response terminates in the one-quantum sector $q=+1$, so only two instructions carrying $\Delta q= -1$ can close it onto a population sector, and they close it on opposite branches of the density operator:

\begin{equation}
  \mathcal{V}_{B_u}\,\hat{\rho}^{(3)} \propto \ket{e}\bra{e},
  \hspace{2cm}
  \mathcal{V}_{K_d}\,\hat{\rho}^{(3)} \propto \ket{g}\bra{g}.
\end{equation}

$B_u$ closes the bra branch and writes the excited-state population, whereas $K_d$ closes the ket branch and writes the ground-state population; the two branches carry opposite prefactors, so their sum conserves the trace. The remaining instructions $K_u$ and $B_d$ drive the response to $q=+2$, for which both population observables vanish identically.

\begin{figure}
    \centering
    \includegraphics[width=1.0\linewidth]{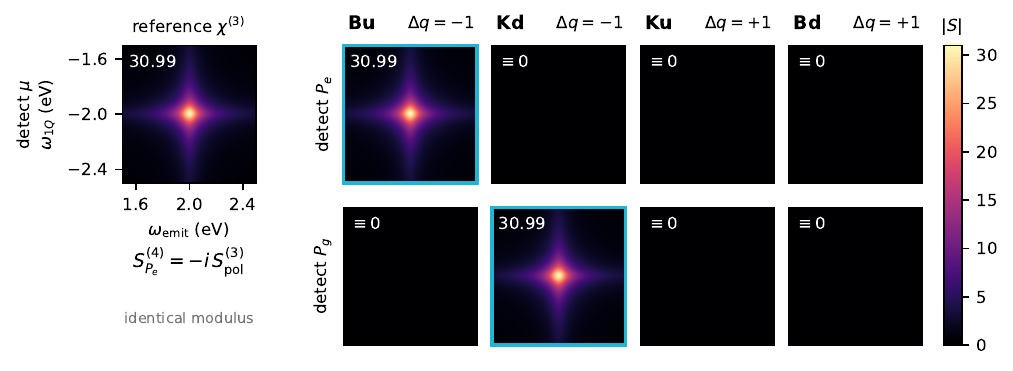}
    \caption{The additional readout interaction selects which population sector is measured. All nine maps come from a single third-order propagation of the rephasing pathways $R_1+R_2+R_3$ at $t_2=10$ eV$^{-1}$, and share one color scale. 
    \textbf{Left}: the reference $\chi^{(3)}$ response, contracted with the transition operator on the final one-quantum coherence, with no projection. 
    \textbf{Right}: the four interaction labels accepted by \texttt{fourth\_interaction}, each combined with the excited- and ground-state population detectors. Only the two labels carrying $\Delta q=-1$ can close the $q=+1$ coherence onto a population, and they close it on opposite branches of the density operator: $B_u$ writes the excited-state population, $K_d$ the ground-state population (highlighted panels). The labels with $\Delta q=+1$ drive the response to $q=+2$, for which both population observables vanish identically ($\equiv 0$). The number in each non-zero panel is $\max|S|$; its equality with the reference panel is the numerical content of Eq.~\eqref{eq:supp-action-pol-identity}, $S^{(4)}_{P_e}=-i\,S^{(3)}_{\rm pol}$: the reference map is third order and the action-detected maps are fourth order. All panels use the axes of the reference panel: $\omega_{\rm emit}$ horizontally, $\omega_{1Q}$ vertically, $241\times241$ points, resolvent broadening $\eta=0.002$ eV.}
    
    \label{fig:qs_projection}
\end{figure}

Figure~\ref{fig:qs_projection} shows this selection rule explicitly. The four projection labels are combined with the two population detectors, and the resulting eight maps are evaluated from a single propagation of $R_1 + R_2 + R_3$ on a shared color scale. Only the $(B_u,\hat{P}_e)$ and $(K_d,\hat{P}_g)$ combinations are non-zero, with identical magnitude; the six remaining maps are identically zero. Because \texttt{fourth\_interaction} also accepts a sequence of interactions, both branches can be summed within a single observable, which is required as soon as a doubly excited manifold contributes.

For the present two-level model the projected response is proportional to the coherent one,
\begin{equation}
  S^{(4)}_{P_e}(\omega_{1Q},t_2,\omega_{\rm emit})
  = -i\,S^{(3)}_{\rm pol}(\omega_{1Q},t_2,\omega_{\rm emit}),
  \label{eq:supp-action-pol-identity}
\end{equation}

an identity satisfied to machine precision over the whole grid, which provides a direct analytical check of the readout-interaction step. Conversely, attaching the same readout interaction to the polarization observable returns exactly zero, since the resulting state carries $q=0$ while the transition operator is purely off-diagonal. The readout interaction and detection observable are therefore independent choices.

\subsection{Radiative observables}
\label{supp:twolevel-radiative}

For population detection, the readout-prepared state is contracted with the excited-state population operator. For fluorescence detection, the same population state is propagated without further field interactions over the specified detection window, and the radiative jump rate is integrated over this interval. With unit detection efficiency, $\varepsilon_{\rm det}=1$, these radiative observables satisfy

\begin{equation}
    I_{rad}^{(4)} = \gamma_1 \text{Tr}[\hat{L}_{\mathrm{rad}}^\dagger \hat{L}_{\mathrm{rad}} \hat{\rho}^{(4)}] = \gamma_1 P_e^{(4)},
\end{equation}
\begin{equation}
    N_{rad}^{(4)}(T) = P_e^{(4)}(1-\exp(-\gamma_1T)).
    \label{eq:supp-integrated-fluorescence}
\end{equation}

The first relation is exact in the numerical implementation. The second is reproduced to machine precision at $T=5/\gamma_1=125$~eV$^{-1}$, since the window integral is evaluated exactly through the integrated detector of Eq.~\eqref{M-eq:integrated-detector}. The trapezoidal option, \texttt{integration="trapezoid"}, samples the window instead; it reaches a relative error of $2.1\times10^{-4}$ with $\texttt{n\_steps}=101$ and converges as $\texttt{n\_steps}^{-2}$. On an infinite window, \texttt{time\_window=(0.0, np.inf)}, the integrated fluorescence of $R_1$ equals the population signal to $10^{-10}$: the stationary state is the ground state, which counts no jump, so the result is the total photon count, the radiative yield of one.

\begin{figure}[t]
    \centering
    \includegraphics[width=1.0\linewidth]{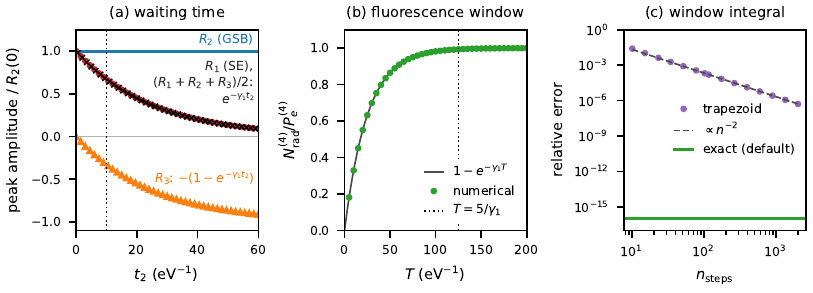}
    \caption{
    Analytical validation of the action-detected observables, evaluated at the peak coordinate $(\omega_{1Q},\omega_{\rm emit})=(-2.000,2.000)$ eV. (a) Peak amplitude of each rephasing pathway versus waiting time, in units of the $t_2=0$ amplitude of $R_2$. The stimulated-emission pathway $R_1$ follows $\exp(-\gamma_1 t_2)$ to machine precision ($0.670320$ at $t_2=10$ eV$^{-1}$), the ground-state bleach pathway $R_2$ is constant, and $R_3$, which vanishes at $t_2=0$, follows $-[1-\exp(-\gamma_1 t_2)]$ because radiative relaxation refills the ground state; the sum $(R_1+R_2+R_3)/2$ follows the same exponential as $R_1$. The dotted line marks the waiting time used in Fig.~\ref{fig:qs_projection}. (b) Integrated radiative signal $N_{\rm rad}^{(4)}(T)/P_e^{(4)}$ for pathway $R_1$ as a function of the detection window, compared with Eq.~\eqref{eq:supp-integrated-fluorescence}; the dotted line marks the window $T=5/\gamma_1=125$ eV$^{-1}$ used in the example. (c) Relative error of (b) at that window: the trapezoidal option converges as $n_{\rm steps}^{-2}$ ($2.1\times10^{-4}$ at $n_{\rm steps}=101$), whereas the default exact evaluation is at machine precision (drawn at $10^{-16}$).}
    \label{fig:qs_validation}
\end{figure}

Figure~\ref{fig:qs_validation} collects the corresponding checks. Panel (a) follows the peak amplitude of each pathway as a function of the waiting time, in units of the $t_2=0$ amplitude of $R_2$: radiative relaxation attenuates the stimulated-emission contribution $R_1$ as $\exp(-\gamma_1 t_2)$, reproduced to machine precision, the ground-state-bleach pathway $R_2$ is constant because the ground state is stationary, and $R_3$ follows $-[1-\exp(-\gamma_1 t_2)]$. Their sum, the complete rephasing response, is $2\exp(-\gamma_1 t_2)$. The pair $R_1+R_2$ alone would give $1+\exp(-\gamma_1 t_2)$, which is 25\% larger at the waiting time $t_2=10$~eV$^{-1}$ used here. Panel (b) compares the integrated fluorescence with Eq.~\eqref{eq:supp-integrated-fluorescence} over the full range of detection windows, and panel (c) compares the exact evaluation with the trapezoidal option.

\newpage
\section{Element-by-element derivation for the open two-level model}
\label{supp:two-level-derivation}

This section details the derivation summarized in \ref{M-app:two-level-analytical-benchmark} of the main text: the action of the radiative and pure-dephasing dissipators on every density-matrix element of the open two-level model of the quickstart, with $\hat{H}=\omega_{eg}\ket{e}\bra{e}$ in the ordered basis $\{\ket{g},\ket{e}\}$ and $\hbar=1$.

The GKSL dissipator of channel \(c\), with jump operator \(\hat{L}_c\), is
\begin{equation}
    \mathcal{D}_c[\hat{\rho}]
    =\hat{L}_c\hat{\rho} \hat{L}_c^\dagger
    -\frac{1}{2}\hat{L}_c^\dagger \hat{L}_c\hat{\rho}
    -\frac{1}{2}\hat{\rho} \hat{L}_c^\dagger \hat{L}_c.
    \label{eq:app-dissipator}
\end{equation}

\subsection{Radiative relaxation}
\label{supp:radiative-relaxation}

Radiative relaxation is described by
\begin{equation}
    \hat{L}_{\mathrm{rad}}=\ket{g}\bra{e},
    \qquad
    \hat{L}_{\mathrm{rad}}^\dagger=\ket{e}\bra{g}.
    \label{eq:app-C-rad}
\end{equation}
The product entering the anticommutator is
\begin{equation}
    \hat{L}_{\mathrm{rad}}^\dagger \hat{L}_{\mathrm{rad}}
    =\ket{e}\bra{g}\ket{g}\bra{e}
    =\ket{e}\bra{e}
    \equiv \hat{P}_e,
    \label{eq:app-C-rad-product}
\end{equation}
where \(\hat{P}_e\) is the projector onto the excited state.  Substitution into
Eq.~\eqref{eq:app-dissipator} gives
\begin{equation}
    \mathcal{D}_{\mathrm{rad}}[\hat{\rho}]
    =\ket{g}\bra{e}\hat{\rho}\ket{e}\bra{g}
    -\frac{1}{2}\hat{P}_e\hat{\rho}
    -\frac{1}{2}\hat{\rho} \hat{P}_e.
    \label{eq:app-rad-dissipator-expanded}
\end{equation}

The first term is the quantum jump term.  Since
\(\bra{e}\ket{g}=0\), its contribution to the optical coherence is zero:
\begin{align}
    \bra{e}\hat{L}_{\mathrm{rad}}\hat{\rho} \hat{L}_{\mathrm{rad}}^\dagger\ket{g}
    &=\bra{e}\ket{g}\bra{e}\hat{\rho}\ket{e}\bra{g}\ket{g}\\
    &=0.
    \label{eq:app-rad-jump-coherence}
\end{align}
For the first anticommutator term,
\begin{align}
    \bra{e}\hat{P}_e\hat{\rho}\ket{g}
    &=\bra{e}\ket{e}\bra{e}\hat{\rho}\ket{g}
    =\rho_{eg}.
    \label{eq:app-rad-left-coherence}
\end{align}
For the second anticommutator term,
\begin{align}
    \bra{e}\hat{\rho} \hat{P}_e\ket{g}
    &=\bra{e}\hat{\rho}\ket{e}\bra{e}\ket{g}
    =0.
    \label{eq:app-rad-right-coherence}
\end{align}
Consequently,
\begin{equation}
    \left(\mathcal{D}_{\mathrm{rad}}[\hat{\rho}]\right)_{eg}
    =0-\frac{1}{2}\rho_{eg}-0
    =-\frac{1}{2}\rho_{eg}.
    \label{eq:app-rad-coherence-result}
\end{equation}

The complete matrix form makes the effect on all density-matrix elements
explicit:
\begin{equation}
    \mathcal{D}_{\mathrm{rad}}[\hat{\rho}]
    =
    \begin{pmatrix}
        \rho_{ee} & -\rho_{ge}/2\\
        -\rho_{eg}/2 & -\rho_{ee}
    \end{pmatrix}.
    \label{eq:app-rad-matrix-dissipator}
\end{equation}
After multiplication by the population-relaxation rate \(\gamma_1\), this
term transfers population from \(\ket{e}\) to \(\ket{g}\), while it damps
the optical coherence at the amplitude rate \(\gamma_1/2\):
\begin{align}
    \left.\dot{\rho}_{ee}\right|_{\mathrm{rad}}&=-\gamma_1\rho_{ee},\\
    \left.\dot{\rho}_{gg}\right|_{\mathrm{rad}}&=+\gamma_1\rho_{ee},\\
    \left.\dot{\rho}_{eg}\right|_{\mathrm{rad}}&=-\frac{\gamma_1}{2}\rho_{eg}.
    \label{eq:app-rad-pop-coherence-rates}
\end{align}

\subsection{Pure dephasing}
\label{supp:pure-dephasing}

The pure-dephasing operator is chosen as
\begin{equation}
    \hat{L}_\phi=\hat{\sigma}_z
    =-\ket{g}\bra{g}+\ket{e}\bra{e},
    \qquad \hat{L}_\phi^\dagger \hat{L}_\phi=\mathbb{1}.
    \label{eq:app-C-phi}
\end{equation}
The dissipator therefore reduces to
\begin{equation}
    \mathcal{D}_\phi[\hat{\rho}]
    =\hat{\sigma}_z\hat{\rho}\hat{\sigma}_z-\hat{\rho}.
    \label{eq:app-dephasing-dissipator}
\end{equation}
Because \(\hat{\sigma}_z\ket{e}=+\ket{e}\) and
\(\hat{\sigma}_z\ket{g}=-\ket{g}\), the off-diagonal element changes sign on
both sides:
\begin{align}
    \bra{e}\hat{\sigma}_z\hat{\rho}\hat{\sigma}_z\ket{g}
    &= (+1)(-1)\bra{e}\hat{\rho}\ket{g}
    =-\rho_{eg}.
    \label{eq:app-dephasing-sandwich}
\end{align}
It follows that
\begin{equation}
    \left(\mathcal{D}_\phi[\hat{\rho}]\right)_{eg}
    =-\rho_{eg}-\rho_{eg}
    =-2\rho_{eg}.
    \label{eq:app-dephasing-coherence-result}
\end{equation}
The example specifies the channel rate as \(\gamma_\phi/2\), so that its
physical contribution to the coherence decay is
\begin{equation}
    \left.\dot{\rho}_{eg}\right|_\phi
    =\frac{\gamma_\phi}{2}
    \left(\mathcal{D}_\phi[\hat{\rho}]\right)_{eg}
    =-\gamma_\phi\rho_{eg}.
    \label{eq:app-pure-dephasing-rate}
\end{equation}
The factor of two is therefore a consequence of the chosen normalization of
the operator \(\hat{\sigma}_z\), not an additional physical process.

\subsection{Full equations of motion and closed-form solution}
\label{supp:full-two-level-eom}

The Hamiltonian contribution to the coherence is
\begin{align}
    -i[\hat{H},\hat{\rho}]_{eg}
    &=-i\bra{e}\hat{H}\hat{\rho}-\hat{\rho} \hat{H}\ket{g}\\
    &=-i\left(\omega_{eg}\rho_{eg}-0\right)
    =-i\omega_{eg}\rho_{eg}.
    \label{eq:app-H-coherence}
\end{align}
Combining the Hamiltonian, radiative, and pure-dephasing contributions gives
\begin{equation}
    \dot{\rho}_{eg}
    =-\left(i\omega_{eg}+\Gamma_2\right)\rho_{eg},
    \qquad
    \Gamma_2=\frac{\gamma_1}{2}+\gamma_\phi.
    \label{eq:app-coherence-eom}
\end{equation}
The corresponding solution is
\begin{equation}
    \rho_{eg}(t)
    =\rho_{eg}(0)\exp\left[-(i\omega_{eg}+\Gamma_2)t\right].
    \label{eq:app-coherence-solution}
\end{equation}
For the parameters of the quickstart model,
\begin{equation}
    \Gamma_2=\frac{0.04}{2}+0.03=0.05~\mathrm{eV}.
    \label{eq:app-Gamma2-number}
\end{equation}

The population equations follow from Eq.~\eqref{eq:app-rad-matrix-dissipator}:
\begin{align}
    \dot{\rho}_{ee}&=-\gamma_1\rho_{ee},\\
    \dot{\rho}_{gg}&=+\gamma_1\rho_{ee}.
    \label{eq:app-population-eom}
\end{align}
The trace is conserved because
\begin{equation}
    \frac{d}{dt}(\rho_{gg}+\rho_{ee})=0.
    \label{eq:app-trace-conservation}
\end{equation}
For an initially excited contribution with weight \(p_0\),
\begin{align}
    \rho_{ee}(t)&=p_0\exp(-\gamma_1t),\\
    \rho_{gg}(t)&=p_0\left(1-\exp(-\gamma_1t)\right).
    \label{eq:app-population-solution}
\end{align}

\newpage
\section{Coupled modes: the six scenarios of the Perspective}
\label{supp:perspective-scenarios}

Section~\ref{M-sec:performance} of the main text compares QuDPy-FDGF with the data deposited with the Perspective of Paiva-Ortega \emph{et al.}~\cite{paivaortega2026detection}, diagram by diagram, for the reference scenario (Sec.~\ref{M-sec:perf-published} of the main text). The deposit holds six scenarios, which differ in the transfer rate $k_{ba}$ between the two modes (0.1, 1, 5, 10, and 20~meV; 5~meV is the reference) and, in the last one, in the relaxation rate $k=20$~meV. In each of them the stored data are the moduli of the response of each Liouville diagram at ten points of the antidiagonal through each of the two diagonal peaks, and the half width of the Lorentzian fitted to them. QuDPy-FDGF is evaluated at the same ten frequencies, with the same estimator. Table~\ref{tab:scenarios} gives, for the coherent sum, the half widths stored and computed, and the largest deviation of the seven individual diagrams in half width and in shape. Over the six scenarios, the 96 half widths of the eight rows (the three coherent diagrams, their sum, and the four fluorescence diagrams, for the two peaks) agree within 1.4\,\% (0.4\,\% on average), the amplitudes within 8\,\% and the shapes within 3.2\,\% of the peak (the origin of the amplitude differences is not isolated; Sec.~\ref{M-sec:perf-published} of the main text).

\begin{table}[h]
\centering
\footnotesize
\setlength{\tabcolsep}{3.5pt}
\begin{tabular}{@{}lrrrrrrrr@{}}
\toprule
& \multicolumn{2}{c}{sum, $a$} & \multicolumn{2}{c}{sum, $b$} & \multicolumn{2}{c}{width (\%)} & \multicolumn{2}{c}{shape (\%)}\\
\cmidrule(lr){2-3}\cmidrule(lr){4-5}\cmidrule(lr){6-7}\cmidrule(lr){8-9}
scenario & stored & FDGF & stored & FDGF & $a$ & $b$ & $a$ & $b$\\
\midrule
$k_{ba}=0.1$~meV & 5.12 & 5.13 & 5.23 & 5.25 & 0.8 & 1.0 & 0.9 & 2.2 \\
$k_{ba}=1$~meV & 5.13 & 5.13 & 5.59 & 5.60 & 0.8 & 1.0 & 0.9 & 2.2 \\
$k_{ba}=5$~meV (ref.) & 5.16 & 5.16 & 7.10 & 7.10 & 0.8 & 0.8 & 0.8 & 2.5 \\
$k_{ba}=10$~meV & 5.18 & 5.19 & 8.91 & 8.90 & 0.8 & 0.7 & 0.7 & 2.7 \\
$k_{ba}=20$~meV & 5.21 & 5.21 & 12.66 & 12.66 & 0.8 & 1.4 & 0.6 & 3.2 \\
$k=20$~meV (relax.) & 13.63 & 13.63 & 15.65 & 15.65 & 0.3 & 1.1 & 1.7 & 3.0 \\
\bottomrule
\end{tabular}
\caption{The six scenarios of the Perspective. Sum: half widths (meV) of the coherent sum $R_1+R_2-R_3$ from the deposited data (stored) and from QuDPy-FDGF, for the two peaks; width and shape: for each peak, the largest deviation over the seven individual diagrams in half width and in shape (percent of the largest sample).}
\label{tab:scenarios}
\end{table}

\section{Cost versus Hilbert-space dimension}
\label{supp:scaling}

\begin{figure}[t]
\centering
\includegraphics[width=\linewidth]{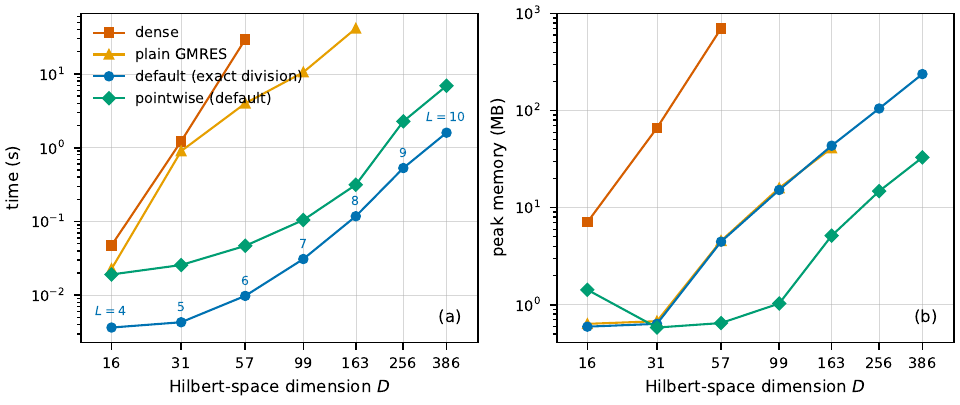}
\caption{Time (a) and peak memory (b) for the spin chain of Example~1 enlarged to $L=4$ to 10 sites ($D=16$ to 386), on a $9\times9$ subgrid. \emph{Dense}: dense backend, not run beyond $D=57$, where one Liouville matrix takes 0.17~GB. \emph{Plain GMRES}: sparse backend without preconditioner, not run beyond $D=163$. \emph{Default}: sparse backend with its defaults, i.e. the exact diagonal solve, a division with no iteration, so that this curve does not measure an iterative solver. \emph{Pointwise}: the default, evaluated point by point. All but the last use the factorized strategy. The time includes model construction; the memory is the peak of the process above its baseline. The numbers are in Table~\ref{tab:scaling}.}
\label{fig:scaling}
\end{figure}

To measure how the cost grows with the size of a model, the spin chain of Example~1 is enlarged from $L=4$ to $L=10$ sites at its published settings ($N_{\max}=4$, canonical state at 4~K, $\eta=0.03$~meV, rephasing ground-state bleach, stimulated emission, and excited-state absorption, $t_2=0$), so that $D=\sum_{N\leq4}\binom{L}{N}$ grows from 16 to 386. The task is a $9\times9$ subgrid of the published window. Figure~\ref{fig:scaling} shows the total time including model construction and the peak memory of the process for the dense backend, the sparse backend with plain GMRES, the sparse backend with its defaults (exact diagonal solve), and the latter with the point-by-point evaluation; the dense and plain-GMRES curves use the factorized strategy. The numbers, with the mean number of GMRES iterations, are in Table~\ref{tab:scaling}.

The dense backend stores the $D^2\times D^2$ generator, $16D^4$ bytes (0.17~GB at $D=57$, 1.5~GB at $D=99$, and 355~GB at $D=386$), together with its factorization, and it needs 0.7~GB and 29~s at $D=57$. The sparse backend stores only density vectors of size $D^2$ and the Krylov basis, so that its memory stays at a few tens of megabytes up to $D=163$ and the calculation reaches $D=386$ in 1.6~s with 238~MB. With plain GMRES its time is set by the number of iterations, which grows from 11 to 1\,787 with $D$ at fixed $\eta$; it is faster than the dense backend by a factor of 7 at $D=57$.

The default solver changes the nature of the solve and the gain it brings must be read accordingly. The model adapter supplies the Hamiltonian in its eigenbasis, so that the generator of this closed chain is diagonal in the dyad basis; the solver detects it and solves each shifted system by the element-wise division by $\eta-i(\omega-\Omega_\nu)$, with no iteration (\texttt{diagonal\_solve=True}). This is the evaluation of a sum over states, exact to rounding, and the large difference between the plain GMRES and default curves of Fig.~\ref{fig:scaling} (a factor 400 at $D=57$) measures how much an iterative solver loses on a system whose inverse is known, not the speed-up of a general-purpose algorithm. The time that remains at large $D$ is spent applying the interactions and the propagations, not solving. With collapse channels that move population or couple coherences, the generator is not diagonal, GMRES runs, and the diagonal preconditioner $\operatorname{diag}[(\eta-i\omega)\mathbb{1}-\mathcal{L}]^{-1}$ is an approximation. On 30 dissipative cases (a three-site chain with its dissipation scaled from 0.1 to 1000, and the six-site spin chain with collective decay or local dephasing from 0.001 to 10~meV, each at two values of $\eta$), it reduces the mean number of iterations to between $3\times10^{-4}$ and 0.76 of the plain count in 28 cases, and it increases it by 20 and 23\% in the two cases of local dephasing at 10~meV (see the API documentation).

The point-by-point column shows the price of the factorized evaluation. At $L=10$ it takes 6.9~s and 33~MB, against 1.6~s and 238~MB: the time is divided by 4 and the memory multiplied by 7, since the vectors of one side of the cut are stored for the whole grid. The time gain is smaller than with GMRES because the solves are now divisions; the factorization mainly saves applications of the interactions.

The growth of the iteration count as $\eta$ decreases is what the position of the poles suggests. When $\mathcal{L}$ is normal, the eigenvalues of $(\eta-i\omega)\mathbb{1}-\mathcal{L}$ are $\eta+\kappa_\nu-i(\omega-\Omega_\nu)$, so that near a resonance the distance to singularity is $\eta+\kappa_\nu$. The solver reports the relative residual of every solve, which is the control on which the results rely, and, as an indicator of the size of the condition number, the quantity $(\eta+|\omega|+\|\mathcal{L}\|)/\eta$. This quantity bounds the condition number only for a normal generator; for a dissipative generator the bound carries an additional factor up to $\sqrt{D_{\mathrm H}}$ (\ref{M-app:numerical-backends}), and we do not claim a quantitative relation between it and the number of iterations, which for a non-normal operator depends on more than the condition number. The spin chain has no dissipative channels, $\kappa_\nu=0$, and $\eta$ alone sets this distance: at $D=57$, the mean number of plain GMRES iterations per solve rises from 111 to 355, 1\,121, 2\,617, and 5\,886 as $\eta$ decreases from 0.3 to 0.1, 0.03, 0.01, and 0.003~meV, with every solve converged and agreement with the dense backend below $4\times10^{-11}$, and the time of the $3\times3$ subgrid rises from 0.14 to 6.2~s. In the dissipative cases we measured, the decay rates $\kappa_\nu$ add to $\eta$ and limit this growth once $\eta$ falls below them. With the diagonal preconditioner every GMRES solve takes one iteration at all of these $\eta$, in 0.01~s, and with the default division no iteration runs, in 7~ms; the agreement with the dense backend is below $4\times10^{-15}$ in both cases.

\paragraph{A model without sectors}
The spin chain conserves the magnon number, so that its Hamiltonian is block diagonal and the sectors keep $D$ small. A static transverse field $h_x\sum_jS^x_j$ couples the sectors $N$ and $N\pm1$, exactly as the probe does, and the model has no sector structure: the whole space, $D=2^L$, is diagonalized and given to the solver as one sector. We tested it on the spin chain of Example~1 with $h_x=0.30$~meV, on a $9\times9$ grid over the window of Example~1 with $\eta=0.03$~meV, on three BLAS threads. The generator of this closed model is diagonal in its eigenbasis, so that the solver divides ($18$ solves), and the calculation takes 16~ms and 6~MB at $L=6$ ($D=64$), 0.7~s and 0.11~GB at $L=8$ ($D=256$), and 21~s and 1.7~GB at $L=10$ ($D=1024$). The memory is that of the stored vectors, of size $D^2=4^L$, and \texttt{memory\_limit\_mb} trades it for time: 0.64~GB with a limit of 400~MB at the same time, and 0.26~GB with 128~MB in 98~s, because the other side of the cut is propagated again for each slab ($486$ solves instead of 18). At $L=11$ ($D=2048$), with a limit of 520~MB, the diagonalization takes 10~s and 0.6~GB, the peak is 1.0~GB, and the grid takes 676~s; $L=12$, whose vectors take 270~MB each, was not attempted on the 16~GB machine. Without the diagonal solve, plain GMRES needs on average 1\,570 iterations per solve at $L=5$ and 2\,005 at $L=6$, 3.5 and 12.6~s, 790 times the time of the division at $L=6$. A model without sectors is therefore limited to about a dozen sites on this machine, by the diagonalization and the vectors of the Liouville space, not by the algorithm of the resolvent.

\begin{table}[h]
\centering
\footnotesize
\setlength{\tabcolsep}{2.5pt}
\begin{tabular}{@{}rr@{\hspace{9pt}}rr@{\hspace{9pt}}rrr@{\hspace{9pt}}rr@{\hspace{9pt}}rr@{}}
\toprule
& & \multicolumn{2}{c}{dense} & \multicolumn{3}{c}{plain GMRES} & \multicolumn{2}{c}{default} & \multicolumn{2}{c}{pointwise}\\
\cmidrule(lr){3-4}\cmidrule(lr){5-7}\cmidrule(lr){8-9}\cmidrule(lr){10-11}
$L$ & $D$ & $t$ (s) & MB & $t$ (s) & MB & it. & $t$ (s) & MB & $t$ (s) & MB\\
\midrule
4 & 16 & 0.047 & 7 & 0.023 & 1 & 11 & 0.0037 & 1 & 0.019 & 1 \\
5 & 31 & 1.22 & 66 & 0.90 & 1 & 425 & 0.0043 & 1 & 0.026 & 1 \\
6 & 57 & 29.3 & 702 & 3.99 & 5 & 1213 & 0.0098 & 4 & 0.047 & 1 \\
7 & 99 & -- & -- & 10.5 & 16 & 1462 & 0.031 & 15 & 0.10 & 1 \\
8 & 163 & -- & -- & 41.3 & 41 & 1787 & 0.12 & 43 & 0.31 & 5 \\
9 & 256 & -- & -- & -- & -- & -- & 0.53 & 105 & 2.27 & 15 \\
10 & 386 & -- & -- & -- & -- & -- & 1.60 & 238 & 6.90 & 33 \\
\bottomrule
\end{tabular}
\caption{Time and peak memory for the spin chain of Example~1 enlarged to $L$ sites, on a $9\times9$ subgrid (Fig.~\ref{fig:scaling}). \emph{Dense}: dense backend. \emph{Plain GMRES}: sparse backend without preconditioner. \emph{Default}: sparse backend with its defaults, i.e. the exact diagonal solve, a division with no iteration. \emph{Pointwise}: the default, evaluated point by point. All but the last use the factorized strategy. $t$ is the total time (s), MB the peak memory above the baseline of the process, and ``it.'' the mean number of GMRES iterations per shifted solve. The dense backend was not run beyond $D=57$, where one Liouville matrix takes 0.17~GB, and plain GMRES not beyond $D=163$.}
\label{tab:scaling}
\end{table}
\clearpage

\section{Agreement of the dense and sparse backends}
\label{supp:backends}

The two backends evaluate the same generator by different numerical routes: the dense backend assembles the matrix of $\mathcal{L}$ and factorizes each shifted matrix, whereas the sparse backend applies $\mathcal{L}$ block by block and solves each shifted system by restarted GMRES. Their agreement therefore tests the matrix-free realization of \ref{M-app:sector-fdgf-realization} on the examples themselves. For the complete calculation of Listing~\ref{M-lst:twolevel}, with radiative decay, pure dephasing, and three observables, the largest difference is $2.6\times10^{-15}$ of the maximum of each map. For Example~1 at its published size ($D=57$, canonical state at 4~K), the backends agree to $3.4\times10^{-16}$ on a $5\times5$ subgrid of the published window with the defaults, where the sparse backend solves the shifted systems by division, and to $5.7\times10^{-11}$ with plain GMRES, a difference set by the GMRES tolerance. For the third-order response of Example~2 ($D=9$, three pathways, $\eta=10^{-5}$~eV), whose dissipative channels make the generator non-diagonal, the sparse backend runs GMRES with the diagonal preconditioner (3.4 iterations per solve on average on a $5\times5$ grid), and the backends agree to $1.9\times10^{-12}$ on a $31\times31$ grid and $2.2\times10^{-12}$ on the published $201\times201$ grid. For its closed fifth-order limit ($D=10$, seven pathways) the difference is $4.7\times10^{-16}$ on a $31\times31$ grid and $5.5\times10^{-16}$ on the $251\times301$ grid of Fig.~\ref{M-fig:cm_chi5_map}. The factorized evaluation of Sec.~\ref{supp:grid} and the evaluation point by point agree to $3.2\times10^{-15}$, $3.4\times10^{-16}$, $5.6\times10^{-12}$ and $2.4\times10^{-16}$ for the sparse backend in these four calculations.

\section{Evaluation of a grid}
\label{supp:grid}

The solver does not evaluate a grid point by point. It cuts each chain between the two scanned intervals, propagates the part before the cut from the initial state once per frequency of the first axis and the part after it backwards from the detection operator once per frequency of the second axis, and obtains the whole map from one matrix product (Sec.~\ref{M-FromLiouville} and \ref{M-app:grid-evaluation}). The number of shifted solves is then proportional to the number of points per axis instead of $2N^2$ for $N\times N$ points; the point-by-point evaluation (\texttt{strategy="pointwise"}) remains available and is the reference of the tests. This is an implementation choice and not a result, and Table~\ref{tab:grid} documents it for the grids of the examples. The quadratic term that remains is the contraction itself, $N^2$ scalar products of length $D^2$, which makes the time grow faster than $N$ when $D$ is large. The saving comes at the price of memory, since the vectors of one side are stored while the other is evaluated, and it applies to the chains that have two scanned intervals; the option \texttt{memory\_limit\_mb} processes the grid in slabs when this storage exceeds a limit, and the chains of the streamed side are then rebuilt for each slab (Sec.~\ref{supp:scaling}). The sparse backend has a cost per solve that dominates for models as small as these, so that for $D\leq10$ the dense backend is the faster one: it evaluates the published $201\times201$ map of the third-order form of Example~2 in 0.055~s and the $251\times301$ map of the closed limit in 0.095~s, against 4.1 and 0.40~s.

Evaluating a pathway point by point takes one chain of resolvents per grid point, that is $2N^2$ shifted solves for two scanned frequency intervals on $N\times N$ points, whereas the factorized evaluation of \ref{M-app:grid-evaluation} takes a number of solves proportional to $N$. The point-by-point evaluation is the reference of the tests and not a competitive method: Table~\ref{tab:grid} documents the solver and is not a result. The grids are those of the examples: 41 and 81 points per axis for the two-level system, 33 for Example~1, and 61 and 151 for Example~2.

\begin{table}[h]
\centering
\footnotesize
\setlength{\tabcolsep}{4pt}
\begin{tabular}{@{}llrrrr@{}}
\toprule
& & \multicolumn{2}{c}{point by point} & \multicolumn{2}{c}{factorized}\\
\cmidrule(lr){3-4}\cmidrule(lr){5-6}
case & grid & solves & $t$ (s) & solves & $t$ (s)\\
\midrule
Two-level, $D=2$, 3 pathways & $41\times41$ & 10\,086 & 3.56 & 82 & 0.22 \\
 & $81\times81$ & 39\,366 & 11.9 & 162 & 0.28 \\
Example~1, $D=57$, 3 pathways & $33\times33$ & 6\,534 & 0.51 & 66 & 0.023 \\
same, plain GMRES & $5\times5$ & 150 & 21.0 & 10 & 2.60 \\
same, dense backend & $5\times5$ & 150 & 929 & 10 & 16.2 \\
Example~2, $D=9$, 3 pathways & $61\times61$ & 22\,326 & 26.6 & 122 & 2.01 \\
 & $151\times151$ & 136\,806 & 163 & 302 & 4.20 \\
Example~2, closed, $D=10$, 7 pathways & $61\times61$ & 52\,094 & 2.68 & 122 & 0.14 \\
 & $151\times151$ & 319\,214 & 16.4 & 302 & 0.29 \\
\bottomrule
\end{tabular}
\caption{Shifted solves and total time for one scanned grid, evaluated point by point and with the factorized strategy (sparse backend with its defaults, i.e. with the exact diagonal solve for Example~1 and for the closed limit of Example~2, and GMRES with the diagonal preconditioner for the third-order Example~2, unless stated). The two-level system is the polarization of its three pathways, and the time is that of the complete spectrum including the propagations. The grids are those of the examples. The $151\times151$ rows of point-by-point evaluation are extrapolated from the $61\times61$ measurement as $N^2$.}
\label{tab:grid}
\end{table}

The time grows with a slope of 0.5 to 1.3 over $N\geq9$ instead of 1.6 to 2.0 (log-log fit of the measurements), the slope above one coming from the contraction. For the closed limit of Example~2, the $151\times151$ map takes 0.29~s with 302 solves, where the point-by-point evaluation would need 319\,214; for its third-order form, 4.2~s with 302 solves instead of 136\,806.

\section{A resolved map at equal resolution}
\label{supp:resolved}

The comparison with a time-domain route must be made at the same resolution and the same accuracy, and the resolution depends on the linewidth. A map is called resolved here when the grid spacing is a quarter of the full width at half maximum, $2\sqrt{3}(\Gamma_2+\eta)$, of the modulus of a Lorentzian pole, so that the number of points per axis follows from the width of the lines and is not chosen freely. Three routes are compared, all with the factorization of Sec.~\ref{supp:grid}: the direct evaluation of the library (dense backend), the sum over states (the generator diagonalized once, each resolvent a division in its eigenbasis), and a time route (the chain sampled with the one-step propagator up to a length $T$ and transformed with the kernel $\exp[(i\omega-\eta)t]$, the cheapest step and length being searched for each target error). The cases are the two-level system and Example~2 in its third-order and closed fifth-order forms; the routes are defined, and Table~\ref{tab:resolved} gives the results, below. For a strongly damped line the direct evaluation and the time route are comparable at a tolerance of $10^{-2}$; the direct evaluation is faster as soon as the tolerance is tighter or the line narrower, by a factor of 2.4 to 190 in the cases tested, and the time route does not reach $10^{-4}$ within 9\,000 samples per axis for lines 0.56~meV wide. This is an advantage in accuracy and in not having to choose $(\Delta t,T)$, and not in a cost independent of the linewidth: the number of points of a resolved map grows as the lines narrow, and the direct evaluation grows with it (14~ms at 84 points per axis and 0.24~s at 818 for the same model). The sum over states is faster than both for these small models; it is the method of choice when the generator can be diagonalized (this is the idea of the Liouvillian-eigenbasis route of UF$^2$ \cite{rose2019uf2,rose2021uf2open}, here without finite pulses), and the direct evaluation is the one that remains available in the sparse backend beyond that limit. The direct evaluation also accepts any list of frequencies, such as a non-uniform grid refined near the peaks, whereas an FFT route is bound to a uniform one.

\paragraph{Scope of the comparison}
Neither time-domain route of this section is tuned. The original QuDPy of Sec.~\ref{M-sec:perf-published} is used as released, and the factorized time route defined below was written for this comparison: it shares the factorization of Sec.~\ref{supp:grid}, but it forms the full $N_t\times N_t$ matrix of samples before transforming it. Transforming the two families of propagated vectors first and contracting them afterwards gives the same maps to $3\times10^{-13}$ of their maximum and, in our tests, a time shorter by a factor of 3 to 12. The speed ratios of Table~\ref{M-tab:perspective-qudpy} of the main text and of Table~\ref{tab:resolved} are therefore those of these implementations. They are not a bound on what a time-domain method can reach, and no general advantage of the frequency domain over the time domain is claimed. What does not depend on the implementation is that the direct evaluation has no window and no time step to choose.

The routes compared above are the following; all of them use the factorization of Sec.~\ref{supp:grid}, so that they propagate once per axis and not once per grid point. A map is resolved when the grid spacing is a quarter of the full width at half maximum, $2\sqrt{3}(\Gamma_2+\eta)$, of the modulus of a Lorentzian pole.
\begin{itemize}
\item \emph{Direct evaluation}: the shifted solves of the library, with the dense backend, which is the relevant one for these small models.
\item \emph{Sum over states}: the generator is diagonalized once and each resolvent is a division in its eigenbasis, $\mathcal{R}_\eta(\omega)=V\,\mathrm{diag}[(\eta-i\omega-\lambda_\nu)^{-1}]\,V^{-1}$. This is the textbook formula; it is limited by the $16D^4$ bytes and the $D^6$ operations of the diagonalization, which is the limit of the dense backend.
\item \emph{Time route}: the vectors of both sides of the cut are sampled with the one-step propagator $\exp(\mathcal{L}\Delta t)$ up to a length $T$, and transformed with the kernel $\exp[(i\omega-\eta)t]$ and trapezoidal weights, by a matrix product on the grid or by a zero-padded two-dimensional FFT, whichever is cheaper. The time step may exceed the Nyquist step of the carrier, since the signal lies in a band of 1~eV at most (bandpass sampling), and the error measured, against the reference map on the same grid, which is the direct evaluation, includes any aliasing. For each target error the cheapest $(\Delta t,T)$ is searched.
\end{itemize}
The two-level system of the listing ($R_1$ and $R_2$, $t_2=10$~eV$^{-1}$, window of 1~eV per axis) is taken at its published rates ($\Gamma_2+\eta=52$~meV, 23 points per axis). Example~2 is taken in its third-order form (three pathways, window of 0.4~eV per axis) with the rates of the example, whose narrowest coherence has a width of 5.6~meV ($84\times84$ points), and with the rates divided by ten (0.56~meV, $818\times818$ points); and in its closed fifth-order form ($D=10$, seven pathways, the window of Fig.~\ref{M-fig:cm_chi5_map}) at $\eta=2$ and $1$~meV, with a spacing equal to $\eta$ ($251\times151$ and $501\times301$ points). Table~\ref{tab:resolved} gives the results.

\begin{table}[h]
\centering
\scriptsize
\setlength{\tabcolsep}{4.5pt}
\begin{tabular}{@{}ll@{\hspace{8pt}}rr@{\hspace{8pt}}rr@{\hspace{6pt}}rr@{\hspace{6pt}}rr@{}}
\toprule
& & \multicolumn{2}{c}{time (ms)} & \multicolumn{6}{c}{time route, for the target error}\\
\cmidrule(lr){3-4}\cmidrule(lr){5-10}
& & & & \multicolumn{2}{c}{$10^{-2}$} & \multicolumn{2}{c}{$10^{-3}$} & \multicolumn{2}{c}{$10^{-4}$}\\
\cmidrule(lr){5-6}\cmidrule(lr){7-8}\cmidrule(lr){9-10}
case & grid & direct & \shortstack[r]{sum over\\states} & $N_t$ & ms & $N_t$ & ms & $N_t$ & ms\\
\midrule
Two-level, 52~meV & $23\times23$ & 0.79 & 0.13 & 84 & 0.45 & 444 & 7.2 & 997 & 34 \\
Ex.~2, 5.6~meV & $84\times84$ & 14 & 3.2 & 195 & 7.3 & 1036 & 126 & 2331 & 521 \\
Ex.~2, 0.56~meV & $818\times818$ & 237 & 71 & 1911 & 565 & 2548 & 1389 & -- & -- \\
Ex.~2 closed, $\eta=2$~meV & $251\times151$ & 59 & 30 & 2160 & 1367 & 2160 & 1367 & 6477 & 11171 \\
Ex.~2 closed, $\eta=1$~meV & $501\times301$ & 153 & 67 & 4318 & 4821 & 5757 & 8536 & 6477 & 11129 \\
\bottomrule
\end{tabular}
\caption{Resolved maps: the direct evaluation (dense backend), the sum over states, and the cheapest of the factorized time routes written for this comparison (scope paragraph above) that reaches a relative error of $10^{-2}$, $10^{-3}$, and $10^{-4}$ with respect to the reference map (for each target error, the number of samples per axis $N_t$ and the time of the route in ms; ``--'': not reached with up to 9\,000 samples per axis). The first row is the two-level system of the listing and the next four are Example~2 (third order with the rates of the example and with the rates divided by ten, then the closed fifth-order limit at two values of $\eta$); the first number of each label is the width of the narrowest coherence. The sum over states agrees with the reference map to $10^{-12}$ or better. At $\eta=2$~meV the samples that reach $10^{-2}$ already reach $10^{-3}$ (error $9\times10^{-4}$). With the sparse backend, the direct evaluation of the same maps takes between 0.1 and 14~s.}
\label{tab:resolved}
\end{table}

For a strongly damped line the two routes are comparable at a tolerance of $10^{-2}$: for the two-level system an FFT on 84 samples per axis takes 0.45~ms, against 0.8~ms for the direct evaluation, and for Example~2 an FFT on 195 samples takes 7.3~ms, against 14~ms. The direct evaluation is faster as soon as the tolerance is tighter or the line narrower: by a factor of 9 to 43 at $10^{-3}$ and $10^{-4}$ for the two-level system, and of 9 to 37 for Example~2. With the rates divided by ten the time route needs 1\,911 samples per axis for $10^{-2}$ and takes 2.4 times as long as the direct evaluation (0.57~s against 0.24~s), 5.9 times for $10^{-3}$, and does not reach $10^{-4}$ with 9\,000 samples. In the closed limit, where the lines are as narrow as the apodization, the time route needs 2\,160 samples per axis and 1.4~s for $10^{-3}$ at $\eta=2$~meV, 23 times the direct evaluation (59~ms), and 6\,477 samples and 11~s for $10^{-4}$, 190 times; at $\eta=1$~meV the factors are 56 and 73. The time route needs a window $T$ long enough for the signal to decay below the tolerance and a step fine enough to avoid aliasing, so that its number of samples per axis grows from 84 to 6\,477 as the lines narrow and the tolerance tightens. The direct evaluation has no such parameter: its result at a frequency is the response with the broadening $\eta$, whatever the width, and its error is that of the linear solve. This is an advantage in accuracy and in not having to choose $(\Delta t,T)$, and not in a cost independent of the linewidth: the number of points of a resolved map grows as the lines narrow, and the direct evaluation grows with it (14~ms at 84 points per axis and 0.24~s at 818 for the same model). The sum over states is faster than both for these small models, since it replaces the solves by divisions (0.1 to 71~ms).

\bibliographystyle{elsarticle-num}
\bibliography{references}